\documentclass[11pt]{article}
\usepackage{amsmath,amsfonts,amssymb,epsfig,color,psfrag}
\usepackage{rotating}
\begin{document}
\title{Geometric Complexity Theory V: Efficient algorithms for Noether Normalization\footnote{To appear in the Journal of the AMS.}}
\author{Dedicated to Sri Ramakrishna \\ \\
Ketan D. Mulmuley \footnote{This work was supported by NSF grant CCF-1017760. This article is the full version of its FOCS 2012 extended abstract \cite{GCT5focs}.}
 \\
The University of Chicago \\ mulmuley@uchicago.edu}

\maketitle

\newtheorem{prop}{Proposition}[section]
\newtheorem{claim}[prop]{Claim}
\newtheorem{goal}[prop]{Goal}
\newtheorem{theorem}[prop]{Theorem}
\newtheorem{hypo}[prop]{Hypothesis}
\newtheorem{guess}[prop]{Guess}
\newtheorem{problem}[prop]{Problem}
\newtheorem{axiom}[prop]{Axiom}
\newtheorem{question}[prop]{Question}
\newtheorem{remark}[prop]{Remark}
\newtheorem{lemma}[prop]{Lemma}
\newtheorem{claimedlemma}[prop]{Claimed Lemma}
\newtheorem{claimedtheorem}[prop]{Claimed Theorem}
\newtheorem{cor}[prop]{Corollary}
\newtheorem{defn}[prop]{Definition}
\newtheorem{ex}[prop]{Example}
\newtheorem{conj}[prop]{Conjecture}
\newtheorem{obs}[prop]{Observation}
\newtheorem{phyp}[prop]{Positivity Hypothesis}
\newcommand{\bitlength}[1]{\langle #1 \rangle}
\newcommand{\ca}[1]{{\cal #1}}
\newcommand{\floor}[1]{{\lfloor #1 \rfloor}}
\newcommand{\ceil}[1]{{\lceil #1 \rceil}}
\newcommand{\gt}[1]{{\langle  #1 |}}
\newcommand{\C}{\mathbb{C}}
\newcommand{\F}{\mathbb{F}}
\newcommand{\N}{\mathbb{N}}
\newcommand{\R}{\mathbb{R}}
\newcommand{\Z}{\mathbb{Z}}
\newcommand{\frcgc}[5]{\left(\begin{array}{ll} #1 &  \\ #2 & | #4 \\ #3 & | #5
\end{array}\right)}

\newcommand{\cgc}[6]{\left(\begin{array}{ll} #1 ;& \quad #3\\ #2 ; & \quad #4
\end{array}\right| \left. \begin{array}{l} #5 \\ #6 \end{array} \right)}

\newcommand{\wigner}[6]
{\left(\begin{array}{ll} #1 ;& \quad #3\\ #2 ; & \quad #4
\end{array}\right| \left. \begin{array}{l} #5 \\ #6 \end{array} \right)}

\newcommand{\rcgc}[9]{\left(\begin{array}{ll} #1 & \quad #4\\ #2  & \quad #5
\\ #3 &\quad #6
\end{array}\right| \left. \begin{array}{l} #7 \\ #8 \\#9 \end{array} \right)}

\newcommand{\srcgc}[4]{\left(\begin{array}{ll} #1 & \\ #2 & | #4  \\ #3 & |
\end{array}\right)}

\newcommand{\arr}[2]{\left(\begin{array}{l} #1 \\ #2   \end{array} \right)}
\newcommand{\unshuffle}[1]{\langle #1 \rangle}
\newcommand{\ignore}[1]{}
\newcommand{\f}[2]{{\frac {#1} {#2}}}
\newcommand{\tableau}[5]{
\begin{array}{ccc} 
#1 & #2  &#3 \\
#4 & #5 
\end{array}}
\newcommand{\embed}[1]{{#1}^\phi}
\newcommand{\comb}[1]{{\| {#1} \|}}
\newcommand{\stab}{{\mbox {stab}}}
\newcommand{\perm}{{\mbox {perm}}}
\newcommand{\trace}{{\mbox {trace}}}
\newcommand{\polylog}{{\mbox {polylog}}}
\newcommand{\sign}{{\mbox {sign}}}
\newcommand{\proj}{{\mbox {Proj}}}
\newcommand{\height}{{\mbox {ht}}}
\newcommand{\poly}{{\mbox {poly}}}
\newcommand{\std}{{\mbox {std}}}
\newcommand{\m}{\mbox}
\newcommand{\formula}{{\mbox {Formula}}}
\newcommand{\circuit}{{\mbox {Circuit}}}
\newcommand{\core}{{\mbox {core}}}
\newcommand{\orbit}{{\mbox {orbit}}}
\newcommand{\sie}{{\mbox {sie}}}
\newcommand{\pie}{{\mbox {pie}}}
\newcommand{\cycle}{{\mbox {cycle}}}
\newcommand{\ideal}{{\mbox {ideal}}}
\newcommand{\qed}{{\mbox {Q.E.D.}}}
\newcommand{\proof}{\noindent {\em Proof: }}
\newcommand{\weight}{{\mbox {wt}}}
\newcommand{\tab}{{\mbox {Tab}}}
\newcommand{\level}{{\mbox {level}}}
\newcommand{\vol}{{\mbox {vol}}}
\newcommand{\vect}{{\mbox {Vect}}}
\newcommand{\val}{{\mbox {wt}}}
\newcommand{\sym}{{\mbox {Sym}}}
\newcommand{\convex}{{\mbox {convex}}}
\newcommand{\spec}{{\mbox {spec}}}
\newcommand{\strong}{{\mbox {strong}}}
\newcommand{\adm}{{\mbox {Adm}}}
\newcommand{\eval}{{\mbox {eval}}}
\newcommand{\for}{{\quad \mbox {for}\ }}
\newcommand{\Q}{\mathbb{Q}}
\newcommand{\mand}{{\quad \mbox {and}\ }}
\newcommand{\invlim}{{\mbox {lim}_\leftarrow}}
\newcommand{\directlim}{{\mbox {lim}_\rightarrow}}
\newcommand{\sformal}{{\cal S}^{\mbox f}}
\newcommand{\vformal}{{\cal V}^{\mbox f}}
\newcommand{\crystal}{\mbox{crystal}}
\newcommand{\conje}{\mbox{\bf Conj}}
\newcommand{\graph}{\mbox{graph}}
\newcommand{\ind}{\mbox{index}}

\newcommand{\rank}{\mbox{rank}}
\newcommand{\id}{\mbox{id}}
\newcommand{\str}{\mbox{string}}
\newcommand{\RSK}{\mbox{RSK}}
\newcommand{\wt}{\mbox{wt}}
\setlength{\unitlength}{.75in}

\begin{abstract} 
We study a basic algorithmic  problem in   algebraic geometry, which we call  NNL,  of constructing
a normalizing map as per Noether's Normalization Lemma.
For general explicit varieties, as formally defined in this paper,
we give a randomized polynomial-time
Monte Carlo algorithm for this problem.
For some interesting  cases of 
explicit varieties, we  give deterministic quasi-polynomial time algorithms.
These  may be contrasted with  the  standard 
EXPSPACE-algorithms for these problems in computational algebraic geometry.

In particular, we show that:
\smallskip

(1)  The categorical quotient  for
any  finite dimensional  representation $V$  of $SL_m$, with constant $m$,
is explicit in characteristic zero. 

(2) NNL for this categorical quotient  can be solved deterministically in time quasi-polynomial in the dimension of 
$V$.

(3)  The categorical quotient of the space of $r$-tuples
of $m \times m$ matrices by 
the simultaneous conjugation action of $SL_m$ is 
explicit  in any characteristic. 

(4) NNL for this categorical quotient  can be solved 
deterministically in time quasi-polynomial in $m$ and $r$ in any characteristic $p \not \in [2,\floor{m/2}]$.

(5) NNL for every  explicit variety in zero or large enough characteristic
can be solved  deterministically in  quasi-polynomial time, assuming 
the  hardness hypothesis for the permanent in geometric complexity theory.

The last result leads to a  geometric complexity theory approach to  put NNL for every explicit variety in P.
\end{abstract}

\section{Introduction} \label{sintro} 
Noether's Normalization Lemma (NNL), proved by Hilbert \cite{hilbert}, is the basis of 
a large number of foundational
results in algebraic geometry, such as  Hilbert's Nullstellensatz. It also lies at the heart of the foundational 
classification problem of algebraic geometry.
For any  projective variety  $W \subseteq P(K^l)$, 
where $K$ is an algebraically closed field   and $P(K^l)$ is 
the projective space associated with $K^l$, the  lemma  says 
that any homogeneous, generic linear map $\psi: K^l \rightarrow K^k$, for any $k \ge \dim(W)+1$, 
induces a regular (well defined) map  on $W$ (this means $\psi$ does not vanish identically on the line through the origin
in $K^l$ corresponding
to any  point in $W$). Furthermore, for any such $\psi$, (1)
$\psi(W) \subseteq P(K^k)$, the image of $W$, is closed in $P(K^k)$, and (2)
the fiber $\psi^{-1}(p)$, for any point $p \in \psi(W)$, is a finite set. 
Accordingly, we call a  homogeneous linear map $\psi: K^l \rightarrow K^k$, $k\ge \dim(W)+1$, that induces 
a regular map on $W$
a {\em normalizing map} for $W$.
In the  context of the main results of this paper, $l$ here will be   exponential in $\dim(W)$, and  $k$ will be polynomial in $\dim(W)$.
In this case,  Noether's Normalization Lemma 
expresses the variety $W$, embedded in the ambient space $P(K^l)$ of exponential dimension, as 
a finite cover of the variety $\psi(W)$,
embedded in the ambient space $P(K^k)$ of polynomial dimension.
This is its main significance from the complexity-theoretic perspective.
We also  refer to the problem of  constructing a normalizing map $\psi$,  with $k=\poly(\dim(W))$,
as NNL in short. 
This is the problem that is studied in this article for the
varieties $W$ that are  given explicitly in a sense that will be made precise.
We do not require $k=\dim(W)+1$ here, and allow a polynomial slack,  for the reasons explained  in Section~\ref{sNNLdetintro}.

In algebraic geometry, the phrase ``explicitly'' is used  informally. 
In this article, it is 
interpreted formally, from the complexity-theoretic perspective, to mean
``using algebraic circuits that can be computed in deterministic polynomial time''.
Thus, we formally introduce in this article the notion of an {\em explicit}
family 
$\{W_n\}$ of 
varieties (Definition~\ref{dexpvariety}) of $\poly(n)$
dimension 
that can be specified succinctly and uniformly by $\poly(n)$-time-computable 
algebraic circuits (cf. Section~\ref{sterminology}) 
of $\poly(n)$ degree having a specification of 
$\poly(n)$ bit-length, 
even though the dimension $l_n$ of the ambient space  containing $W_n$ 
can be  exponential in $n$. If $W_n$ is projective,  we let $l_n$  be one plus   the dimension of the ambient space.
If the family $\{W_n\}$ is explicit,  we also 
say that the variety $W_n$ is explicit, with the understanding that
$n\rightarrow \infty$ in all complexity bounds.
It turns out that (cf. Section~\ref{sexample}) a large class of varieties that arise in practice
are  explicit in this formal complexity-theoretic sense.

The problem NNL for such explicit varieties $W_n$ (cf. Definition~\ref{desopexp}) 
is the problem of constructing a specific kind (cf. Sections~\ref{sNNLdetintro} and \ref{simpliexp}) of 
a normalizing map $\psi_n: K^{l_n} \rightarrow K^{k_n}$ for $W_n$,
with $k_n=\poly(n)$, having  a  
succinct specification of $\poly(n)$ bit-length.

For general   explicit varieties $W_n$, 
the standard  algorithm for NNL  (cf. Section~\ref{sexpunconddet}), 
based on Gr\"obner basis theory \cite{mayr2},
takes in the worst case  work-space  that is polynomial in $l_n$,
and time that is exponential in $l_n$.
In our context $l_n$, in general,  is exponential in $n$, and hence,  this work-space bound is  exponential in $n$, i.e., $O(2^{\poly(n)})$, 
and the time bound  is double exponential in $n$.
This  shows that   NNL for explicit varieties is in EXPSPACE.
Assuming the Generalized  Riemann Hypothesis, it can be shown to be in EXPH (cf. Section~\ref{sexpunconddet}).
Here EXPSPACE denotes 
the  complexity class of problems that can be solved using exponential work-space in double exponential time,
and EXPH denotes  the exponential hierarchy \cite{arora}.
Informally, these stand for the classes 
 of problems  that are computationally highly intractable (far more intractable
than the problems in the class  NP, which  can be solved in exponential time and 
polynomial space).
Thus, on the basis of the existing literature in computational algebraic geometry,
it may appear that NNL for explicit varieties is highly intractable, and perhaps,
even inherently so.

The  algorithmic results in this article indicate that this is not  the case.

First, it is shown in this article (cf. Theorem~\ref{tmonteintroexplicitnew}) that  NNL for any explicit variety $W_n$ can be solved 
by a $\poly(n)$-time randomized Monte Carlo algorithm, whose output is correct with a high probability. This means, in practice, NNL for explicit varieties 
can be solved efficiently and correctly with
a high probability. 
But this  does not show that  NNL for explicit varieties 
is in $\mbox{BPP} \subseteq \mbox{PSPACE}$, for the reasons explained in Section~\ref{smonteintro}.
Hence, it does not affect the  current EXPSPACE-status of NNL, or the EXPH status assuming the Generalized Riemann Hypothesis.

So we ask if NNL for any explicit variety 
can be solved  {\em deterministically} in  polynomial
time, thereby bringing it down from EXPSPACE to P.  We say that NNL for an explicit variety has an
{\em explicit solution}, in  the complexity-theoretic sense, if it can be solved in deterministic polynomial time.
The   motivation for  such an explicit solution comes from the 
foundational  classification problem of algebraic geometry
(cf. Section 1.8 in \cite{hartshorne}). Its  goal \cite{dieudonne}
is to classify a given  algebraic variety by transforming it regularly
into some   {\em canonical} normal form. Without any
relaxation, this goal may be infeasible, since it is not even known
at present if the isomorphism problem for algebraic varieties 
is decidable \cite{totaro}. 
Hence, our goal is to do the best that we can from the complexity-theoretic perspective.
As a first step in this direction,
one would like
an ``explicit'' normalizing map for an ``explicitly'' given
variety. A random normalizing map is not enough  in this context, 
since randomness is the opposite of canonicity.
Solving NNL in deterministic polynomial time is 
this first step towards   the classification problem of algebraic
geometry, interpreted 
from the complexity-theoretic perspective. 
For the reasons explained in  Section~\ref{sdiscuss}, 
this  turns out to be  far  harder than solving NNL in randomized 
polynomial time by a Monte Carlo algorithm.
We turn to this harder problem  next. 

It is  shown in this article that, for some interesting cases of explicit varieties $W_n$, NNL can indeed 
be solved deterministically
in quasi-$\poly(n)$-time, i.e., in $O(2^{(\log n)^c})$ time, for some 
constant $c>1$. (Here it is assumed
that the dimension $k_n$ of the
target space of the constructed normalization map $\psi_n: K^{l_n} \rightarrow K^{k_n}$ is also quasi-polynomial in $n$.)
Thus, for these explicit varieties,  NNL can be brought down from EXPSPACE to quasi-P,
the class of problems that can be solved deterministically in quasi-polynomial time.

The first such case of an explicit variety is the categorical quotient \cite{mumford}  $V/G$ associated with 
any finite dimensional  (rational) representation $V$
of $G=SL_m$, with constant $m$, in characteristic zero. (By a rational 
representation, we mean that 
the entries of the representation matrix are rational
functions of the coordinates of $G$. We will only be concerned with such representations in this article.)
Here $V/G=\spec(K[V]^G)$ \footnote{By 
abuse of notation, $\spec$ in this article  really means $\mbox{max-spec}$; cf.
\cite{eisenbudbook} (page 54).}
 is the variety 
whose  coordinate ring is $K[V]^G \subseteq K[V]$,  where $K[V]$ 
denotes the coordinate ring of $V$, and $K[V]^G$ its subring of $G$-invariants.
Explicitness of this variety  is by itself a  key result in this article.
It means (cf. Theorem~\ref{tintroexplicitconstantnew})  that a succinct encoding, in the form of 
a symbolic determinant,  of a set of  (exponentially many) generators
for this invariant ring can be constructed 
in $\poly(n)$ time, where $n$ is the dimension of $V$.

This succinct and efficient encoding of generators 
is in the spirit of the 
encodings that were  used in the so-called symbolic method of classical invariant theory (cf. Chapter 8 A in \cite{weyl}).
For example, the First Fundamental Theorem  for the ring of vector invariants 
proved by Weyl \cite{weyl} (cf. Theorem 2.6 A therein) 
implies  such a polynomial-time-computable succinct encoding, in the form of a symbolic 
determinant, of a set of (exponentially
many) generators for this
invariant ring. The problem of proving similar  First Fundamental Theorems
for invariant rings
has been  studied intensively in the last century; cf. Section 9
in \cite{popovbook} for a survey. 
This classical problem 
is interpreted in this article (cf. Definition~\ref{dexpffthilbert} (d)), from the complexity-theoretic perspective, as 
the problem of constructing an explicit 
encoding, in the form of a symbolic determinant or  a circuit, of a set of generators
of the invariant ring, where  {\em explicit} means polynomial-time-computable.
Classical invariant theory
did not  specify formally  what   ``explicit'' means. 

For the variety $V/G$
associated with  any  $n$-dimensional representation $V$ 
of $G=SL_m$, with constant $m$, in characteristic zero,  it is shown in this article that NNL can be solved deterministically 
in $O(n^{O(\log \log n)})$ time; cf. Theorem~\ref{thilbertnew}.

Noether's Normalization Lemma  was, in fact,  proved by Hilbert
\cite{hilbert} to give an algorithm 
for  constructing a finite set of generators 
for  the invariant ring $K[V]^G$ in this context.
Hilbert did  not prove any explicit upper bound on its  running time, 
or on the degrees of the generators.
Such a bound on the degrees 
was proved in  Popov \cite{popov} a century later,
and improved significantly in Derksen \cite{derksen}.
This improved analysis yields  an  exponential-time algorithm for computing a set of generators 
for the ring $K[V]^G$ of invariants 
for  any finite dimensional representation $V$ of $G=SL_m$ and, in conjunction with Gr\"obner basis theory
\cite{mayr2},  an  EXPSPACE-algorithm for  NNL for this invariant ring.
This algorithm for constructing
a set of generators requires  time that is exponential in the dimension of $V$, 
and the algorithm for NNL requires  exponential work-space 
and double exponential time,  even when  $m$  is constant.
Hilbert's paper focused mainly on  the case when $m$  is three, since
an algorithm to construct a finite set of generators
was not known before even in this case; cf. Section~\ref{sgeneralintro}.

Explicitness  for constant $m$ of the categorical quotient  associated with 
this invariant ring $K[V]^G$ (Theorem~\ref{tintroexplicitconstantnew})
implies that 
the problem  of computing an  encoding, in the form of a symbolic determinant,
of a set of generators for this invariant ring  is in P.
Thus there has been a rather remarkable change in  the status of this fundamental problem 
of invariant theory over  the course of a century from a problem 
that was not even known to be computable 
before Hilbert to a problem that is now in P, as shown in this article;  cf. Section~\ref{sgeneralintro}.

The quasi-polynomial-time deterministic algorithm in this article for NNL for this invariant ring
(cf. Theorem~\ref{thilbertnew}),
for constant $m$,   brings  the  original instance of NNL in Hilbert's paper in this case 
from EXPSPACE to quasi-P. 
Analogous results hold for any connected, reductive, algebraic group of constant dimension (cf. Theorem~\ref{tsemisimple}).

The second case of an explicit variety that we consider is the categorical quotient $V/G$ \cite{mumford}  associated with
the space $V=M_m(K)^r$ of $r$-tuples of $m\times m$ matrices over $K$, with 
the  simultaneous conjugation-action  of $G=SL_m$ (without any restriction on $m$ this time). 
It is shown in this article that this variety is  explicit in characteristic
zero, and is  explicit in a relaxed sense in positive  characteristic (cf. Theorem~\ref{tintroexplicitmatrixnew}).

Furthermore (cf. Theorem~\ref{tmatrixNew}), 
NNL for this variety can be solved deterministically 
in quasi-$\poly(m,r)$  time
in any  characteristic $p \not \in [2,\floor{m/2}]$, thereby bringing NNL in this case too from EXPSPACE to quasi-P.
This extends the same result in characteristic zero that is implied,  as pointed out by Forbes and
Shpilka \cite{fs3},  by  a variant of a conditional result 
in the preliminary version \cite{GCT5focs} of this article, in conjunction with
their earlier work  \cite{fs2} on arithmetic circuits;
cf. Remark 1  in Section~\ref{sprooftech}.

More generally  (cf. Theorems~\ref{tintronnlexplicitnew}, \ref{tnnlexpfull},  and Section~\ref{sposcharequi}), 
NNL for any explicit variety in zero or large enough characteristic
can be solved deterministically in quasi-polynomial time, thereby bringing it from
EXPSPACE to quasi-P, assuming the  hardness hypothesis for the permanent 
in   geometric complexity theory.
This hypothesis proposed in  \cite{GCT1} (or rather its stronger variant)  is that 
the permanent of $n\times n$ matrices 
cannot be approximated infinitesimally closely  by symbolic determinants over $K$
of  $O(2^{n^\epsilon})$ size, for some constant $\epsilon >0$, as $n\rightarrow \infty$.
It is an algebraic geometric  strengthening 
of the fundamental $\mbox{VP} \not = \mbox{VNP}$ conjecture in the work of  Valiant \cite{valiant2}.

In B\"urgisser \cite{burg2}, 
some other consequences of this hardness hypothesis 
have been derived, which also crucially
rely on the fundamental result of Kaltofen \cite{kaltofen,kaltofen2}, 
as in this paper. Consulting \cite{GCT1,burg2,GCThyderabad,GCT2,manivel}
(and especially Section 9.3  in \cite{manivel} that  explains
in detail the precise relationship of
the work in \cite{burg2} with the earlier work in \cite{GCT1}) 
may help the reader to   understand    this hypothesis better.

\subsection{Geometric complexity theory approach to the  basic algorithmic  problems in algebraic geometry and invariant theory} \label{sgctapproach}
The results described above lead to the following  geometric complexity theory
approach to the  basic algorithmic  problems of  algebraic geometry and invariant theory under consideration, namely, 
(1) the problem NNL, and
(2) the  problem of constructing a  set of generators for the  ring of invariants 
of a reductive group. Both these problems are motivated by  Hilbert \cite{hilbert}.

The  goal of the approach in the context of the first problem is to show that it 
is in P for every explicit variety. The approach is to (1)  first prove  the  hardness hypothesis \cite{GCT1} for the
permanent in geometric complexity theory (or its weaker form, cf. Theorem~\ref{tequivnnlnew}), 
then  (2) use the results in this article 
to show that NNL for every explicit variety is in quasi-P, in zero or large enough characteristic
(cf. Theorem~\ref{tnnlexpfull} and Section~\ref{sposcharequi}), and (3) finally, 
remove the quasi-prefix and the characteristic restriction  by proving 
a stronger form of the hardness hypothesis (cf. Sections~\ref{sremove} and \ref{sposcharequi}).
An  approach to prove the required hardness hypothesis in geometric complexity theory
will be  given in the sequel to this article
(the revised version of \cite{GCT6}). 

(N.B. The current version of \cite{GCT6} on the arxiv has become   outdated in view of
the recent result  \cite{ikenmeyer} that the occurrence-based obstructions 
in \cite{GCT2},   based on the vanishing of the rectangular Kronecker coefficients,
cannot be used to prove superpolynomial lower bounds for the permanent. The approach in
the revised version of \cite{GCT6} will be based on the far more powerful multiplicity-based obstructions,
to which this negative result does not apply.)

To put NNL in quasi-P, one does not need the hardness   hypothesis in geometric complexity theory,
or even its weaker form, in full strength  for all explicit varieties.
For explicit varieties of intermediate difficulty, such as explicit 
categorical quotients,  one only needs 
weaker  complexity-theoretic hypotheses for  the classes of circuits depending on the varieties.
Thus one can  approach this   goal  step by step,
increasing the hardness   of the varieties  in tandem 
with the strength of the circuits; cf. Sections~\ref{sexplicit} and \ref{sextensions}.

The  goal  of the approach  in the context of   the second  problem  is to show that it  is  in P 
for every  invariant ring $K[V]^G$, for  any finite dimensional representation $V$
of a connected reductive group $G$ in characteristic zero, and after a relaxation, for any reductive group 
in any characteristic,  allowing  encoding of a  set of generators by 
algebraic circuits;  cf. Conjecture~\ref{cexplicitcategorical} and  Section~\ref{sgeneralize}.  
The results in this article (Theorems~\ref{tintroexplicitmatrixnew} and \ref{tintroexplicitconstantnew}) 
show this for  some important rings of invariants.
The class of encoding circuits depends on the ring of invariants. 
For example, when $G=SL_m(\C)$, with constant $m$, one only needs  depth four
circuits; cf. Theorem~\ref{texplicitconst}.
Thus one can again approach the  goal  step by step, increasing the hardness   of the ring of invariants in tandem 
with the strength of the  circuits; cf. Section~\ref{snnlconstant}.

If  the goals  for both the problems (1) and (2) are achieved in a  stronger  form, it would follow
that NNL for every categorical quotient $V/G$ is in P, and  moreover, 
that the closed $G$-orbits in $V$
have an explicit (polynomial-time-computable) parametrization,
for any finite dimensional representation $V$ of a  reductive group $G$ in 
any characteristic;  cf. Sections~\ref{sgeneralize} and \ref{sclosed}.

There is  a fundamental  difference
between this  approach to the basic algorithmic problems of    algebraic geometry and invariant theory
and the standard approaches 
in  computational algebraic geometry \cite{mayr2} and  computational  invariant theory \cite{derksenbook}.
The difference lies in how the
basic objects of algebraic geometry--namely, the varieties--are specified in the computer. 
Computational algebraic geometry, based  on  Gr\"obner basis theory \cite{mayr2} and the theory of
solving polynomial equations \cite{kollar,koiran}, and  computational invariant theory \cite{popov,derksenbook} 
use the standard specification of the varieties  in terms of 
their defining equations. If one uses this standard specification, then 
the basic algorithmic problems of  algebraic  geometry and invariant theory are inherently intractable. For example,
Gr\"obner basis computation is EXPSPACE-hard \cite{mayr2},  solving
polynomial equations (Hilbert's Nullstellensatz) is NP-hard \cite{portier},
NNL is NP-hard (cf. Section~\ref{snoether}), 
the problem of constructing a finite set of generators for the ring $K[V]^G$ of  invariants  is inherently intractable,
because  the number of generators of  this  ring can be exponential in $\dim(V)$ even
when $\dim(G)$ is constant (cf. the proof of Proposition~\ref{pfftandsfthilbert}),
and hence, the standard \cite{mumford}  parametrization of the closed $G$-orbits in $V$ 
by the points of the categorical quotient $V/G$ is also inherently intractable.

But this article illustrates that a large class of algebraic varieties that arise in practice 
can be specified  {\em explicitly} using circuits, the basic objects of complexity theory. If 
one uses instead this    explicit complexity-theoretic specification of the basic geometric objects (varieties), 
as in the  approach  here,  then the results in this article indicate that the basic algorithmic problems 
(1) and (2) in algebraic geometry and invariant theory, along with the basic problem in geometric invariant theory
\cite{mumford} of parametrizing closed orbits in representations
of reductive groups,
which  are inherently intractable in the standard specification, are 
tractable in the explicit specification. The formal notion of explicitness introduced in this article 
(Definition~\ref{dexpvariety}), which is thus the
fundamental  difference between the geometric complexity theory approach to these basic problems and the 
standard approaches, is the driving theme of this article.

This article belongs to a series \cite{GCT1,GCT2,GCT3,GCT4}  of articles on
geometric complexity theory. 
See \cite{GCTcacm, GCTjacm} for an overview  of the earlier articles in this series,
and \cite{manivel} for an overview of the mathematical issues therein.
Preliminary versions of the results here were announced  in \cite{GCT5focs}.

We now  state   the main results of this article in more detail.

\noindent {\bf Notation:} Till  Section~\ref{sextensions}, $K$ will henceforth denote  
an algebraically closed base field of characteristic zero, unless mentioned otherwise. 
We  use the standard notation for the complexity classes, such as P (the class of problems that can be solved in 
polynomial time), BPP (the class of decision problems that can be solved by polynomial time Monte Carlo algorithms),
NC (the class of problems that can be solved in poly-logarithmic parallel time using polynomial number of 
processors), DET (the class of  problems LOGSPACE-reducible to computation of the 
determinant of integer matrices),
EXP (the class of problems that can be solved in exponential time), EXPSPACE (the class of problems that can be solved in
exponential work-space), PSPACE (the class of problems that can be solved in polynomial work-space),
$AC^0$ (the class of problems that can be solved by constant depth Boolean circuits of polynomial size), 
PH (the polynomial hierarchy), 
EXPH (the exponential hierarchy), and so on. 
See \cite{arora,cook} for their formal definitions.

\subsection{The problem  NNL} \label{sNNLdetintro}
We now define the problem NNL 
for the  explicit variety associated with 
the determinant in the first article \cite{GCT1} in this series. 

This explicit variety,  denoted as $\Delta[\det,m]$, is 
defined as follows. 
Let $X=(x_1,\ldots,x_r)$ be a tuple of $r$ variables.
For convenience, let us assume that $r=m^2$, so that $X$ can be thought of as an $m\times m$ variable matrix,
identifying $x_i$'s with the entries of $X$ in any way.
By a homogeneous symbolic determinant of size $m$ over $X=(x_1,\ldots,x_r)$, we mean 
the determinant of a symbolic $m \times m$ matrix, whose each entry is a homogeneous linear function over $K$ of $x_1,\ldots,x_r$.
Let ${\cal X}$ be the 
vector space over $K$ of homogeneous polynomials of degree $m$ in the variables $x_1,\ldots,x_r$, and
$P({\cal X})$ the projective space associated with ${\cal X}$.
Let $\Sigma[\det,m] \subseteq P({\cal X})$ be the set of all points in $P({\cal X})$ 
that correspond to nonzero homogeneous polynomials in ${\cal X}$ that can be expressed as homogeneous 
symbolic determinants of size $m$ over $X$. 
Then $\Delta[\det,m] \subseteq P({\cal X})$ is  the Zariski-closure $\overline{\Sigma[\det,m]}$ of
$\Sigma[\det,m]$. Its dimension is $\le m^4$. 
Informally, $\Delta[\det,m]$  is explicit because it can be specified {\em succinctly} by a small   circuit over $\Q$ 
of $\poly(m)$ total bit-size for computing $\det(X)$, and for a given $m$,  the specification of 
such a circuit  can be computed in $\poly(m)$ time \cite{skyum}.
For a formal proof of explicitness, see Section~\ref{sorbitclosuredet}.
Since the circuit \cite{skyum,malod} for computing the determinant is very special (namely, weakly skew, cf. \cite{malod} and
Section~\ref{sterminology}),
we call $\Delta[\det,m]$ {\em strongly explicit}.

It has to be stressed here that $\Delta[\det,m]$ is {\em not} 
specified by giving its equations in the ambient space $P({\cal X})$. This is not even possible using $\poly(m)$ bits, since
the dimension of $P({\cal X})$  is exponential in $m$.
All complexity bounds for the results below for $\Delta[\det,m]$ 
are in terms of the $O(\poly(m))$ bit-length of its
succinct specification. Thus an EXPSPACE-algorithm  means an algorithm that takes work-space that is exponential in $m$, 
a P-algorithm   means an algorithm that takes time that is polynomial  in $m$, and so on.

Let  $\hat \Delta[\det,m] \subseteq {\cal X}$ denote the affine cone of $\Delta[\det,m]$. This is defined to be
the union of all lines through the origin in ${\cal X}$ that correspond to the points of $\Delta[\det,m] \subseteq P({\cal X})$.
Let $R(\det,m)$ denote the homogeneous coordinate ring of $\Delta[\det,m]$. This is the same as  the coordinate ring
of $\hat \Delta[\det,m]$.

By Noether's Normalization Lemma (Lemma~\ref{lnnl0}), there exists a homogeneous linear map $\psi: {\cal X} \rightarrow K^k$,
for any $k> \dim(\Delta[\det,m])$, such that $\psi$ does not vanish on any nonzero point
in  $\hat \Delta[\det,m] \subseteq {\cal X}$.
Hence, $\psi$ yields a regular (well-defined) map 
from $\Delta[\det,m]$ to $P(K^k)$, which we denote by $\psi$ again.
We  call such a $\psi$ a {\em normalizing map} (for $\Delta[\det,m]$). 
Any  generic $\psi$ for such  $k$ is a normalizing 
map. But deterministic construction or even verification of a normalizing map,
as we shall see below, is very difficult.

Let $x_i$, $1 \le i \le k$,  denote the coordinates of $K^k$, and given a normalizing map $\psi$, let 
$\psi^*(x_i): {\cal X} \rightarrow K$ denote the pullback of $x_i$ via $\psi$. We also denote its restriction to  
$\hat \Delta[\det,m]$ by $\psi^*(x_i)$.
If $k=\dim(\Delta[\det,m])+1$, the minimum possible value, then we call the subset $\{\psi^*(x_i)\ | \ 1 \le i \le k\} \subseteq R(\det,m)$ 
an {\em h.s.o.p. (homogeneous system of parameters)} for $\Delta[\det,m]$.
Existence of such 
an h.s.o.p. is a classical  fact that holds for any variety; cf. 
Section~\ref{snoether}.

An h.s.o.p. for $\Delta[\det,m]$   can be constructed 
in work-space that is exponential in $m$, and in time that is double exponential in $m$ (cf. Theorem~\ref{tnnldetexpspace}),
by first computing the equations of $\Delta[\det,m]$ as a subvariety of $P({\cal X})$ using Gr\"obner basis theory \cite{mayr2}.
This space requirement is exponential in $m$, because 
the number of variables in the equations of $\Delta[\det,m]$ as a subvariety of $P({\cal X})$ is equal to the dimension of ${\cal X}$,
which is exponential in $m$, and  Gr\"obner basis computation
\cite{mayr2} takes work-space that is at least polynomial in the number of variables.
If we insist on an h.s.o.p., then this is the best that can be done at present.
However, if we do not insist on the  optimal $k=\dim(\Delta[\det,m]) +1 \le m^4$, but 
allow a slack, and only require that $k$ be $\poly(m)$, then we can do much better.

Accordingly, we define {\em the problem NNL for $\Delta[\det,m]$}
as  the problem of constructing  a normalizing map $\psi$ for $k=\poly(m)$, not necessarily optimal,
with a {\em succinct} specification of $\poly(m)$ bit-length. Thus we let go of optimality but insist on succinctness.
We have to now explain what we mean by  succinct.
Obviously,  the standard specification of $\psi$ as a linear map from
${\cal X} \to K^k$ is not succinct, since the dimension of ${\cal X}$ is exponential in $m$. 
Hence, we  confine ourselves to normalizing maps which have a succinct specification as follows.

For any $m \times m$  matrix $B$ with rational entries, let
$\psi_{B}$ denote the homogeneous, linear, evaluation map on  ${\cal X}$, which maps a polynomial
$p(X) \in {\cal X}$ to $p(B)$. We denote its restriction to $\hat \Delta[\det,m]$ by $\psi_B$ again.
Given any set ${\cal B}=\{B_1,\ldots,B_k\}$ of $m\times m$  matrices with 
rational entries, 
let $\psi_{{\cal B}}: {\cal X}  \rightarrow K^k$ denote the homogeneous linear map that maps 
$p=p(X) \in {\cal X}$ to $(\psi_{B_1}(p),\ldots,\psi_{B_k}(p))$. 
Let $S({\cal B})=\{\psi_{B_i} \ | \ 1 \le i \le k \} \subseteq R(\det,m)$.

We call $S({\cal B})$ an {\em  s.s.o.p. (small system of parameters)}\footnote{Such an s.s.o.p. is later called a strict s.s.o.p.,
as per the 
terminology in Section~\ref{simpliexp}, wherein we introduce a more general definition of an s.s.o.p. But we shall not worry about this  issue in this section.}
for $\Delta[\det,m]$ if  (1) the total bit-length of   $B_i$'s
is $\poly(m)$, and (2) the homogeneous linear map
$\psi_{{\cal B}}$ does not vanish on any non-zero point in $\hat \Delta[\det,m] \subseteq {\cal X}$. Hence, 
$\psi_{{\cal B}}$  yields  a regular (well-defined) map  from  $\Delta[\det,m]$ to $P(K^k)$, which we denote by
$\psi_{{\cal B}}$ again. We specify the s.s.o.p. $S({\cal B})$ {\em succinctly}
by  giving the matrices in ${\cal B}$.
We call  $\psi_{{\cal B}}$ the  {\em succinct normalizing map} corresponding to this s.s.o.p.

It can be shown that an s.s.o.p.
exists (Corollary~\ref{cssopdet}). However, an
s.s.o.p. with the optimal cardinality equal to
$\dim(\Delta[\det,m])+1$ may  not exist. This is why we allowed the slack above. 

A $\poly(m)$-time-constructible s.s.o.p. is  called 
an {\em e.s.o.p. (explicit system of parameters)}, where  explicit means $\poly(m)$-time-constructible.
Quasi-s.s.o.p. and quasi-e.s.o.p. are 
defined  by replacing $\poly(m)$ by  $\mbox{quasi-poly}(m):= 2^{\polylog(m)}$ 
throughout in the definitions.

The {\em problem NNL for $\Delta[\det,m]$}
is to construct an  s.s.o.p. for $\Delta[\det,m]$, given
the succinct specification of $\Delta[\det,m]$ in the form  a circuit 
for computing $\det(X)$.
We say that  NNL for $\Delta[\det,m]$ has an {\em explicit solution} if $\Delta[\det,m]$ has an e.s.o.p.

The current best, unconditional,
deterministic algorithm for constructing an s.s.o.p. for $\Delta[\det,m]$, 
based on  Gr\"obner basis theory \cite{mayr2},
also takes work-space  that is exponential in $m$, and time that is double exponential
in $m$ (cf. Theorem~\ref{tssopverifydet}),  as in the case of an h.s.o.p., again because 
the dimension of the ambient space $P({\cal X})$ containing $\Delta[\det,m]$ is exponential in $m$.

\subsection{A Monte Carlo algorithm} \label{smonteintro}
The following result shows that, if we are satisfied with Monte Carlo algorithms, then
an s.s.o.p. for $\Delta[\det,m]$ can be constructed efficiently and correctly, with a high probability.

\begin{theorem} \label{tmonteintronew}
An s.s.o.p. for $\Delta[\det,m]$ can be constructed  by  a $\poly(m)$-time randomized 
Monte Carlo algorithm, whose output is correct with a high probability.
\end{theorem}

Hilbert's original paper \cite{hilbert} itself gives a randomized Monte Carlo algorithm
to construct a normalizing map for any variety. For $\Delta[\det,m]$, the algorithm is the following: Just choose 
a random, homogeneous, linear map from $P({\cal X})$ to $P(K^k)$, with $k>\dim(\Delta[\det,m])$. 
It can be shown using  Gr\"obner basis theory \cite{mayr2} (cf. the proof of Theorem~\ref{tnnldetexpspace}) that
it  is a normalizing map with a high probability,
if  the entries of the matrix specifying this map are large enough
randomly chosen  integers of  bit-length exponential  in $\dim({\cal X})$.
Since $\dim({\cal X})$  is exponential
in $m$ in our context, the number of random bits used by this algorithm and 
its running time are thus double exponential in $m$. 
In contrast, the randomized algorithm in Theorem~\ref{tmonteintronew}
uses  only $\poly(m)$ random bits  and  $\poly(m)$ time. This is possible because the normalizing
map constructed by this algorithm has a succinct specification.
Obviously, the usual matrix representation 
of a linear map from ${\cal X}$ to $K^k$ is not succinct,  since
$\dim({\cal X})$ is  exponential in $m$.

The Monte Carlo algorithm in Theorem~\ref{tmonteintronew} is not a BPP-algorithm, 
since NNL is not a decision problem, but rather a construction problem, whose output
is not uniquely defined. More importantly, BPP is known to be in $\mbox{PH} \subseteq \mbox{PSPACE}$ \cite{arora}. In contrast, Theorem~\ref{tmonteintronew}
does {\em not} imply that NNL for $\Delta[\det,m]$ is in PSPACE. 
As already mentioned, at present we can only show unconditionally that it is in EXPSPACE.
This is because the problem of verifying correctness of  the output 
of the Monte Carlo algorithm in Theorem~\ref{tmonteintronew}, a potential s.s.o.p.,
turns out to be very difficult. If the problem of verifying an s.s.o.p.  were in PSPACE, then it would have followed
from Theorem~\ref{tmonteintronew} that NNL for $\Delta[\det,m]$ is in PSPACE.
But the current best algorithm for this verification  requires exponential work-space (cf. Theorem~\ref{tssopverifydet}). 

Further results on NNL for $\Delta[\det,m]$ will be given in Section~\ref{snnlvshard}.

\begin{theorem} \label{tmonteintroexplicitnew}
Theorem~\ref{tmonteintronew} holds with any explicit variety (cf. Definition~\ref{dexpvariety})
in place of $\Delta[\det,m]$.
\end{theorem}

Theorems~\ref{tmonteintronew} and \ref{tmonteintroexplicitnew} hold in arbitrary characteristic;  cf. Section~\ref{sposcharequi}.

We next turn  to some exceptional instances of  explicit varieties for which NNL can be 
solved deterministically in quasi-polynomial time using the existing techniques.

\subsection{The ring of matrix invariants} \label{snnlmatrixnew}
The  first  such instance   is the categorical quotient \cite{mumford} 
associated with the ring of matrix invariants.

Let $M_m(K)$ be the space of $m\times m$ matrices over $K$. Let
$V=M_m(K)^r$, the direct sum of $r$ copies of $M_m(K)$, with the adjoint (simultaneous conjugate)
action of $G=SL_m(K)$.

Let $U=(U_1,\ldots,U_r)$ be an $r$-tuple  of variable $m\times m$ matrices. 
The variable entries of $U_i$'s can be thought of as the coordinates of
$V$, and  the coordinate ring $K[V]$ of $V$ can be identified with
the ring $K[U_1,\ldots,U_r]$ generated by the variable entries of $U_i$'s. 
An invariant in  $K[V]$ is  a polynomial $f(U_1,\ldots,U_r)$ in the variable entries of $U_i$'s such that
\[ f(U_1,\ldots,U_r)=f(P^{-1} U_1 P,\ldots, P^{-1} U_r P),\] 
for all $P \in G$. Let $n=\dim(V)= r m^2$.
Let $K[V]^G \subseteq K[V]$ be the subring of  invariants. It is  finitely generated \cite{hilbert,procesimatrix}.
Hence, by a general construction of algebraic geometry,  one can associate with it  the variety  $V/G=\spec(K[V]^G)$,
called the {\em categorical quotient} \cite{mumford}.  

We say that  $V/G$ is 
{\em strongly explicit} 
if, given $m$ and $r$,   one can construct in $\poly(n)$ time a 
symbolic matrix 
$A(U,y)$ of $\poly(n)$ size such that:  (1) each entry of $A(U,y)$ is a (possibly non-homogeneous) linear function, with rational
coefficients,  of
the entries of $U_i$'s and auxiliary variables $y=(y_1,\ldots,y_k)$, $k=\poly(n)$, and 
(2) the coefficients of $\det(A(U,y))$, considered as a polynomial in $y$,  belong to and generate
$K[V]^G$.

\begin{theorem} \label{tintroexplicitmatrixnew}
The categorical quotient $V/G$ is strongly explicit.  It is strongly explicit in
a relaxed sense (cf. Definition~\ref{dexpffthilbert}) if $K$ is an algebraically closed field of positive characteristic.
\end{theorem}

We specify $V/G$ and $K[V]^G$  {\em succinctly} by simply specifying $V$ and $G$, which can be done  by giving $m$ and $r$ in unary.
This succinct specification is polynomial-time-equivalent  to the succinct specification of $V/G$ as an explicit variety in terms of the circuit for $\det(A(U,y))$, 
since, by Theorem~\ref{tintroexplicitmatrixnew},  this circuit can be computed in $\poly(n)$ 
time, given $m$ and $r$. 
The pair $(m,r)$ in unary  will thus  be the input in all the  problems for $V/G$ and $K[V]^G$ described below.
The bit-length of this succinct specification of $V/G$  is $O(m+r)=O(n)$, 
even though the dimension of the ambient space containing $V/G$ 
(cf. Section~\ref{sproofexpmatrix}) is 
exponential in $m$. All  space and time bounds for the algorithms below with this input will be in terms of $n$. 

By Noether's Normalization
Lemma (Lemma~\ref{lnoether20}), there exists a set $S \subseteq K[V]^G$ of $\poly(n)$
homogeneous invariants such that 
$K[V]^G$ is integral over the subring generated by $S$.\footnote{A ring $R$ is said to be integral  over its subring $T$ if every  $r \in R$ satisfies 
a monic polynomial equation of the form $r^l + b_{l-1} r^{l-1} + \ldots + b_1 r + b_0=0$, where each $b_i \in T$.}
(This statement of Noether's Normalization Lemma is equivalent to the one given in the beginning of this introduction.)
In fact, there even exists such an $S$ of optimal cardinality equal to $\dim(K[V]^G)$ (which is less than $n$). 
It is known that any generically chosen   $S$ of this cardinality has the required property. 
Such an $S$ of optimal cardinality is called an {\em h.s.o.p. (homogeneous system of parameters)} of $K[V]^G$. (Existence
of such an h.s.o.p. is again a classical fact, cf. Section~\ref{snoether}, that holds  for any  finitely generated  $K$-algebra.) 
It is shown here (cf. Theorem~\ref{thsop})
that the problem of constructing an h.s.o.p. for $V/G$ is  in EXPH (the exponential hierarchy),  assuming the Generalized Riemann Hypothesis.
The hierarchy is exponential, and not polynomial, because the dimension of the ambient space containing $V/G$ is exponential in $m$,
and hence the current best PH-algorithm for Hilbert's Nullstellensatz in Koiran \cite{koiran} becomes an EXPH-algorithm in our context.
If we insist on an h.s.o.p., then this is the best that we can do   at present.
However, we can do much better if, as in Section~\ref{sNNLdetintro}, 
we relax  the optimality constraint on the  cardinality,
but insist on succinctness of specification in exchange. We are thus led  to the following notion of an s.s.o.p.

We call a set $S \subseteq K[V]^G$ an {\em s.s.o.p. (small   system of parameters)} for $K[V]^G$ if 
(1) $S$ contains $\poly(n)$ homogeneous  invariants of $\poly(n)$ degree, 
(2)  $K[V]^G$ is integral over the subring generated by $S$, and 
(3)  each invariant $s=s(U_1,\ldots,U_r)$ in $S$  can be expressed as a symbolic determinant of $O(\poly(n))$ size, i.e., 
as the determinant of a symbolic matrix $M_s$ of $O(\poly(n))$ size, 
whose  entries are linear (possibly non-homogeneous) functions, with rational coefficients,
of the variable 
entries of $U_i$'s, and (4) for each $s \in S$, 
the total bit-size of the specification of the symbolic matrix $M_s$ (including the total bit-size of 
the constants therein)  is    $O(\poly(n))$.

We call $S$  an {\em e.s.o.p.(explicit system of parameters)} if, in addition,
given $m$ and $r$, the specification of $S$,
consisting of a symbolic matrix $M_s$ 
as above for each $s \in S$, can be computed in $\poly(n)$ time. This is a specialization to $V/G$ of the general definition 
of an e.s.o.p. for strongly explicit varieties (Definition~\ref{desopexp}) given later.
Quasi-s.s.o.p. and quasi-e.s.o.p. are defined by replacing 
$\poly(n)$  by quasi-$\poly(n)$ throughout in the definitions.

It can be shown that an  s.s.o.p. exists (cf. Corollary~\ref{corssopexistsgen}). 

By {\em the problem NNL for $K[V]^G$ or $V/G$}, we mean  the problem of constructing an 
s.s.o.p. for $K[V]^G$, given the succinct specification of
$K[V]^G$ in terms of the unary pair $(m,r)$. This definition of NNL is a specialization and simplification of the 
general definition of NNL for explicit varieties (Definition~\ref{desopexp}) given later. We say that  NNL 
for $K[V]^G$ has an {\em explicit solution}  if $K[V]^G$ has an e.s.o.p.

\begin{theorem} (For characteristic zero, 
see \cite{GCT5focs}, \cite{fs2,fs3}, and Remark 1 in Section~\ref{sprooftech}) \label{tmatrixNew} 
Let $V=M_m(K)^r$, and $G=SL_m(K)$ as above. Then
$K[V]^G$  has a quasi-e.s.o.p., assuming that 
$K$ is an algebraically closed field of characteristic $p \not \in [2,\floor{m/2}]$.
\end{theorem}

A stronger form of this result  (Theorem~\ref{tmatrixarbi})  implies quasi-explicit
(i.e., quasi-polynomial-time computable)
parametrization of the closed $G$-orbits in $V$ 
for any characteristic $p \not \in [2,\floor{m/2}]$; cf. Theorem~\ref{tquasiexplicitorbits}. Analogous results 
hold for the invariant ring associated with any quiver; cf. Theorem~\ref{tquiver}.

\subsection{The general ring of invariants} \label{sgeneralintro}
We now describe another exceptional explicit variety for which NNL can be solved deterministically in quasi-polynomial time with the existing
techniques.
This is the categorical quotient
associated with any invariant ring of $SL_m(K)$, with constant $m$, in characteristic zero. 

Let $V$ be any   finite dimensional  representation  of $G=SL_m(K)$, with arbitrary
$m$ for the moment. 
Since $G$ is reductive \cite{fultonrepr}, $V$  can be decomposed as a direct sum of irreducible 
representations of $G$:
\begin{equation} \label{eqweyl0}
 V=\sum_\lambda m(\lambda) V_\lambda(G).
\end{equation}

Here $\lambda:\lambda_1 \ge \ldots \ge \lambda_l$, $l <m$, is a partition, i.e., a 
non-increasing sequence
of non-negative integers,  $V_\lambda(G)$ is the irreducible representation of $G$
(Weyl module \cite{fultonrepr}) labelled by $\lambda$, and $m(\lambda)$ is its multiplicity.
Fix the  standard monomial basis \cite{doubillet,raghavan} for each $V_\lambda(G)$, and thus a standard monomial basis for $V$.
Let $v=(v_1,\ldots,v_n)$, $n=\dim(V)$,  be the coordinates of $V$ in this basis. 
Let $K[V]=K[v_1,\ldots,v_n]$ be the coordinate ring of $V$. Let  $K[V]^G$ be 
its  subring of $G$-invariants. We call a polynomial $f(v) \in K[V]$ a {\em $G$-invariant} if $f(\sigma^{-1} v)=f(v)$ 
for all $\sigma \in G$.  By Hilbert \cite{hilbert}, 
$K[V]^G$ is finitely generated. Hence, one can associate with it the  categorical quotient \cite{mumford}
$V/G=\spec(K[V]^G)$.
We specify  $V/G$ and $K[V]^G$  {\em succinctly} by just giving 
the specification $\langle V, G \rangle$ of $V$ and $G$, consisting of 
 $n$ and $m$  (in unary), and
the  multiplicities   $m(\lambda)$'s (in unary) for all $\lambda$'s  that occur with
nonzero multiplicity in the decomposition (\ref{eqweyl0}). The bit-length of this succinct specification  is $O(n+m)$, though the dimension of the ambient space containing $V/G$ is exponential in $n$, even when $m$ is constant; 
cf. the proof of  Proposition~\ref{pfftandsfthilbert}.

We call $V/G$  {\em strongly explicit} if, given $\langle V, G \rangle$, one can compute in $\poly(n,m)$ time 
a symbolic matrix $A(v,y)$ such that: (1) each entry in $A(v,y)$ is a (possibly non-homogeneous)  linear function, with rational coefficients,
of  the coordinates $v=(v_1,\ldots,v_n)$ and auxiliary variables $y=(y_1,\ldots,y_k)$,
$k=\poly(n,m)$, and  (2) the coefficients of
$\det(A(v,y))$, considered  as a polynomial in $y$, 
belong to and generate $K[V]^G$.

\begin{theorem} \label{tintroexplicitconstantnew}
Let  $V$  be  a finite dimensional representation of $G=SL_m(K)$,  with constant $m$, as above.
Then  $V/G$ is strongly explicit.
\end{theorem}

S.s.o.p., e.s.o.p., quasi-s.s.o.p., quasi-e.s.o.p.  for strongly explicit $V/G$ are
defined just as for the ring of matrix invariants in Section~\ref{snnlmatrixnew}, 
except that we use the coordinates $v_1,\ldots,v_n$ of
$V$ in place of the  coordinates for $M_m(K)^r$ used earlier, and the succinct specification 
$\langle V, G\rangle$ of $K[V]^G$ in 
place of the succinct specification of the ring of matrix invariants  by the pair $(m,r)$ earlier; cf. Definition~\ref{dexplicit} for details.

For constant $m$,  we define a {\em  near-e.s.o.p.} 
for $K[V]^G$ by replacing $\poly(n,m)$ in the definition of an e.s.o.p. 
by $O(n^{O(\log \log n)})$ throughout.

By {\em the problem NNL for $K[V]^G$ or $V/G$}, we mean  the problem of constructing an s.s.o.p. for $K[V]^G$,
given  $\langle V, G\rangle$ as above.
We say that  NNL for $K[V]^G$ has {\em an explicit solution} if $K[V]^G$ has an e.s.o.p.

\begin{theorem} \label{thilbertnew}
Let  $V$  be  a finite dimensional representation of $G=SL_m(K)$, with constant $m$, as above.
Then $K[V]^G$ has a near-e.s.o.p. 
\end{theorem} 

Analogues of Theorems~\ref{tintroexplicitconstantnew} and \ref{thilbertnew} hold for  any connected, reductive,
algebraic group of constant dimension (cf. Theorem~\ref{tsemisimple}).

By Theorems~\ref{tmarbitrary} and \ref{trussell},
the ring $K[V]^G$ has a quasi-e.s.o.p.,  without  any restriction on $m$,
if (1) the permanent of an $n \times n$ variable matrix $X$
cannot be computed  by symbolic determinants over $X$   of 
$O(2^{n^\epsilon})$ size\footnote{Here the entries of the
symbolic matrices are possibly  non-homogeneous linear functions of $X$.},
for some constant $\epsilon >0$, as $n\rightarrow \infty$ (cf. Valiant \cite{valiant2}), 
and (2) $V/G$ is explicit (cf. Conjecture~\ref{cexplicitcategorical} and the
remark thereafter).
Analogous results  hold for any reductive, algebraic group (possibly disconnected)
in zero or large enough characteristic (cf. Sections~\ref{sgeneralize} and \ref{sposcharequi}).

Classical invariant theory mainly studied the invariant ring $K[V]^G$  for constant  $m$,
as in Theorem~\ref{thilbertnew}, 
because of Gordan's seminal work (cf.  \cite{gordon}
and  Section 3.7 in \cite{sturmbook}) that gave an algorithm for 
constructing a finite set of generators for  the ring of invariants
of binary forms.
In this case, $V$ is the space of binary forms with the natural action of $SL_2(K)$, and  $m=2$. 
In the modern terminology, Gordan showed that the problem
of constructing finitely many generators for the ring invariants 
of binary forms is computable,
though the formal notion of computability  
was developed much later. 
It was not known before Hilbert if this  holds for general $m$,  or even for  $m=3$. 
It was not even known that finitely many generators exist  when $m=3$. 
This was shown by Hilbert  in his first paper \cite{hilbert0} for any $m$. 
But this proof was non-constructive.
It was severely criticized by Gordan (cf. Section 3.7 in \cite{sturmbook} for this story),  since 
it did not give an algorithm for constructing a set of generators. In
the modern terminology, it did not yield a proof,  as Gordan sought, for computability
of  the problem of constructing a set of generators for $K[V]^G$.
Such a proof was given by Hilbert in his second paper \cite{hilbert}, as a response to Gordan's criticism.
For these reasons,  the second paper  mainly focused on the
case when $m=3$. 
Theorem~\ref{tintroexplicitconstantnew}, or rather its stronger form (Theorem~\ref{texplicitconst}),   implies 
that the problem of constructing
a set of generators for $K[V]^G$ for constant $m$ is, in fact,
in  $\mbox{DET} \subseteq \mbox{NC} \subseteq \mbox{P}$, allowing 
encoding of the set by  a symbolic determinant.
Theorem~\ref{thilbertnew} shows that the original instance
of NNL  in Hilbert \cite{hilbert},  with constant $m$, is in quasi-P.

 \subsection{Noether normalization  vs.  hardness} \label{snnlvshard}
Next we ask if NNL for $\Delta[\det,m]$ can be solved deterministically
in  $\poly(m)$ time. For the reasons explained later in Section~\ref{sdiscuss}, this turns out to be a much harder problem 
than the analogous problems for the special cases of explicit varieties addressed in Theorems~\ref{tmatrixNew} and 
\ref{thilbertnew}.
At present, we only have a conditional result:

\begin{theorem} \label{tintronnldetnew}
The variety   $\Delta[\det,m]$ has a quasi-e.s.o.p.,
if the  permanent of an $n \times n$ variable matrix $X$  cannot be 
approximated infinitesimally closely by 
symbolic determinants over  $X$ of   size $\le 2^{n^\epsilon}$, for some constant $\epsilon >0$, as $n\rightarrow \infty$.
\end{theorem}

Entries of the symbolic matrices here are allowed to be non-homogeneous linear functions of the entries of $X$, with coefficients in $K$.
When $K=\C$, by {\em infinitesimally close approximation} of the permanent, we mean that, for any $\delta >0$, 
there exists a symbolic determinant over $X$  of  size $\le 2^{n^\epsilon}$, 
such that the distance  between the coefficient vectors of $\perm(X)$ and the symbolic determinant 
in the $L_2$-norm  is less than $\delta$.

The lower bound assumption for  the permanent in the result above  is 
a stronger form of the hardness hypothesis for the permanent in geometric complexity theory 
(cf. Conjecture 4.3 in \cite{GCT1}), with $\Omega(2^{n^\epsilon})$ lower bound in place of
the superpolynomial lower bound. 

Actually, we prove a stronger result (Theorem~\ref{tnnldetlower}) that, under this 
lower bound assumption,  NNL for $\Delta[\det,m]$  can even
be solved  fast in parallel.

\begin{theorem}\label{tintronnlexplicitnew}
Theorem~\ref{tintronnldetnew}  holds with   any explicit variety (cf. Definition~\ref{dexpvariety}) in place of $\Delta[\det,m]$.
\end{theorem}

By these results, NNL for  any explicit variety can be brought down from
EXPSPACE to quasi-P, assuming the hardness hypothesis \cite{GCT1} for the permanent in geometric complexity theory.
The quasi-prefix can be removed under a stronger assumption (cf. Theorem~\ref{tnnldetblack}).

Theorem~\ref{tintronnldetnew} is a consequence of the following stronger  result.
It shows that solving NNL for $\Delta[\det,m]$  in deterministic polynomial time is in fact
equivalent, ignoring a quasi-prefix,   to proving a weaker implication of the 
hardness hypothesis in Theorem~\ref{tintronnldetnew}.

\begin{theorem}  \label{tequivnnlnew} 
The variety   $\Delta[\det,m]$ has an e.s.o.p. iff, 
ignoring a quasi-prefix, there exists a family $\{f_n(x_1,\ldots,x_n)\}$ of 
exponential-time-computable, integral, multi-linear polynomials such that $f_n$ cannot be 
approximated infinitesimally closely  by symbolic determinants over $(x_1,\ldots,x_n)$
of  size $\le 2^{n^\epsilon}$, for some constant $\epsilon >0$, as $n \rightarrow \infty$.
\end{theorem}

By exponential-time-computable, we mean that  the  polynomial can be computed, given an integral input, in
time  that is exponential in the total bit-length of the  input. 

Theorems~\ref{tintronnldetnew}, \ref{tintronnlexplicitnew},  \ref{tequivnnlnew}, and their analogues for
explicit varieties also  hold in large enough  positive characteristics (cf. Section~\ref{sposcharequi}). Furthermore, the largeness restriction on
the characteristic can be removed assuming a slight
extension of the hardness hypothesis; cf. Remark 1 at the end of Section~\ref{sextensions}.

These results establish an essential equivalence between
the problem NNL for explicit varieties and 
the weaker form of the hardness hypothesis \cite{GCT1} 
in geometric complexity
theory. 

\subsection{Proof technique} \label{sprooftech}
We now briefly explain how the main results in this article  are proved.

The formal notion of explicitness introduced in this article (Definition~\ref{dexpvariety}) lies 
at the heart of  the proofs, along with the fundamental work  
\cite{hilbert,mumford,popov,derksen,procesimatrix,razmyslov,formanek1,raghavan}
in algebraic geometry 
and geometric invariant theory, the fundamental work \cite{valiant2,skyum,strassen,kaltofen,kaltofenlec,kaltofen2,malod} in
algebraic complexity theory, and the fundamental work \cite{schnorr,nisan,impa,russell,amir,manindra,fs2,fs1}
on a   derandomization problem in complexity theory, called black-box polynomial identity testing.
Derandomization  means converting a randomized efficient
algorithm into a deterministic  efficient algorithm by removing random bits. 
Theorems~\ref{tmatrixNew} and Theorems~\ref{thilbertnew}--\ref{tequivnnlnew} are proved by derandomizing the 
Monte Carlo algorithm in Theorem~\ref{tmonteintroexplicitnew} for the explicit varieties under consideration,
unconditionally or assuming a suitable hardness hypothesis.
Derandomization of this Monte Carlo algorithm for a given explicit variety 
amounts to  bringing NNL for that variety from EXPSPACE, where it is 
by the general result (Theorem~\ref{tssopexplicit}), to P. 
This EXPSPACE vs. P gap  in the complexity of NNL that  needs to be bridged 
to derandomize this Monte Carlo algorithm for a given explicit variety is 
the basic  difference between derandomization in this article and derandomization in the earlier 
articles \cite{nisan,impa,russell,manindra,fs2,amir} in complexity theory, wherein such a  gap is absent.
The use of  geometric invariant theory \cite{mumford} in derandomization,
as in the proofs of Theorems~\ref{tmatrixNew} 
and \ref{thilbertnew},  is another basic  difference.
In contrast, the earlier works \cite{nisan,impa,russell,manindra,fs2,amir} on derandomization in complexity theory
do not use any    invariant theory.

The  efficient  Monte Carlo algorithm for NNL for explicit varieties 
in Theorems~\ref{tmonteintronew} and \ref{tmonteintroexplicitnew}  
is based on the classical results in  algebraic geometry due to Hilbert and others, and  the  fundamental work 
in Heintz and Schnorr \cite{schnorr} on black-box polynomial identity
testing; cf. Section~\ref{smontedet}.

Theorem~\ref{tintroexplicitmatrixnew} on 
explicitness of the categorical quotient associated with the ring of matrix invariants
is proved    in characteristic zero using the 
First and Second Fundamental Theorems  for matrix invariants 
due to  Procesi and Razmyslov \cite{procesimatrix,razmyslov}; cf. Section~\ref{sexpmatrixring}.

The situation in positive characteristic turns out to be much harder.
The analogous First Fundamental Theorem for matrix invariants in positive characteristic in  
Donkin \cite{donkin}  is  too weak for 
the proof of Theorem~\ref{tintroexplicitmatrixnew} in positive characteristic, 
since the only known upper bound \cite{domokos} for the degrees of 
the generators in \cite{donkin} is exponential in the size $m$ of the matrices.
The crux  of the proof of Theorem~\ref{tintroexplicitmatrixnew} in positive characteristic
is the  {\em geometric  First Fundamental Theorem}  (cf. Theorem~\ref{tprocesigen}) proved in this article, 
which  provides a set of separating \cite{derksenbook} matrix-invariants of polynomial degree  in  arbitrary  characteristic.
This  is proved here using the criterion for stability in  arbitrary characteristic due to Hilbert \cite{hilbert}, 
Mumford et al. \cite{mumford}, and King \cite{king}, 
and the fundamental Brauer-Nesbitt theorem \cite{brauer,eggermont} in modular representation theory.

The explicitness result in Theorem~\ref{tintroexplicitmatrixnew} lies at the heart 
of  the proof of Theorem~\ref{tmatrixNew}.

Theorem~\ref{tintroexplicitmatrixnew}, in conjunction with Theorem~\ref{tmonteintroexplicitnew} 
(which holds in arbitrary characteristic, cf. Section~\ref{sposcharequi}), implies that
NNL for the ring of matrix invariants  has a polynomial-time Monte Carlo algorithm  in arbitrary characteristic.

Theorem~\ref{tmatrixNew} is proved by derandomizing this Monte Carlo algorithm, up to a quasi-prefix; cf. Sections~\ref{sstit}
and \ref{smatrixpos}.
This is done in two steps.

The first  crucial step (for the reasons that will become clear in Section~\ref{sdiscuss})
is to show that this Monte Carlo algorithm can be 
derandomized assuming the standard black-box derandomization
hypothesis for symbolic determinant identity testing \cite{schnorr,impa,russell,agrawal}, 
which is recalled in Section~\ref{ssymbdet} here.
This is shown using the 
fundamental work in Hilbert \cite{hilbert} and Mumford et al. \cite{mumford}; cf. Theorem~\ref{tclosedimage}, Remark 3 
thereafter, and Remark 2 in Section~\ref{sposcharequi}.
This implies that NNL for the ring of matrix invariants has a deterministic polynomial-time algorithm 
assuming the standard black-box derandomization hypothesis
for symbolic determinant identity testing.

\noindent {\em Remark 1 (on the second step of derandomization)}:  In the preliminary version 
\cite{GCT5focs} of this article, only this conditional result was proved in characteristic zero.
Subsequently it was pointed out by Forbes and Shpilka  \cite{fs3}  that 
the step in the proof of this result wherein 
symbolic determinant identity testing  enters  can be  modified,
as  explained in  Section~\ref{sforbes} here, so as to use 
instead the polynomial identity testing for read-once oblivious algebraic branching 
programs (cf. Section~\ref{sterminology}). 
A quasi-polynomial-time deterministic black-box algorithm  for this  problem was  already known from their
earlier work \cite{fs2}. Thus 
this instance of NNL can be solved  deterministically in quasi-polynomial time
in  characteristic zero  using the existing techniques. This is contrary to what was suggested in 
the preliminary version \cite{GCT5focs} of this article, 
because of the  relationship of this problem
with the  wild (``impossible'') problem \cite{drozd} 
of classifying matrix tuples (though  this instance of NNL itself is not wild).

This proof of Theorem~\ref{tmatrixNew} in characteristic zero can be extended to any 
characteristic $p \not \in [2,\floor{m/2}]$,  using
the  refined form 
of the Geometric First Fundamental Theorem for matrix invariants (cf. Theorem~\ref{tprocesigen}) proved in this article,
in place of the First Fundamental Theorem for matrix invariants due to Procesi and Razmyslov \cite{procesimatrix,razmyslov}; 
cf. Section~\ref{smatrixpos}.

Theorem~\ref{tintroexplicitconstantnew} on explicitness of the categorical quotient associated with 
the general ring of invariants of $SL_m$, for constant $m$, is proved 
using  the fundamental works in geometric invariant theory due to Hilbert
\cite{hilbert}, Mumford et al. \cite{mumford}, and Derksen and Kemper \cite{derksen,derksenbook}, 
in conjunction with standard monomial theory \cite{raghavan},
and the fundamental works  in algebraic complexity theory
due to Strassen \cite{strassen}, Valiant \cite{valiant2}, Malod and Portier \cite{malod}, and others; cf. Section~\ref{sexpconstant}.

The explicitness result in Theorem~\ref{tintroexplicitconstantnew} lies at the heart 
of  the proof of Theorem~\ref{thilbertnew}.

Theorem~\ref{tintroexplicitconstantnew}, in conjunction with Theorem~\ref{tmonteintroexplicitnew}, implies that 
NNL for the general ring of invariants of $SL_m$, for constant $m$, has a  polynomial-time Monte Carlo algorithm
in characteristic zero.

Theorem~\ref{thilbertnew} is proved by derandomizing this Monte Carlo algorithm, up to a quasi-prefix; cf. 
Section~\ref{snnlconstant}.
This is again done in two steps.

The first  crucial step is, again,  to show that this Monte Carlo algorithm can be derandomized assuming the standard 
black-box derandomization hypothesis for symbolic determinant identity testing. 
This can be done (just as in the case of Theorem~\ref{tmatrixNew})
using the work of Hilbert \cite{hilbert} and Mumford et al. \cite{mumford}; cf. Theorem~\ref{tclosedimage} and Remark 3 thereafter.
This implies that NNL for the general ring of invariants of $SL_m$, for constant $m$,  has a deterministic polynomial-time algorithm 
in characteristic zero,
assuming the standard black-box derandomization hypothesis (cf. Section~\ref{ssymbdet})
for symbolic determinant identity testing.

Using a refined form (Theorem~\ref{texplicitconst}) of Theorem~\ref{tintroexplicitconstantnew} in the first step, it follows
that NNL for the general ring of invariants of $SL_m$, for constant $m$,  has a deterministic polynomial-time algorithm 
assuming a  weaker  black-box derandomization hypothesis for diagonal depth three circuits \cite{saxena}.

This  hypothesis was already known to hold, up to a quasi-prefix, from 
the earlier work of Shpilka and Volkovich \cite{amir}, and  Agrawal, Saha, and Saxena \cite{manindra}.
Thus it follows that NNL for $V/G$ as in Theorem~\ref{thilbertnew}
can be solved in quasi-polynomial time deterministically. This was the result that was stated in the preliminary version
\cite{GCT5focs} of this article.
The 
stronger $O(n^{O(\log \log n)})$-time bound
stated in Theorem~\ref{thilbertnew} follows
in view of  the recent  result in Forbes, Saptharishi, and Shpilka
\cite{fs1}, which  gives an $O(s^{O(\log \log s)})$-time-computable black-box derandomization
of polynomial identity testing for diagonal depth three circuits of size $\le s$.

Let us now turn to Theorem~\ref{tequivnnlnew}, Theorem~\ref{tintronnldetnew} being its corollary.

The first step is to show that 
the  Monte Carlo polynomial-time algorithm for NNL for $\Delta[\det,m]$ in Theorem~\ref{tmonteintronew} can 
be derandomized assuming a strengthened form, introduced in this article (cf. Section~\ref{sstrongblack}),
of the standard black-box derandomization hypothesis \cite{schnorr,impa,russell,agrawal} for 
symbolic determinant  identity testing; cf. Section~\ref{sdetpoly}.

By Kabanets and Impagliazzo \cite{russell},  
the standard  hypothesis  holds, up to a quasi-prefix, 
assuming a sub-exponential symbolic determinant
lower bound for some family of   exponential-time-computable, integral, multi-linear polynomials.

It is similarly shown in this article (cf. Theorem~\ref{trussell2} and the remark thereafter)
that the  strengthened hypothesis 
holds, up to a quasi-prefix, if
there exists 
a family $\{f_n(x_1,\ldots,x_n)\}$ of exponential-time-computable, integral, multi-linear 
polynomials such that $f_n$ 
cannot be approximated infinitesimally closely by symbolic determinants 
of  size sub-exponential in $n$.
This is proved  using the fundamental work on 
black-box factorization of multivariate polynomials 
in Kaltofen and Trager \cite{kaltofen2}, which lies at the heart of this proof, in conjunction with the fundamental hardness
vs. randomness principle in Nisan and Wigderson \cite{nisan},  and Kabanets and Impagliazzo \cite{russell}.

This implies 
the  reduction from NNL to hardness stated in Theorem~\ref{tequivnnlnew}. 
The reduction in the other direction is easy (cf. Lemma~\ref{lnnldetblack} and Proposition~\ref{pconverserussell}).

Theorem~\ref{tintronnlexplicitnew} and the generalization of Theorem~\ref{tequivnnlnew} for general explicit
varieties (Theorem~\ref{tequisdit}) 
follow by systematically extending the proofs of Theorems~\ref{tintronnldetnew} and \ref{tequivnnlnew}; cf. Section~\ref{sexplicit}.

The proofs of these results can be extended to large enough positive characteristics
using the standard techniques of algebraic geometry and
algebraic complexity theory; cf. Section~\ref{sposcharequi}.

Conditional generalizations of Theorem~\ref{thilbertnew} to explicit  categorical quotients associated with representations
of general  reductive algebraic  groups are  given  in Section~\ref{sgeneralize}. These  can be proved 
by extending the proof for $SL_m$ using the standard techniques in geometric invariant theory \cite{mumford} 
and representation theory.

\subsection*{Organization of the paper}
The rest of this paper is organized as follows.

\noindent {\bf Logical structure of the proofs:}
The proofs of the main results are presented  in the following steps. (1) 
The variety under consideration is shown to be explicit. (2) An 
EXPSPACE-algorithm is given   for  constructing an h.s.o.p. for the variety. 
For the  categorical quotient associated with the ring of matrix invariants, 
a more efficient EXPH-algorithm is given, assuming the Generalized Riemann Hypothesis.
(3) An efficient Monte Carlo algorithm is given for constructing an s.s.o.p. 
for the variety. (4) This algorithm is   derandomized using
the strengthened or the standard form of the black-box derandomization hypothesis for
an appropriate class of circuits.  Which form is used depends on the  closure properties
of the variety. The class of circuits also depends on the variety. (5) If this class
is sufficiently restricted, as  happens for  the categorical quotients
associated with the ring of matrix invariants and the general ring of invariants of $SL_m$ with
constant $m$, then this black-box derandomization 
is  carried out unconditionally, up to a quasi-prefix. (6) Otherwise, it is shown that 
the black-box derandomization hypothesis holds assuming an appropriate hardness hypothesis.

\noindent {\bf Organization of the sections:}
In Section~\ref{sblackboxhyp}, we introduce the strengthened form of the standard \cite{schnorr,impa,russell,agrawal} 
black-box derandomization hypothesis for polynomial identity testing, and 
prove  the essential equivalence between
strengthened black-box derandomization 
and sub-exponential algebraic circuit size 
lower bounds for infinitesimally close approximation.
This is a  key ingredient in the proofs of 
Theorems~\ref{tintronnldetnew}, \ref{tintronnlexplicitnew}, and \ref{tequivnnlnew}.
In Section~\ref{snoether}, we recall Noether's Normalization Lemma, and show that 
the problem of constructing an h.s.o.p. 
for a general variety, specified  in the  standard fashion by  its
defining equations, belongs to PH, assuming the Generalized Riemann Hypothesis.
In Section~\ref{snnldeter}, we  study  the basic prototype $\Delta[\det,m]$ of an
explicit variety, and prove Theorems~\ref{tmonteintronew}, \ref{tintronnldetnew}, and \ref{tequivnnlnew}.
In Section~\ref{sexplicit}, we  formulate the general notion of an explicit variety  motivated
by its basic prototype $\Delta[\det,m]$,  define the problem NNL for explicit varieties, and 
prove Theorems~\ref{tmonteintroexplicitnew}, \ref{tintronnlexplicitnew}, and the generalization of Theorem~\ref{tequivnnlnew} for 
explicit varieties.
In Section~\ref{sexpmatrixring}, we prove Theorem~\ref{tintroexplicitmatrixnew} in characteristic zero.
Theorem~\ref{tmatrixNew} in characteristic zero
is  proved  in Section~\ref{sstit}. 
In Section~\ref{sexpconstant}, we prove Theorem~\ref{tintroexplicitconstantnew}.
Theorem~\ref{thilbertnew} is  proved in   Section~\ref{snnlconstant}.
Theorems~\ref{tintroexplicitmatrixnew} and \ref{tmatrixNew} in arbitrary  characteristic, their  generalizations to quivers, and 
extensions of Theorems~\ref{tintronnldetnew}, \ref{tintronnlexplicitnew}, and \ref{tequivnnlnew}
to large enough 
positive characteristics are proved in Section~\ref{sextensions}. It is also 
explained in Section~\ref{sextensions} how Theorems~\ref{tmonteintronew} and  \ref{tmonteintroexplicitnew} 
can be extended to  arbitrary characteristics. Furthermore,
implications  of the results in this article to explicit parametrization of closed orbits
and explicit parametrization of semi-simple representations of finitely generated algebras
are also given in Section~\ref{sextensions}.
Finally, in Section~\ref{sdiscuss}, 
we discuss the difficulties that need to be overcome  to improve
the current best bound for NNL for $\Delta[\det,m]$
in  Theorem~\ref{tssopverifydet}.

\section{Black-box polynomial identity testing} \label{sblackboxhyp}
In this section, we introduce  the strengthened  black-box derandomization hypothesis 
for polynomial identity testing (cf. Section~\ref{sstrongblack}), and prove 
the essential equivalence between strengthened black-box derandomization 
and sub-exponential  lower bounds for infinitesimally close approximation
(cf. Theorem~\ref{trussell2} and  Proposition~\ref{pconverserussell}). 
This  is a key ingredient  in the proofs of 
Theorems~\ref{tintronnldetnew}, \ref{tintronnlexplicitnew}, and \ref{tequivnnlnew}.

\subsection{Circuits and symbolic determinants}  \label{sterminology}
We begin by recalling  \cite{malod,yehu} the  circuit classes for which we need these
hypotheses.

By a {\em circuit} over the field $K$
\cite{malod}, we mean a directed acyclic graph with
vertices of in-degree zero or two, in which each node of in-degree $2$ is labelled with
$*$ or $+$, and each node of in-degree $0$ is labelled with a variable or a constant in $K$.
By the polynomial computed by the circuit, we mean the polynomial computed at the root,
in the obvious way. By {\em the  size of the circuit}, we  mean the total number of edges in it. 
If the constants in the circuit are in $\Q$ or its finite extension, then by {\em the bit-size or the bit-length
of the circuit}, we mean the size of the circuit plus the total bit-size of the specification of all the constants.

We say that the circuit $C=C(x_1,\ldots,x_n)$ over the variables $x_1,\ldots,x_n$ has 
{\em low or small degree} if the degree of the polynomial computed 
by it  is $O(s^a)$, for some fixed constant $a>0$, where $s$ is the size of $C$.

By a  {\em weakly skew circuit}, we mean  \cite{malod}
a circuit whose each node $v$ labelled with $*$ 
has at least one child $u$ such that 
the sub-circuit rooted at $u$ is connected to the rest of the circuit by just the edge 
$(u,v)$. 

By a {\em symbolic determinant}
over $x_1,\ldots,x_n$ of size $m$, we mean the determinant of a symbolic 
$m\times m$ matrix, whose each entry is a linear combination (possibly non-homogeneous) over $K$ of 
of  $x_1,\ldots,x_n$.
Weakly skew circuits are polynomially equivalent  \cite{malod} to  symbolic determinants.

Weakly skew circuits are also polynomially equivalent to algebraic branching programs
\cite{yehu,fs2}. In this article we will only use  a special class of such programs
called {\em read-once oblivious algebraic 
branching programs} \cite{fs2}. 
Such a  program can be specified, for some $n,l$, and $d$, 
by a tuple $(M_1, M_2,\ldots, M_l)$ of $n \times n$
matrices such that every entry of $M_i$, $ 1 \le i \le l$,  is a 
uni-variate polynomial of degree $\le d$
over $K$ in the distinct variable $z_i$ associated with $M_i$.
The uni-variate polynomials are  specified by giving all their coefficients.
The size of this program is $O(n^2 l  d)$.
The polynomial computed by this program  is defined to be $\trace(\prod_i M_i)$. 
Clearly, this polynomial  can also be computed by a weakly skew circuit 
of $\poly(n,l,d)$ size. Hence,
such programs can be viewed as restricted classes of weakly skew circuits, or equivalently,
symbolic determinants.

A {\em diagonal depth three circuit} \cite{saxena} $C$
over the variables $x_1,\ldots,x_n$  is a circuit that computes  a sum 
$\sum_{i=1}^k f_i^{e_i}$
of powers of linear functions,
where each $f_i$ is a possibly non-homogeneous linear function of
 $x_i$'s with coefficients in $K$.
Here $k$ is called the {\em top fan-in} of the circuit, and 
$e=\max\{e_{i}\}$  its {\em degree}. The size of this circuit is $O(n e k)$.

By a {\em circuit with oracle gates}
for a function $f(y_1,\ldots,y_r)$, we mean a circuit
in which some  gates are labelled with $f$. These gates have in-degree $r$. 
The computation of $f$ at any such gate  is assigned unit cost.

\subsection{Polynomial identity testing}\label{sblackstd}
Next, we   recall    the standard black-box derandomization hypothesis 
for   polynomial identity testing  \cite{schnorr,impa,russell,agrawal} over the  algebraically
closed base field $K$.

The {\em polynomial identity testing}    problem over $K$  is 
the problem of deciding if a given  circuit $C(x)$,
$x=(x_1,\ldots,x_r)$, over $K$ 
computes an identically zero polynomial. 
By {\em the polynomial identity testing problem for  small degree circuits}, or {\em the low-degree polynomial identity testing problem}, we mean the polynomial identity testing problem wherein
the degree of the polynomial computed by $C(x)$ is assumed to be $O(s^a)$, for some constant $a>0$, where $s$ is the size of $C(x)$.

There is a simple randomized polynomial-time algorithm \cite{ibarra}  for deciding if a given 
 circuit with rational constants computes an identically zero polynomial: just 
substitute large enough random integer values for the variables, and test if the circuit
evaluates to zero.

The {\em white-box derandomization problem} \cite{russell}   for polynomial identity testing is
to find a deterministic polynomial time algorithm 
for deciding if a given  circuit with rational constants computes an identically
zero polynomial.

The harder {\em black-box derandomization problem}  for polynomial identity testing 
\cite{schnorr,impa,russell,agrawal} over $K$  is to construct  
a {\em hitting set} against all circuits over $K$ with size  $\le s$  on $r\le s$ variables, given $r$ and $s$ in unary.
By a hitting set,  we mean a set $S_{r,s} \subseteq \N^r$  of test inputs 
such that, for every circuit $C$ over
$K$ and $x=(x_1,\ldots,x_r)$  with size $\le s$ computing a non-zero polynomial $C(x)$,  $S_{r,s}$ contains 
a test input $b$ such that $C(b)\not =0$.  
The  {\em standard black-box-derandomization hypothesis for polynomial identity testing}  \cite{schnorr,impa,russell,agrawal}  is that there exists a  $\poly(s)$-time-computable hitting set.
We call such a  hitting set {\em explicit}.
More generally, if a hitting set
is computable  in $O(T(s))$ time, 
we say that the polynomial identity testing for circuits of size $\le s$ 
has $O(T(s))$-time-computable black-box derandomization. 

The standard  black-box derandomization   hypotheses for the restricted   circuit classes 
in Section~\ref{sterminology} are defined similarly.

\subsection{Symbolic determinant identity testing} \label{ssymbdet}
We now describe such a  hypothesis for symbolic determinants
(cf. Section~\ref{sterminology}) in more detail, since it plays a crucial role in this paper.

Let $Y=Y(x_1,\ldots,x_n)$ be any $m\times m$ symbolic matrix, whose
each entry is a homogeneous linear form over $K$ in the variables $x_1,\ldots,x_n$. 
By {\em symbolic determinant identity testing}, we mean the 
problem of deciding, given $Y$,  if  the  symbolic determinant $\det(Y)$ is an identically zero polynomial
in $x_i$'s.  We call  a set $S_{n,m} \subseteq \N^n$  of test inputs a {\em hitting set} in this context 
if,  for every non-zero symbolic determinant $\det(Y)$ over $x_1,\ldots,x_n$ of size $m$,
$S_{n,m}$ contains a test input  on which that symbolic determinant does not vanish. 
The {\em standard black-box derandomization hypothesis   for symbolic determinant identity testing}  
\cite{schnorr,impa,russell,agrawal} 
is that, given $n$ and $m$, one can construct such a hitting set in $\poly(n,m)$ time. 
This is a weaker form of the standard black-box derandomization hypothesis for polynomial identity testing 
described in Section~\ref{sblackstd}.

\subsection{Black-box polynomial identity testing vs. hardness}
The following result is a variant of Theorem~7.7  in \cite{russell}. This is why 
polynomial identity testing is expected to have efficient black-box derandomization. 

\begin{theorem}[Kabanets and  Impagliazzo] \label{trussell} (cf.  Theorem~7.7  in  \cite{russell})
Suppose there exists a family $\{f_m(x_1,\ldots,x_m)\}$ of
exponential-time-computable,  multi-linear,  integral
polynomials   such that 
$f_m$  cannot be evaluated by a circuit  over $K$ of  
$O(2^{m^a})$ size, for some constant
$a>0$, as $m \rightarrow \infty$.  
Then polynomial identity testing for small degree 
circuits over $K$  of size $\le s$ has $O(2^{\polylog(s)})$-time-computable black-box derandomization. 
\end{theorem}

Here by an exponential-time-computable, integral polynomial $f_m(x_1,\ldots,x_m)$, 
we mean a polynomial such that, given an integral input $a=(a_1,\ldots,a_m)$, 
$f_m(a)$  can be computed in time that is exponential in the total 
bit-length of $a$. Since $f_m$ here is multi-linear, this is equivalent to saying that
the coefficient vector of $f_m$ can be computed in time exponential in $m$.

The proof of Theorem~\ref{trussell} is  similar to that of
Theorem 7.7 in \cite{russell} (which works in the black-box model).
Since we are going to prove its stronger form (Theorem~\ref{trussell2}) later, we only point out here how to take care of the 
main difference between the setting in \cite{russell} and the one here. 
The difference  is that in \cite{russell} the size of the circuit is defined to be the total number of  edges in it plus the total
bit-length of the constants, whereas here the size  means the total number of edges. 
A key ingredient in  the  proof in \cite{russell} is 
an efficient algorithm in \cite{kaltofen} for factoring multivariate polynomials 
(cf. Lemma 7.6. in \cite{russell}). In its place we use instead the following result  
in \cite{kaltofen,kaltofen2} that 
does not depend on the bit-lengths of the  constants  in the circuit.

\begin{theorem}[Kaltofen and Trager] (cf. Corollary 6.2. in \cite{kaltofen},  Theorem 1 in \cite{kaltofen2}, and Theorem 2.21 in B\"urgisser \cite{burgbook})

\label{tkaltofen}

Suppose $\{g_n(x_1,\ldots,x_n)\}$ is a $p$-computable family \cite{valiant2} 
of polynomials over $K$. This means
$g_n$ is a polynomial of $\poly(n)$ degree that can be computed by a 
nonuniform circuit over $K$ of $\poly(n)$ size. 
Then each factor  of $g_n$ in  $K[x_1,\ldots,x_n]$  
can also be computed by a 
nonuniform circuit over $K$ of $\poly(n)$ size. 

More generally, given any families $\{g_n(x_1,\ldots,x_n)\}$,
$\{f_n(x_1,\ldots,x_n)\}$ of polynomials over $K$, 
with  $f_n$ dividing $g_n$, 
there exists for every $n$  a nonuniform circuit over $K$ of 
$\poly(n,\deg(g_n))$ size, with oracle gates for $g_n$, that computes $f_n$. 
\end{theorem}

A simpler 
proof of the first statement in Theorem~\ref{tkaltofen}   can be found in Section 2.3 in
B\"urgisser \cite{burgbook} (cf. Theorem 2.21 therein). 
B\"urgisser also proves a stronger statement in \cite{burg2} (cf. Theorem 1.3 therein)
concerning   complexity of  infinitesimally close approximation.
For the  converse of  Theorem~\ref{trussell},  see \cite{schnorr,agrawal}.

\subsection{The strengthened black-box derandomization hypothesis} \label{sstrongblack}
Next,  we formulate the {\em strengthened   black-box derandomization hypothesis   for polynomial identity testing}.

Let  $x=(x_1,\ldots,x_r)$ be a tuple of $r$ variables.
The {\em strengthened black-box derandomization problem for small degree circuits} is to construct in $\poly(s)$ time 
a {\em  hitting set}  against all nonzero polynomials $f(x) \in K[x]$ of degree $\le d=O(s^a)$, $a>0$ a constant, 
that can be approximated infinitesimally  closely  by   circuits over $K$ and $x$  of size $\le s$,
given $r,s$, and $d$ in unary.

When $K=\C$, by {\em infinitesimally close approximation}, we mean that,  for any $\epsilon >0$, there exists such a circuit 
$C_\epsilon$ over $K$  of size $\le s$ such that the distance $||C_\epsilon(x)-f(x)||_2$ between the coefficient vectors of 
$C_\epsilon(x)$ and 
$f(x)$ in the $L_2$-norm  is less than $\epsilon$; cf. \cite{burg2} for the definition for general $K$.
By a {\em hitting set},  we mean a set $S_{r,s} \subseteq \N^r$  of test inputs
such that,  for every nonzero $f(x)$ of  degree $\le d$  that can be approximated
infinitesimally closely by circuits over $K$ of size $\le s$,
$S_{r,s}$ contains a test input $b$ such that $f(b)\not =0$.

The following result implies  that such a hitting set exists.
For any positive integer $u$, let $[u]:=\{1,\ldots,u\}$.

\begin{theorem} [Heintz and Schnorr] (cf. Theorem 4.4 in \cite{schnorr} and its proof) \label{theintz}
A randomly chosen subset 
$B \subseteq [u]^r$, $u=2 s (d+1)^2$,  of size $q=6 (s+1+r)^2$ is with a high probability a hitting set 
against all non-zero polynomials that can be approximated infinitesimally closely by 
 circuits over $K$ and $r$ variables of size $\le s$ and degree $\le d$.
Specifically, at least $(1-u^{-q/6})$-th fraction of the sequences $(b_1,\ldots,b_q)$, $b_i \in [u]^r$, are hitting.
\end{theorem}

The  {\em strengthened black-box-derandomization hypothesis} for polynomial identity testing for small degree circuits is that there exists  a $\poly(s)$-time-computable  hitting set
$S_{r,s}$.  We call such a hitting set {\em explicit}.
More generally, if a hitting set  is computable  in $O(T(s))$ time, 
we say that the polynomial identity testing for small degree circuits of size $\le s$
has $O(T(s))$-time-computable strengthened black-box derandomization. 
The {\em strengthened black-box derandomization hypothesis for general polynomial identity testing without any degree restrictions} is defined similarly.

The similar 
{\em strengthened black-box derandomization hypothesis for symbolic determinant identity testing} is that,  given $n$ and $m$, 
one can construct in  $\poly(n,m)$ time a hitting set 
against all nonzero homogeneous polynomials $h(x_1,\ldots,x_n)$'s over $K$ of degree $m$ that 
can be approximated infinitesimally closely
by symbolic determinants (cf. Section~\ref{ssymbdet}) of size $m$ over $x_1,\ldots,x_n$.

The strengthened black-box derandomization hypothesis 
is   counter-intuitive unlike the  standard   hypothesis in  Section~\ref{sblackstd}.
Conjecturally (cf.  Section~\ref{sdiscuss}), there  exist
integral polynomials of small degree that can be approximated infinitesimally closely by small  circuits over $K$
but  cannot be computed exactly by such circuits. Hence, a priori, there is no reason why there should exist easy-to-compute 
hitting sets against such hard-to-compute polynomials.

\subsection{Equivalence between strengthened black-box derandomization and  lower bounds for infinitesimally close approximation}
\label{sequivalencestrong}
The following strengthening  of Theorem~\ref{trussell} 
says that one can still compute efficiently in quasi-polynomial time a hitting set against such polynomials,
assuming a sub-exponential lower bound for infinitesimally close approximation for 
a family $\{p_m\}$ of  exponential-time-computable, multi-linear, 
integral polynomials.
A good candidate for $p_m$  is the permanent. It cannot be approximated infinitesimally closely by small 
algebraic  circuits as per the hardness hypothesis  \cite{GCT1}
of geometric complexity theory.
The result below   is the main reason
why the strengthened black-box derandomization hypothesis  is  expected to hold. 

\begin{theorem} \label{trussell2}
Suppose there exists  a family $\{p_m(x_1,\ldots,x_m)\}$ 
of  exponential-time-computable,  multi-linear, integral
polynomials such   that $p_m$ 
cannot be approximated infinitesimally closely  by   circuits  over $K$ of  
$O(2^{m^\epsilon})$ size, for some   
constant $\epsilon >0$, as $m\rightarrow \infty$.  Then polynomial identity testing for small degree 
circuits  over $K$ 
with size $\le s$ and $n \le s$ variables  has $O(2^{\polylog(s)})$-time-computable strengthened black-box derandomization. 
\end{theorem}

This result also holds if we use symbolic determinants instead of  circuits in the lower bound hypothesis;
cf.  the proof of Theorem~\ref{tnnldetlower0}.

\proof We extend  the proof  of Theorem 7.7 in Kabanets and Impagliazzo \cite{russell}
 using Theorem~\ref{tkaltofen}, 
which lies at the heart of this proof.

We want to  construct in quasi-$\poly(s)$ time a hitting set 
for strengthened black-box derandomization of polynomial identity testing  for small degree 
circuits  with size $\le s$ and $n \le s$ variables.

Let  $m=(\log s)^e$, for a large enough constant $e$ to be fixed later.
Construct an $NW$-design \cite{nisan} for this $n$ (the number of variables) with this choice of  $m$. 
By the NW-design, we mean a family of sets $R_1,\ldots,R_n \subseteq [l]$,
$l \le m^2=(\log s)^{2 e}$, each of cardinality $m$, such that $|R_i \cap R_j| \le \log n$, for all $i\not = j$.
By \cite{nisan} (cf. Lemma 2.23 in \cite{russell}),
such a set system can be constructed in $\poly(n, 2^l)=O(2^{\polylog(s)})$ time.

This set system and  the given hard multi-linear polynomial  $p(x_1,\ldots,x_m)$ 
together  yield an arithmetic NW-generator $NW^p$. By this we mean the function

\begin{equation} \label{eqNW}
NW^p:  \quad x=(x_1,\ldots,x_l) \in \N^l \rightarrow (p(x|_{R_1}),\ldots, p(x|_{R_n})) \in \Z^n, 
\end{equation}
where $x|_{R}$ denotes the tuple of the elements in $x$ indexed by $R$. 

\begin{claim} The set $H=\{ NW^p(a) \ | \ a \in [D]^l\}$,  $D = d m +1$, is a  hitting set against every 
nonzero polynomial $f(y)$, $y=(y_1,\ldots,y_n)$,  of degree $\le d=O(s^t)$, $t>0$ a constant,
that can be approximated infinitesimally closely  by circuits over $K$ 
and $y=(y_1,\ldots,y_n)$ of size $\le s$ (assuming that the constant $e$ above is chosen to be large enough).
\end{claim} 

Since $p$ is exponential-time-computable, $H$ is  $O(2^{\polylog(s)})$-time computable. So  it remains to prove the claim.

\noindent {\em Proof of the claim:} 
Suppose to the contrary that $f(b) = 0$,  for every $b \in H$, for some nonzero polynomial 
$f(y)$  of degree $\le d$ 
that can be approximated infinitesimally closely  by  circuits over $K$ of size $\le s$.

Let  $g_0(x_1,\ldots,x_l,y_1,\ldots,y_n):=f(y_1,\ldots,y_n)$. For $1 \le i \le n$, let $g_i(x_1,\ldots,x_l,y_{i+1},\ldots y_n)$
be the polynomial obtained from $f$ by replacing $y_1,\ldots,y_i$ by the polynomials $p(x|_{R_j})$, $1 \le j \le i$. 
Then $g_n=f(NW^p(x))$, and the  degree of each $g_i$ is $\le d m < D$.  Since $f(b)=0$ for all $b \in H$, $g_n(a)=0$ for all $a \in [D]^l$.
Since $\deg(g_n) < D$, by the  Schwarz-Zippel lemma \cite{schwarz},   $g_n$ is identically zero.
But $g_0=f$ is not identically zero. So there exists a smallest $0 \le i < n$ such that $g_i$ is not identically zero but 
$g_{i+1}$ is identically zero. Fix this $i$. Since $g_i$ is not identically zero, we can set $y_{i+2},\ldots,y_n$ and $x_j$, $j \not \in R_{i+1}$,
to some integer values so that the restricted polynomial $\tilde g_i(x_{j_1},\ldots,x_{j_m},y_{i+1})$ remains a non-zero polynomial,
where $R_{i+1}=\{x_{j_1},\ldots,x_{j_m}\}$.
Let us denote this polynomial by renaming the variables as $g(x_1,\ldots,x_m,y)$. 

Then  $g(x_1,\ldots,x_m,y)$ is a non-zero polynomial  with degree  $\le d m$, but
$g(x_1,\ldots,x_m, p(x_1,\ldots,x_m))$ is identically zero. By Gauss's Lemma, $h(x_1,\ldots,x_m,y)= p(x_1,\ldots,x_m) - y$
is a factor of $g(x_1,\ldots,x_m,y)$.  By Theorem~\ref{tkaltofen}, 
$h(x_1,\ldots,x_m,y)$ has a circuit over $K$ of $\poly(m,\deg(g))=\poly(s)$ size, with oracles gates for $g$. 
Setting $y=0$ in this circuit, we get a circuit for  $p(x_1,\ldots,x_m)$  of $\poly(s)$ size with oracle gates for $g$.

But $g$ has a circuit of size $O(n^2 \log n)$ with one oracle gate for $f$. 
This is because $|R_j \cap R_{i+1}|$, $j \le i$, is at most $\log n$, by the property of the $NW$-design.
Hence, after the specialization of  the variables $y_{i+2},\ldots,y_n$ and $x_j$, $j \not \in R_{i+1}$, as above, 
each $p(x|_{R_j})$, $j \le i$, gets restricted to a multi-linear polynomial in at most $\log n$ variables. This restricted polynomial
can be computed brute-force by a circuit $C_j$ of size at most $O(\log n 2^{\log n}) = O(n \log n)$ size.
We get a circuit for $g$, as desired, by  connecting  the  inputs $y_1,\ldots,y_i$ of the oracle for $f$ to the outputs of $C_1,\ldots, C_i$,
respectively, and  specializing the variables $y_{i+2},\ldots,y_n$ to their integer values chosen above.

It follows that  $p(x_1,\ldots,x_m)$ can be computed 
by a circuit $C$ over $K$ of size $O(s^c)$ with oracle gates for $f$,
for some constant $c>0$ independent of $e$. Given any circuit $D_\delta$ of size $\le s$ 
for approximating $f$ within precision $\delta >0$, 
let $C_\delta$ denote the circuit obtained from $C$ by substituting $D_\delta$ for $f$.  
Since $f$ can be approximated infinitesimally closely by circuits of size $\le s$,  by choosing $\delta$ small
enough,  $C_\delta$ can  approximate $p(x_1,\ldots,x_m)$ to any precision.
The size of $C_\delta$ is $O(s^{c+1})$.  Choosing $e$ large enough, the size of $C_\delta$ 
can be made $\le 2^{m^\epsilon}$ for any $\epsilon >0$. This contradicts hardness of infinitesimally close approximation of  $p$.
\qed 

If the polynomial $p$ in Theorem~\ref{trussell2} is the permanent, 
the following stronger result holds.

\begin{theorem} \label{trussellperm}
Suppose the permanent of $k\times k$ matrices 
cannot be approximated infinitesimally closely  by   circuits  over $K$ of  $O(2^{k^\epsilon})$ size, for some   constant
$\epsilon >0$, as $k\rightarrow \infty$. 
Then a hitting set for strengthened black-box derandomization of small-degree circuits over $K$ of size $\le s$
can be constructed in $O(\polylog(s))$ parallel time using  $O(2^{\polylog(s)})$ processors.
\end{theorem}

\proof The proof is  like that of Theorem~\ref{trussell2},
letting $m=k^2$ and $p_m(x)=\perm(x)$, and thinking  of $x=(x_1,\ldots,x_m)$ 
as a $k\times k$ matrix.  We follow the same notation as in the proof of
Theorem~\ref{trussell2}. 
We only need to explain
why the construction of a hitting set can be efficiently parallelized. 

The arithmetic NW-generator, cf. (\ref{eqNW}),  
based on the permanent is the function 

\begin{equation} 
NW^{\perm}:  
\quad x=(x_1,\ldots,x_l) \in \N^l \rightarrow 
(\perm(x|_{R_1}),\ldots, \perm(x|_{R_n})) \in \Z^n, 
\end{equation}
where $x|_{R}$ denotes the tuple of the elements in $x$ indexed by $R$ with cardinality 
$m=k^2$. 

Since  $n \le s$, we can compute each $\perm(x_{R_j})$ in parallel. 
Thus, it suffices to explain 
why each $\perm(x_{R_j})$ can be computed fast in parallel. 
Fix one $R_j$. Without of loss generality, assume that the elements in $R_j$
are $x=(x_1,\ldots,x_m)$. Think of $x$ as a $k\times k$ matrix. Then
$perm(x)=\sum_{\sigma} \prod_i x_{i \sigma(i)}$,
where $\sigma$ ranges over all permutations of $k$ letters. Since $m=\polylog(s)$,
the number of terms in this expansion is $O(2^{\polylog(s)})$. 
So we can assign a processor 
to each monomial in the expansion. The processor can compute that monomial 
in $\poly(m)=\polylog(s)$ time. 

It follows that $NW^{\perm}$ can be computed in $O(\polylog(s))$ parallel time using
$O(2^{\polylog(s)})$ processors.
\qed 

\noindent {\em Remark 1:} The
crucial fact used in the proof of Theorem~\ref{trussellperm} is 
that the permanent of $k \times k$ matrices can be computed in 
$O(\polylog(s))$ parallel time using
$O(2^{\polylog(s)})$ processors, if $k=O(\polylog(s))$. This need not hold, in general,
for the  exponential-time-computable $p_m$ in the statement of Theorem~\ref{trussell2}. 

\noindent {\em Remark 2:} 
Theorem~\ref{trussellperm}
also holds, with a similar proof, if we replace the permanent in its statement by any 
PSPACE-computable, integral, multi-linear polynomial  satisfying a similar  lower bound 
assumption. 

The following result is the easy converse of Theorem~\ref{trussell2}, ignoring the quasi-prefix.

\begin{prop} \label{pconverserussell}
Suppose the strengthened black-box derandomization hypothesis for  polynomial identity testing for small degree 
circuits over $K$ holds. Then
there exists a family $\{p_m(x_1,\ldots,x_m)\}$ of
exponential-time-computable,   multi-linear, integral 
polynomials such that  $p_m$  cannot be approximated infinitesimally closely  by   circuits  over $K$ of  
$O(2^{m/a})$ size, for some constant $a>0$, as $m \rightarrow \infty$.  
\end{prop}

\proof The proof is similar that of Theorem 51 in \cite{agrawal}.

Choose $s=2^{m/a}$, where $a>0$ is a large  enough constant to be chosen later.
Suppose there exists an $O(s^b)$-time-computable, integral hitting set 
$T$ of size $\le s^b$ against all nonzero multi-linear polynomials in $m$ variables that 
can be approximated infinitesimally closely by  circuits over $K$ of size $\le s$.

Let $p_m(x)$, $x=(x_1,\ldots,x_m)$, be a multi-linear polynomial such that 
\begin{equation} \label{eqlinsystem}
p_m(t) =0, \quad \forall t  \in T.
\end{equation}

Each condition here is a linear constraint on $2^m$ coefficients of $p_m(x)$.
The number of these constraints is $|T| \le s^b =2^{m b /a}< 2^m$ if $a > b$. 
Hence, as $m \rightarrow \infty$, 
there is a non-zero integral $p_m(x)$ satisfying these constraints. One such  
$p_m(x)$ can be computed in $2^{O(m)}$ time 
by solving the linear system (\ref{eqlinsystem}). By (\ref{eqlinsystem}), this exponential-time computable $p_m(x)$ cannot be approximated
infinitesimally closely by  circuits over $K$ of size $\le s$,  since $T$ is a hitting set. 
\qed

By Theorem~\ref{trussell2} and Proposition~\ref{pconverserussell}, strengthened black-box derandomization and sub-exponential
lower bounds for infinitesimally close approximation of exponential-time-computable, multi-linear, integral polynomials 
are essentially equivalent notions.

\subsection{The EXPSPACE-bound for strengthened black-box derandomization} \label{sstdvsstrong}
The following is the currently best unconditional deterministic upper bound for strengthened black-box 
derandomization.

\begin{theorem} \label{tstrongexp}
The strengthened black-box derandomization problem for general polynomial identity testing belongs to EXPSPACE.
It belongs to EXPH (the exponential hierarchy)
assuming  the Generalized Riemann Hypothesis.
\end{theorem}

In contrast:

\begin{prop} \label{pstandard}
The standard black-box derandomization problem for polynomial identity testing over $K$ belongs to PSPACE unconditionally, and 
to PH assuming the Generalized Riemann Hypothesis.
\end{prop} 

This proposition can be proved using  Theorem~\ref{theintz} and the following result.

\begin{theorem}[cf. Koiran \cite{koiran}]  \label{tkoiran}
The problem {\em Hilbert's Nullstellensatz}
of deciding if a given system of multi-variate integral polynomials, specified as 
circuits, has a complex 
solution is in PSPACE unconditionally, and in $AM \subseteq RP^{NP} \subseteq \Pi_2$ assuming the Generalized Riemann Hypothesis. The same also holds 
for the homogeneous variant of Hilbert's Nullstellensatz, 
namely, the problem of
deciding if a given system of homogeneous, multi-variate, integral polynomials has a nontrivial complex solution.
\end{theorem}


\noindent {\em Proof of Theorem~\ref{tstrongexp}:} 
For simplicity, we  only prove the result for symbolic determinant identity testing.
The proof for  general polynomial identity testing 
is similar, using the universal circuit polynomial $H(Y)$ introduced in \cite{GCT1} 
(and recalled in Section~\ref{shy} here) in place of the symbolic determinant.

We want to construct a hitting set against all non-zero homogeneous 
polynomials of degree $m$ that can be approximated infinitesimally closely
by symbolic determinants of size $m$ on $r$ variables. Without loss of generality, we can assume that $r=m^2$ (by adding more variables
or increasing the size of the determinant). We can identify these $m^2$ variables with the entries of a variable $m \times m$ 
matrix $X$. Then 
all such nonzero polynomials correspond to the nonzero points of the variety 
$\hat \Delta[\det,m] \subseteq {\cal X}$ constructed in 
Section~\ref{sNNLdetintro}, since the closure in the Zariski topology coincides with the closure
in the complex  topology; cf. Theorem 2.33  in \cite{mumfordalg}. We  now follow the same terminology as in Section~\ref{sNNLdetintro}. 

A symbolic determinant of size $m$ over $m^2$ variables can be computed \cite{malod}
by a  circuit of size $s=O(\poly(m))$. 
Hence, by Theorem~\ref{theintz}, 
there exists a subset of $[u]^{m^2}$, $u=2 s (m+1)^2$,  of size $q=6 (s+1+m^2)^2$ that is  a hitting set 
against all non-zero polynomials that can be approximated infinitesimally closely by 
symbolic determinants of size $m$ over the $m^2$ variable entries of $X$.

We can enumerate  all subsets of $[u]^{m^2}$ of size $q$, and for each enumerated subset 
$B \subseteq [u]^{m^2}$ of size $q$,  check if it is a hitting set. The enumeration  can be done using  $\poly(m)$ work-space. 

However, checking if a given $B \subseteq [u]^{m^2}$ of polynomial size $q$ 
is a hitting set turns out to be much more difficult for the reasons
that will be explained in more detail  in Section~\ref{sdiscuss}.
This is the main difficulty, since finally we have to output a correct 
$B$ of polynomial size. This check can be done using exponential space as follows.

As in Section~\ref{sNNLdetintro}, for any $m \times m$ rational matrix $b$, let 
$\psi_b$ be the 
homogeneous linear evaluation function  on ${\cal X}$, which maps $p(X) \in {\cal X}$ to $p(b)$.
Let $H(b)$ denote the hyperplane that is the zero set 
of $\psi_{b}$. Then $B$ is a hitting set iff $\hat \Delta[\det,m] \cap \bigcap_{b \in B}
H(b) = \{0\}$. To carry out this test, we first compute 
the defining 
equations of $\hat \Delta[\det,m] \subseteq {\cal X}$. Using Gr\"obner basis theory (cf. Theorem 1  in \cite{mayr2}), this 
can be done in work-space that is polynomial in the dimension of ${\cal X}$ and exponential in the dimension of $\Delta[\det,m]$. 
This  work-space requirement is exponential in $m$. The total bit-length of the specification of the 
resulting defining equations of $\hat \Delta[\det,m]$ is at most exponential in the work-space requirement,
and thus, at most double exponential in $m$.
After this, we again use Gr\"obner basis theory (cf. Theorem 1 in \cite{mayr2})
to carry out the test. This  takes work-space that is polynomial in the dimension of ${\cal X}$, 
exponential in the dimension of $\Delta[\det,m]$, and poly-logarithmic in
the total bit-length of the specification of the defining equations.
This space requirement is again exponential in $m$, i.e., $O(2^{\poly(m)})$.
Overall, this is an EXPSPACE-algorithm. 

In the preceding proof, one can use Theorem~\ref{tkoiran} in place of Gr\"obner basis theory.
Specifically, given  $B \subseteq [u]^{m^2}$ of polynomial size $q$, 
one can construct in exponential time,
using Theorem 5.7 in \cite{burg2}, a system of polynomial equations over $\Q$ in exponentially many variables
with the  specification in terms of circuits  of  exponential total bit-length, such that 
$B$ is a hitting set iff this system does not have a complex solution. 
Using Theorem~\ref{tkoiran}, the latter test can be carried out by an EXPSPACE-algorithm
unconditionally, and by an EXPH-algorithm, assuming  the Generalized Riemann Hypothesis.
This is an EXPH-algorithm and not a PH-algorithm, since the number of variables in the system
is exponential.

Assuming the Generalized Riemann Hypothesis, this gives an EXPH-algorithm for the verification of a hitting
set, and hence,  an EXPH-algorithm for strengthened black-box derandomization.
\qed

\section{Noether's Normalization Lemma} \label{snoether}
In this section we recall Noether's Normalization Lemma
and show that the problem of constructing an h.s.o.p. for a general
variety, given by 
the standard specification (defined below)
in terms of its  defining equations, belongs to $\mbox{PH}$.

\begin{lemma} [Noether's Normalization Lemma] (Cf. page 36 in \cite{mumfordalg}) \label{lnnl0}
Let $X \subseteq P(K^k)$ be  a  projective variety   of dimension $n$.
Let $\psi: K^k \rightarrow K^m$, $m \ge n+1$,  be a homogeneous 
linear map  that does not vanish on any line through the origin in $K^k$ corresponding to any point of  $X$. This means $\psi$ 
induces a regular (well defined) linear map from $X$ to $P(K^m)$, which we denote by $\psi$ again.
Then the homogeneous coordinate ring $R(X)$   of $X$ is integral over the subring generated by the pullbacks $\psi^*(x_i)$'s 
of the coordinate functions $x_i$'s, $1 \le i \le m$, on  $K^m$. This implies that 
(1) $\psi(X) \subseteq P(K^m)$, the image of $X$, is closed in $P(K^m)$, and (2)
the fiber $\psi^{-1}(p)$, for any point $p \in \psi(X)$, is a finite set. 

Conversely,  if 
$R(X)$  is integral over the subring generated by  $\psi^*(x_i)$'s, then $\psi$ is regular on $X$.
\end{lemma}

Any  $\psi$ chosen uniformly  at random has the regularity property stated above if $m \ge n+1$.

The following  graded version of Noether's Normalization Lemma  is implicit in its  proof.

\begin{lemma}[Graded Noether Normalization] \label{lnoether20}
(cf. Theorem 13.3. in \cite{eisenbudbook}, Corollary 2.29 in
 \cite{mumfordalg}, and also the proof of Theorem 1.5.17 in \cite{bruns})
Let $R$ be any positively graded, finitely generated  $K$-algebra. 
Let  $f_1,\ldots,f_t$ be any  non-constant, homogeneous generators of $R$, and
$H \subseteq R$ any set of homogeneous elements such that, letting $I(H)$ denote the ideal generated by $H$, 
$f_i^{e_i} \in I(H)$ for some positive integer $e_i$, for every $i$.
Then $R$ is integral over the   subring generated by $H$. 
\end{lemma}

\begin{defn} \label{dhsop} \cite{eisenbudbook}
Let $R$ be any positively graded, finitely generated  $K$-algebra. 
A set $H$ of homogeneous invariants of  cardinality equal to $\dim(R)$ such that $R$ is integral over the subring
generated by $H$ is called an {\em h.s.o.p. (homogeneous system of parameters)}  
of $R$.
\end{defn}

Thus $\psi^*(x_i)$'s, $1 \le i \le m$, in
Lemma~\ref{lnnl0} form an h.s.o.p. of the homogeneous coordinate ring $R(X)$ of $X$, if $m=n+1$.

Let $Z \subseteq K^t$ be a variety consisting of the common zeroes  of a set of homogeneous 
integral polynomials $f_1(z), f_2(z),\ldots$, $z=(z_1,\ldots,z_t)$.  Assume that 
$Z$ is specified  by 
giving circuits  for  $f_i$'s, and that the constants in these circuits 
are rational.
We call such a specification of $Z$ {\em standard}.
Its {\em bit-length}  is defined to be 
the total bit-length of the specification of the circuits  for  $f_i$'s.

The following result shows  that for  general varieties  over $K$,
given by the standard specification as above, 
the   problem of constructing an h.s.o.p. is  in  PH assuming the Generalized Riemann Hypothesis.  The succinct specification  of $\Delta[\det,m]$ (cf. 
Section~\ref{sNNLdetintro}) in terms of a small circuit for computing the determinant
is not standard, since it does not specify defining equations for 
the variety.
Hence the  following result does {\em not} apply to $\Delta[\det,m]$ given in the succinct specification. The current best EXPSPACE-bound
for $\Delta[\det,m]$ given in  the succinct  specification will be proved later (cf. Theorem~\ref{tnnldetexpspace}).

\begin{theorem} \label{tnoetherph}
The  problem of constructing an h.s.o.p.  for a general variety over $K$,
given by  the standard specification  in terms of circuits for the 
defining equations,   belongs to PH assuming the Generalized Riemann Hypothesis,
and to PSPACE unconditionally.
\end{theorem}

Here by PH, we  really mean its functional analogue, since the problem under consideration is a construction problem, not a 
decision problem.
The PSPACE-bound holds in arbitrary characteristic.

\proof 
Let $Z \subseteq K^t$ be a variety consisting of the common zeroes  of a set of homogeneous 
integral polynomials $f_1(z), f_2(z),\ldots$, $z=(z_1,\ldots,z_t)$, specified 
in the standard fashion  by the  circuits  for  $f_i$'s with rational constants.
Let $N$ be the total  bit-length of this specification.

Testing if $\dim(Z)=0$ is 
the complement  of  the homogeneous
Hilbert's Nullstellensatz problem  in Theorem~\ref{tkoiran}. Hence, 
by Theorem~\ref{tkoiran}, 
we can test if $\dim(Z)=0$ by  a  PSPACE-algorithm, and also by a $\Sigma_2$-algorithm
assuming the Generalized Riemann Hypothesis. If $\dim(Z)=0$, then h.s.o.p. for $Z$ is empty,
and we are done.

So let us assume   that $\dim(Z) \ge 1$. 

Let $s \le \dim(Z)$ be any positive  integer.
Consider  random linear forms $L_r(z)=\sum_k b_{k,r} z_k$, $1 \le r \le s$, where 
$b_{k,r}$'s are random integers of large enough $\poly(N)$ bit-length. 
Let  $H_r \subseteq K^t$ be  the hyperplane defined by $L_r(z)=0$. 

We claim that, if $s=\dim(Z)$, then 
$Z\cap \bigcap_{r} H_r = \{0\}$ with a high probability. If $s< \dim(Z)$, then clearly 
$Z\cap \bigcap_{r} H_r \not =  \{0\}$, since it has non-zero dimension.

By Hilbert's Nullstellensatz and Lemma~\ref{lnoether20}, it  the follows that,
 if $s=\dim(Z)$, then
the homogeneous coordinate ring of $Z$ is integral over the subring generated by $L_r(z)$'s, and hence $\{L_r(Z)\}$ is  an h.s.o.p. for $Z$.

So, let us first prove the claim.
Accordingly, assume that $s=\dim(Z)$.
Let $d=\max\{\deg(f_i)\}$. Clearly, $d \le 2^M$, where $M$ denotes the maximum 
number of multiplication gates  in the circuit for any $f_i$. 
Since $M \le N$, it follows
that $d \le 2^N$. 
By raising $f_i$'s to appropriate powers, we can assume that all of them have the same degree $D \le 2^{N^2}$. 
Consider generic  linear combinations of $f_i$'s and generic linear forms
\begin{equation} \label{eqsystemnoether} 
\begin{array}{l} 
F_j(z)=\sum_i y_{i,j} f_i(z), \quad 1 \le j \le t-\dim(Z), \\
L_r(z)=\sum_k w_{k,r} z_k, \quad 1 \le r \le s=\dim(Z), 
\end{array}
\end{equation}
where $y_{i,j}$'s and $w_{k,r}$'s are indeterminates. 
Let $R$ denote the multi-variate resultant of $F_j$'s and $L_r$'s. It is a polynomial in $y_{i,j}$'s and $w_{k,r}$'s 
of degree $\le D^t$. By  Noether's Normalization Lemma (cf. Lemma~\ref{lnnl0} and the remark thereafter), 
the system of equations (\ref{eqsystemnoether}) has only $\{0\}$ as its
solution for some rational  values for  $y_{i,j}$'s and $w_{k,r}$'s. Hence $R$ is not identically zero as a polynomial
in $y_{i,j}$'s and $w_{k,r}$'s. 
By the Schwarz-Zippel lemma \cite{schwarz}, we can specialize $y_{i,j}$'s randomly to some integers
of $O(\log (D^t))=\poly(N)$ bit-length  so that the resulting specialization $R'$ of $R$ is not
identically zero. Then $R'$ is  a nonzero polynomial in $w_{k,r}$'s of degree $\le D^t$.
By the Schwarz-Zippel lemma again, $R'$ does not vanish  identically if we let $w_{k,r}=b_{k,r}$ for 
randomly chosen integers  of   $O(\log (D^t))=\poly(N)$ bit-length.   For such $b_{k,r}$'s, 
$Z\cap \bigcap_r H_r = \{0\}$. This proves the claim. 

Next, we show  that $\dim(Z)$ and a  specification of  $L_r(z)$'s, $1 \le r \le \dim(Z)$, 
such that $Z\cap \bigcap_{r} H_r = \{0\}$ can be computed in $\poly(N)$ work-space.

We begin by letting $s=1$, the first guess for $\dim(Z)$.
With this choice of $s$, 
choose $b_{k,r}$'s as above randomly of large enough 
$\poly(N)$ bit-length and  test if $Z \cap \bigcap_r H_r = \{0\}$. The latter test  can be carried out in PSPACE unconditionally (cf. Theorem~\ref{tkoiran}).
If the test fails, we increase $s$ by one and repeat the test.
The test succeeds with a high probability when 
$Z \cap \bigcap_r H_r = \{0\}$ and $s=\dim(Z)$.
Randomization in this algorithm can be removed, since $\mbox{RPSPACE}=\mbox{NPSPACE}=\mbox{PSPACE}$. 
This yields a PSPACE-algorithm for computing $\dim(Z)$ and $L_r(z)$'s, $1\le r \le \dim(Z)$,
such that $Z \cap \bigcap_r H_r = \{0\}$, as desired.

Assuming the Generalized Riemann Hypothesis, 
whether $Z \cap \bigcap_r H_r = \{0\}$  can be tested  by a $\Sigma_2$-algorithm, by Theorem~\ref{tkoiran}. 
This gives a  $\Sigma_2$-algorithm for testing if there exist
$L_r(z)$'s, for the given choice of $s$, such that $Z \cap \bigcap_r H_r = \{0\}$: 
guess $y_{i,j}$'s and $w_{k,r}$'s, and  test 
if $Z \cap \bigcap_r H_r = \{0\}$ using
the $\Sigma_2$-algorithm (Theorem~\ref{tkoiran}). 

Using this $\Sigma_2$-algorithm for testing the existence of $L_r(z)$'s in place of the
PSPACE-algorithm before, we get a PH-algorithm for computing $\dim(Z)$
and  $L_r(z)$'s, $1 \le r \le \dim(Z)$,  such that $Z \cap \bigcap_r H_r = \{0\}$.
\qed

The proof of Theorem~\ref{tnoetherph}  shows that the problem of constructing an 
h.s.o.p.  for general varieties specified by their equations is Turing-reducible in randomized 
polynomial time to the complement of 
the homogeneous Hilbert's Nullstellensatz problem in Theorem~\ref{tkoiran}.
Conversely, the complement of the 
homogeneous Hilbert's Nullstellensatz problem
can be reduced to the
problem of constructing an h.s.o.p.  for general varieties (since a projective variety $X$ is
empty iff its h.s.o.p.   is empty).
Since the Hilbert's Nullstellensatz  problem in Theorem~\ref{tkoiran}
is NP-hard \cite{portier}, 
it follows that the problem of constructing an h.s.o.p.
for general varieties is co-NP-hard. 
In analogy with the problem NNL for $\Delta[\det,m]$ in Section~\ref{sNNLdetintro}, we can define the problem NNL for general varieties
$X$, specified in the standard fashion by their  equations, as the problem of constructing a small homogeneous set $S \subseteq R(X)$
of cardinality polynomial in the dimension of $X$ (but not necessarily of optimal cardinality equal to $\dim(X)+1$) such that
$R(X)$ is integral over the subring generated by $S$. Even this problem is  co-NP-hard. 

In contrast, we shall prove in the next two sections that the problem 
NNL of constructing an s.s.o.p. 
for $\Delta[\det,m]$, with a succinct specification, and more generally, the problem NNL 
for any explicit variety  can be solved 
in quasi-polynomial time, assuming a lower bound for infinitesimally
close approximation.

\section{NNL for $\Delta[\det,m]$} \label{snnldeter}
In this section we prove Theorems~\ref{tmonteintronew}, \ref{tintronnldetnew},  and \ref{tequivnnlnew}.
We follow the same notation as in Section~\ref{sNNLdetintro}.

The variety $\Delta[\det,m]$, defined in Section~\ref{sNNLdetintro},  can alternatively be defined as follows.
Let $X$ be a variable  $m\times m$ 
matrix.  Let ${\cal X}$ be the 
vector space over $K$ of homogeneous polynomials of degree $m$ in the 
variable entries of $X$, and $P({\cal X})$ the projective space associated with ${\cal X}$.
Thus $g=\det(X)$ is an element of ${\cal X}$.  Furthermore, ${\cal X}$ is a 
representation of $G=GL_{m^2}(K)$, where $\sigma \in GL_{m^2}(K)$ maps $h(X) 
\in {\cal X}$ to $h(\sigma^{-1} X)$, thinking of $X$ 
as an $m^2$-vector. 
Then $\Delta[\det,m] \subseteq P({\cal X})$ is the Zariski-closure 
of the orbit $G g \subseteq P({\cal X})$, thinking of $g$  as also a point in $P({\cal X})$. 
(We can also use  $SL_{m^2}(K)$ here
instead of $GL_{m^2}(K)$, since that does not change $\Delta[\det,m]$.)
As in Section~\ref{sNNLdetintro}, let $\hat \Delta[\det,m] \subseteq {\cal X}$ be the affine cone of $\Delta[\det,m]$, and 
$R(\det,m)$ the homogeneous coordinate ring of $\Delta[\det,m]$.

We assume that 
$\Delta[\det,m]$ is specified succinctly as in Section~\ref{sNNLdetintro}.
This can be done either by  giving a small uniform circuit of 
$\poly(m)$ bit-length  for computing $\det(X)$, or alternatively, by 
just giving $m$ in unary (from which a circuit for the determinant can 
be computed in $\poly(m)$ time). 
The bit-length of this succinct specification 
is $\poly(m)$. All complexity bounds in this section
will be in terms of this bit-length, or equivalently, in terms of $m$. 

The problem NNL  for $\Delta[\det,m]$, given in  this succinct specification, is to 
construct an   s.s.o.p. of the form $S({\cal B})$,  as defined in Section~\ref{sNNLdetintro}, 
for some set ${\cal B}$ of $m\times m$ rational matrices of $\poly(m)$ total bit-length.

\noindent {\em Remark:} Later (cf. Definition~\ref{desopexp}) we  define  a more general s.s.o.p.,
which need not be of the form $S({\cal B})$. But  s.s.o.p.'s  of this form are  most natural. They are 
called {\em strict s.s.o.p.}  in Definition~\ref{dstrict}.
In this section, we assume, as in Section~\ref{sNNLdetintro},
that an s.s.o.p. for
$\Delta[\det,m]$ is always of  the form $S({\cal B})$. 
Thus NNL here is   strict NNL as per the  terminology in  Section~\ref{simpliexp}.

We call an s.s.o.p. $S({\cal B})$ {\em separating} if, for  any two distinct points $p,q \in \hat \Delta[\det,m]$, 
$\psi_{{\cal B}}(p) \not = \psi_{{\cal B}}(q)$, with $\psi_{{\cal B}}$ as in Section~\ref{sNNLdetintro}.  
Thus a separating s.s.o.p. denotes
a {\em dimension-reducing map}, with a succinct specification, from ${\cal X}$ to $K^k$, 
$k=\poly(m)$, that is injective on $\hat \Delta[\det,m]$.  
By the {\em strong form of NNL} for $\Delta[\det,m]$, we mean the problem of constructing a separating s.s.o.p. for
$\Delta[\det,m]$.
A $\poly(m)$-time-constructible, separating s.s.o.p. is  called 
a   {\em separating e.s.o.p. (explicit system of parameters)}.
Separating quasi-s.s.o.p. and quasi-e.s.o.p. are  defined  by replacing $\poly(m)$ by  $2^{\polylog(m)}$.

\subsection{Unconditional upper bound for the problem of constructing an h.s.o.p.} 
Before we turn to the construction of an s.s.o.p., 
we  address the construction of an
h.s.o.p. for $\Delta[\det,m]$ (cf. Section~\ref{sNNLdetintro}).
By Theorem~\ref{tnoetherph}, the problem of constructing an h.s.o.p. for a general  variety, given by the standard specification 
in terms of  defining equations,
is in PSPACE. This does {\em not} imply that the same problem for $\Delta[\det,m]$ is in PSPACE, 
since $\Delta[\det,m]$ is not specified 
in the  standard fashion by its  defining equations, 
but rather succinctly by a small uniform circuit for the determinant.
The current best algorithm based on Gr\"obner basis theory \cite{mayr2} for 
converting the  succinct specification of $\Delta[\det,m]$ to its  standard specification takes  space that is  exponential  in $m$.  
Hence,  we only get the following EXPSPACE-bound for the succinct specification.

\begin{theorem} \label{tnnldetexpspace}
The problem of constructing an h.s.o.p. for $\Delta[\det, m]$, specified succinctly,
belongs to EXPSPACE. (This means it can be solved in work-space that is
exponential in $m$). 
Assuming the Generalized Riemann Hypothesis, it belongs to EXPH,  if $\Delta[\det,m] \subseteq P({\cal X})$ has defining equations
that can be computed in time that is exponential in $m$.
\end{theorem}

\proof Given the succinct specification of $\Delta[\det,m]$, we first compute the
equations defining it as a subvariety of
$P({\cal X})$,  using Gr\"obner basis theory as in the proof of Theorem~\ref{tstrongexp}, 
in  work-space that  is exponential in $m$.
The total degree and the bit-length of  the specification of these
equations is at most double-exponential in $m$.

We  apply Gr\"obner basis  theory (cf. Theorem 1 in \cite{mayr2}) again 
to compute an h.s.o.p. for $\Delta[\det,m]$, using these defining equations.
This takes work-space  that is polynomial in $\dim({\cal X})$, exponential in $\dim(\Delta[\det,m])$, and 
poly-logarithmic in the total bit-length of the specification of the defining equations.
This work-space requirement 
is single-exponential in $m$, i.e., $O(2^{\poly(m)})$. The total running time as well as the bit-length of the output h.s.o.p.
is double exponential in $m$. 
This  gives 
an EXPSPACE algorithm for computing an h.s.o.p. for  $\Delta[\det,m]$.

If $\Delta[\det,m]$ has  defining 
equations that can be computed in exponential time, then we 
can skip the first step above of computing defining equations, and use these equations instead. 
After this, we can use the PH-algorithm
for general varieties in Theorem~\ref{tnoetherph} for computing an h.s.o.p., assuming the Generalized Riemann Hypothesis. 
Since the dimension of ${\cal X}$ is exponential in $m$, this PH-algorithm 
becomes an EXPH-algorithm in our context.
\qed

The bit-length of the specification of the h.s.o.p. constructed in Theorem~\ref{tnnldetexpspace} is double-exponential in $m$.
If we insist on an h.s.o.p. then Theorem~\ref{tnnldetexpspace} is the best that we can do at present.
However, if we are willing to settle for an s.s.o.p. (which need not have the optimal cardinality like an h.s.o.p.),
then Theorem~\ref{tequivnnlnew}, proved in this section (cf. Theorem~\ref{tequisditdet}), 
says that the double exponential time bound in 
Theorem~\ref{tnnldetexpspace} can be brought down to quasi-polynomial,
assuming that there exists a family $\{f_n(x_1,\dots,x_n)\}$ 
of exponential-time-computable, integral, multi-linear polynomials such that $f_n$   cannot be 
approximated infinitesimally closely by symbolic determinants over $K$ of 
sub-exponential size.

\subsection{A Monte Carlo algorithm} \label{smontedet}
We begin by proving  the following  stronger form of Theorem~\ref{tmonteintronew}.

\begin{theorem} \label{tmontedet}
A separating  s.s.o.p.  for $\Delta[\det,m]$ 
can be constructed by  a $\poly(m)$-time Monte-Carlo algorithm  that is correct with a high probability.
\end{theorem}
\proof 
Since the determinant can be specified by a circuit with rational constants,
it follows from Gr\"obner basis theory \cite{mayr2} that $\Delta[\det,m]$ has 
defining equations with rational coefficients. For any rational $m \times
m$ matrix $B$, let $H_B$ denote the set of the zeroes of 
the associated homogeneous linear map $\psi_B$ 
(cf. Section~\ref{sNNLdetintro}). 
Given any set ${\cal B}=\{B_1,\ldots,B_k\}$ of 
$m \times m$ rational matrices,  the associated homogeneous linear 
map $\psi_{\cal B}$  (cf. Section~\ref{sNNLdetintro}) does not vanish on 
any non-zero point in $\hat \Delta[\det,m]$ iff 
the variety 
$\hat \Delta[\det,m] \cap \bigcap_{B \in {\cal B}} H_B$ does not have a
non-trivial solution
over $K$. Since $\Delta[\det,m]$ has 
defining equations with rational coefficients,
this  variety also has defining equations 
with rational coefficients. 
By Hilbert's Nullstellensatz,  this variety does not have  a non-trivial
solution over any algebraically closed field of characteristic zero  iff it does not 
have a non-trivial solution over $\C$. 
So, without loss of generality, we can assume that $K=\C$. 

As explained in the beginning of this section,  $\Delta[\det,m] \subseteq P({\cal X})$ is the Zariski-closure 
of the $GL_{m^2}(\C)$-orbit of $\det(X)$, where $X$ is an $m\times m$ variable matrix, and 
${\cal X}$ is  the vector space over $\C$ of homogeneous polynomials of degree $m$ in the 
variable entries of $X$.

Since the Zariski-closure coincides with the closure in the complex topology (cf. Theorem 2.33 in Mumford \cite{mumfordalg}),
it follows that $\Delta[\det,m]$ is the closure 
of the $GL_{m^2}(\C)$-orbit of $\det(X)$ in the complex topology on $P({\cal X})$. 
In concrete terms, this means that every point in the affine cone $\hat \Delta[\det,m]$ of $\Delta[\det,m]$ can be approximated 
infinitesimally closely by  symbolic determinants of size $m$ over the $m^2$ variables  entries of $X$.

Since the determinant has a small circuit, it now follows from Heintz and Schnorr (Theorem~\ref{theintz}) that one can compute 
by a  $\poly(m)$-time Monte Carlo algorithm
a hitting set ${\cal B}=\{B_1, \ldots, B_k\}$, $k=\poly(m)$,  of  integral $m\times m$ 
matrices, with $\poly(m)$ total bit-length,  such that, with a high probability,  (1) for every non-zero polynomial $p(X) \in
\hat \Delta[\det,m]$, there exists a  matrix $B_i \in {\cal B}$ such that $p(B_i)$ is not zero,
and (2) more generally, given any two distinct polynomials $p_1(X), p_2(X) 
\in \hat \Delta[\det,m]$, there exists a matrix $B_i \in {\cal B}$ such that 
$p_1(B_i) \not = p_2(B_i)$. 

We  assume that ${\cal B}$ constructed  above  is  a hitting set with this property. 
Let $S({\cal B})=\{\psi_{B_i}\}$ be the associated subset 
of the homogeneous coordinate ring of $\Delta[\det,m]$, as defined in Section~\ref{sNNLdetintro}.

\begin{claim} \label{csepssop} 
The set  $S({\cal B})$ is a separating s.s.o.p. 
for $\Delta[\det,m]$. 
\end{claim}

Theorem~\ref{tmontedet}  follows from the claim.

\noindent {\em Proof of the claim:} 
First, we prove that the set  $S({\cal B})$ is an  s.s.o.p.
(as defined in  Section~\ref{sNNLdetintro}) for $\Delta[\det,m]$. 

Let $\psi_{{\cal B}}: {\cal X} \rightarrow K^k$ be the homogeneous linear map associated with ${\cal B}$ 
as   in Section~\ref{sNNLdetintro}. The total bit-length of $B_i$'s is $\poly(m)$. So 
it suffices  to show that $\psi_{{\cal B}}$ does not vanish on any non-zero point  in $\hat \Delta[\det,m]$.
By Noether's Normalization Lemma (Lemma~\ref{lnnl0}), it then follows that the homogeneous coordinate ring of $\Delta[\det,m]$ is
integral over the subring generated by $S({\cal B})$.

Suppose to the contrary that $\psi_{{\cal B}}$ 
does vanish on some non-zero polynomial $p=p(X) \in \hat \Delta[\det,m]$. 
Then $p(B_i)=0$ for all $i \le k$. Since $p(X)$ can be approximated infinitesimally closely by symbolic determinants over $X$ of size $m$
and  ${\cal B}$ is a hitting set, this implies that $p(X)$ is identically zero; a contradiction. 

It remains to  show that  $S({\cal B})$ is separating.
Consider any  two distinct polynomials  $p_1(X),p_2(X) \in \hat \Delta[\det,m]$. 
By our assumption about the hitting set ${\cal B}$, $p_1(B_i) \not = p_2(B_i)$ for some $i$. 
This means  $\psi_{\cal B}(p_1) \not = \psi_{\cal B}(p_2)$. 
It follows that $S({\cal B})$ is separating.
\qed

\begin{cor} \label{cssopdet}
A separating s.s.o.p. for $\Delta[\det,m]$ exists.
\end{cor}
\proof This follows from Theorem~\ref{tmontedet}. \qed

\subsection{Reduction of NNL to strengthened black-box derandomization} \label{sdetpoly}
The Monte Carlo algorithm in Theorem~\ref{tmontedet} can  be derandomized assuming a suitable derandomization hypothesis.

\begin{theorem} \label{tnnldetblack}
The variety $\Delta[\det,m]$ has a separating  e.s.o.p.,
assuming the strengthened black-box derandomization hypothesis for symbolic determinant identity testing.
\end{theorem}
\proof 
By the strengthened black-box derandomization hypothesis for symbolic determinant identity testing, we can compute in $\poly(m)$ time a hitting set 
${\cal B} \subseteq M_m(\Z)$  against all non-zero polynomials of degree $m$ that can be approximated infinitesimally closely by
symbolic determinants of size $m$ over the entries of $X$, an $m\times m$ variable matrix. More generally, we can also assume that,
given any two distinct polynomials $p_1(X)$ and $p_2(X)$ that can be approximated infinitesimally closely by
symbolic determinants of size $m$ over  $X$, there exists $b\in {\cal B}$ such that $p_1(b) \not = p_2(b)$.
It then follows, as in the proof of Theorem~\ref{tmontedet}, that the associated set 
$S({\cal B})$ is  a  separating  s.s.o.p. for $\Delta[\det,m]$.
\qed 

\subsection{Reduction of  NNL  to a  lower bound hypothesis} \label{snnldetquasi}
The strengthened black-box derandomization hypothesis in Theorem~\ref{tnnldetblack} can  be traded with a lower bound
hypothesis  as in the following result.

\begin{theorem} \label{tnnldetlower0}
The variety  $\Delta[\det,m]$ has a  separating quasi-e.s.o.p., assuming
that there exists a family $\{f_n(x_1,\ldots,x_n)\}$ of
exponential-time-computable, integral, multi-linear polynomials such that $f_n$
cannot be approximated infinitesimally closely by  circuits 
over $K$ of  $O(2^{n^\epsilon})$ size, for some constant $\epsilon >0$, as $n\rightarrow \infty$.
Alternatively, we can assume that $f_n$ 
cannot be approximated infinitesimally closely by symbolic determinants 
over $K$ of  $O(2^{n^{\epsilon'}})$ size, for some constant $\epsilon' >0$, as $n\rightarrow \infty$.
\end{theorem}

\proof The first statement   follows  from the proof of
Theorem~\ref{tnnldetblack} and Theorem~\ref{trussell2},
since symbolic determinant identity testing is a special case 
of low-degree polynomial identity testing. By \cite{skyum}, any 
circuit over $K$ of degree $d$ and size $s$ can be simulated by a circuit over $K$
of  $O(\log d (log d + \log s))$ depth, and hence \cite{valiant2}, by a 
symbolic determinant over $K$  of  $O(2^{O(\log d (\log d + \log s)})$ size. 
The second statement follows from the first statement, in conjunction with this fact,
letting $d=n$, $s=2^{n^\epsilon}$, and $\epsilon'=2 \epsilon$. 
\qed

Assuming a lower bound for the permanent, we  get  the following stronger result.
This proves a stronger form of Theorem~\ref{tintronnldetnew}.

\begin{theorem} \label{tnnldetlower}
A separating  s.s.o.p. for $\Delta[\det,m]$ can be constructed in $O(\polylog(m))$ 
parallel time using 
$O(2^{\polylog(m)})$ processors,
assuming that the permanent of $n \times n$ matrices 
cannot be approximated infinitesimally closely  by  symbolic determinants  over $K$ of  $O(2^{n^\epsilon})$ size, for some
constant $\epsilon >0$, as $n\rightarrow \infty$.
\end{theorem}

\proof This  follows from Theorem~\ref{tnnldetblack} and Theorem~\ref{trussellperm},
since low-degree algebraic circuits of sub-exponential size are equivalent  to
symbolic determinants of sub-exponential size; cf. 
the proof of Theorem~\ref{tnnldetlower0}. \qed

Define the variety $\Delta[\perm,n,m]$, just as we defined $\Delta[\det,m]$ at the beginning of this section, replacing 
$\det(X)$   by $z^{m-n}\perm(Y)$, where $Y$ is some $n\times n$ sub-matrix of $X$, and $z$ is a variable entry in $X$ outside $Y$.
Then the lower bound assumption in Theorem~\ref{tnnldetlower} in the terminology of \cite{GCT1}  is that 
$\Delta[\perm,n,m] \not \subseteq \Delta[\det,m]$, if $m=O(2^{n^\epsilon})$, for some small enough constant $\epsilon >0$.
This is a stronger form of Conjecture 4.3 in \cite{GCT1}.
If we assume instead (as in Conjecture 4.3 in 
\cite{GCT1}) that $\Delta[\perm,n,m] \not \subseteq \Delta[\det,m]$,
if $m=O(\poly(n))$, then it can be proved  similarly that NNL for $\Delta[\det,m]$ can be solved in  $O(2^{n^\epsilon})$-time
(after  replacing $\poly(n)$ by $2^{n^\epsilon}$ in the definition of an s.s.o.p.), for every constant $\epsilon >0$.

\subsubsection{Equivalence} \label{sequivalencedet}
The following result implies the easy converse to  Theorem~\ref{tnnldetblack}. 

\begin{lemma} \label{lnnldetblack}
Suppose ${\cal B} \subseteq M_m(\Z)$ specifies an  s.s.o.p. $S({\cal B})$ for $\Delta[\det,m]$. 
Then ${\cal B}$ is a hitting set against all non-zero polynomials of degree $m$ that can be approximated infinitesimally closely by
symbolic determinants
of size $m$ over the $m^2$  entries of $X$, an $m\times m$ variable matrix. 
Hence, existence of an e.s.o.p. for  $\Delta[\det,m]$ implies the strengthened black-box derandomization
hypothesis for  symbolic determinant identity testing. 
\end{lemma}

\proof
By the definition of $\Delta[\det,m]$, every non-zero polynomial $p(X)$ of degree $m$ that can be approximated infinitesimally closely
by symbolic determinants over $X$ of size $m$ corresponds to a non-zero point in $\hat \Delta[\det,m]$. 
Since $S({\cal B})$ is an s.s.o.p., it follows that $\psi_{\cal B}$ (as defined in Section~\ref{sNNLdetintro}) does not vanish 
on any non-zero point in $\hat \Delta[\det,m]$. This implies that 
${\cal B}$ is a hitting set against all non-zero polynomials of degree $m$ that can be approximated infinitesimally closely by
symbolic determinants of size $m$ over the $m^2$ entries of $X$.

In symbolic determinant identity testing, we can assume, without loss of generality, that the number of variables is at most quadratic in the size of the matrix, increasing the size otherwise.
Hence the last statement follows.
\qed 

The following result proves  Theorem~\ref{tequivnnlnew}.

\begin{theorem}  \label{tequisditdet}
\noindent (a)The strengthened black box derandomization hypothesis for  symbolic determinant identity testing holds iff 
$\Delta[\det,m]$ has an e.s.o.p.

\noindent (b) A sub-exponential lower bound for a family of 
exponential-time-computable, integral, multi-linear polynomials as in 
Theorem~\ref{tnnldetlower0} holds
iff, ignoring  quasi-prefixes,  $\Delta[\det,m]$ has an e.s.o.p.
\end{theorem}
\proof 
(a) This follows from Theorem~\ref{tnnldetblack} and Lemma~\ref{lnnldetblack}.

(b) This follows from (a),  Theorem~\ref{trussell2}, and Proposition~\ref{pconverserussell}. \qed 

\subsection{The current best unconditional deterministic upper bound for NNL} \label{scurrentdet}
The following result gives the current best unconditional deterministic bound for NNL for $\Delta[\det,m]$.
It does not follow from Theorem~\ref{tnnldetexpspace}, since an h.s.o.p. constructed there 
need not have a succinct specification of  $\poly(m)$ bit-length.

\begin{theorem} \label{tssopverifydet}
The problem of constructing or verifying an  s.s.o.p. for $\Delta[\det,m]$ belongs to EXPSPACE unconditionally. It belongs  to 
EXPH assuming the Generalized Riemann Hypothesis.
\end{theorem}

\proof 
The statement for construction follows from Theorem~\ref{tstrongexp} 
and  Theorem~\ref{tequisditdet} (a). The proof for
verification  is implicit in the proof for  construction. \qed

\section{Explicit algebraic varieties} \label{sexplicit}
In this section we formulate a general notion of 
an explicit algebraic variety,  motivated by the concrete example of
$\Delta[\det,m]$ studied in the preceding section, and  define the problem NNL in this context (cf. Section~\ref{simpliexp}).
We then generalize the results for $\Delta[\det,m]$ in the preceding section systematically to general explicit varieties.

\begin{defn} \label{dexpvariety}
\noindent (a) A family $\{W_n\}$, $n\rightarrow \infty$,  of affine varieties is called 
{\em explicit} 
if there exist families of positive integers $\{r_n\}$, $\{m_n\}$,  a family $\{\psi_n\}$ of maps
$\psi_n: K^{r_n} \rightarrow K^{m_n}$:
\begin{equation} \label{eqpsiexp}
v=(v_1,\ldots,v_{r_n}) \rightarrow (f_1(v),\ldots,f_{m_n}(v)),
\end{equation}
with $r_n=\poly(n)$, $m_n=n^{\Omega(1)}$, $\log m_n = O(\poly(n))$, 
and each $f_j$  a homogeneous polynomial of $\poly(n)$ degree, and there also exist
homogeneous polynomials $g_j(x)$, $x=(x_1,\ldots,x_n)$,  $1\le j \le m_n$, of $\poly(n)$ degree,
such that:
\begin{enumerate} 
\item  $W_n$ is the Zariski-closure of the image $\mbox{Im}(\psi_n)$ of $\psi_n$. This means $W_n \cong \spec(R)$, where
$R$ is the subring of $K[v_1,\ldots,v_{r_n}]$ generated by $f_1(v),\ldots,f_{m_n}(v)$.
\item The polynomial $F_n(v,x)=\sum_j f_j(v) g_j(x)$ is uniformly
$p$-computable \cite{valiant2}. This means one can compute in $\poly(n)$ time a circuit $C_n$,
with rational constants,  over the variables $v=(v_1,\ldots,v_{r_n})$ and 
$x=(x_1,\ldots,x_n)$ of $\poly(n)$  total bit-size, including the bit-sizes of 
the constants,   that computes $F_n(v,x)$, and the  total degree  $\deg(F_n)$ of $F_n$ 
is $\poly(n)$.
\item The polynomials $g_j(x)$'s are linearly independent.
\end{enumerate}

We call $\psi_n$ the map defining $W_n$, and $F_n$ the polynomial defining 
$W_n$. We specify $W_n$ succinctly by the circuit $C_n$. 
Alternatively, we can specify $W_n$ by the circuits $C_{n,c}$'s, $1 \le c \le 
\deg(F_n)$,  where $C_{n,c}$  computes the degree $c$-component in $v$ of $F_n$. The total bit-length of this succinct 
specification of $W_n$ is  $\poly(n)$.

We say that $\{W_n\}$ is {\em strongly explicit} if the circuit $C_n$ is weakly skew (cf.
Section~\ref{sterminology}).

\noindent (b) A family of projective varieties is called explicit (strongly explicit) 
if the family of the affine cones
of these varieties is explicit (respectively, strongly explicit).

\noindent (c) An {\em explicit  family of  affine or projective varieties
without  degree restrictions} 
is defined just as in (a) and (b), 
but without putting any restriction on the degrees of $f_j,g_j$, and $F_n$. 

\noindent (d) Quasi-explicit families  are defined by replacing $\poly(n)$ by $2^{\polylog(n)}$. 
\end{defn}

We denote the coordinate ring of $W_n$ by $K[W_n]$.
If $\{W_n\}$ is explicit, by abuse of terminology, we also say that the variety $W_n$ is explicit.

\subsection{Examples} \label{sexample}
We now give a few examples of explicit varieties.

\subsubsection{The orbit closure of the determinant} \label{sorbitclosuredet}
The orbit-closure $\Delta[\det,m] \subseteq P({\cal X})$ studied in Section~\ref{snnldeter} is explicit. Specifically, 
following the same notation as in Section~\ref{snnldeter}, 
the affine cone $\hat \Delta[\det,m]$ of  $\Delta[\det,m]$ is explicit, with the defining
map $\phi: M_{m^2}(K) \rightarrow {\cal X}$ that maps $v \in M_{m^2}(K)$ to 
$\det(v X)$, thinking of $X$ as an $m^2$-vector. 
The polynomial $F=F(v,X)$ defining $\Delta[\det,m]$  is  $\det(v X)$.
The monomials in the entries of $X$ of degree $m$ play the role of $g_j$'s  in Definition~\ref{dexpvariety}, and
$f_j$'s are the coefficients of  $\det(v X)$ considered 
as a polynomial in $X$.

\subsubsection{Explicit varieties associated with depth three  circuits} \label{lexporbitclosuresthree}
Let $S^d_n$ be the space of  homogeneous forms in $n$ variables of degree $d$, and $P(S^d_n)$ the associated projective space.
Let $Y(d,k,n) \subseteq P(S^d_n)$ be
the projective closure of the set of polynomials that can be expressed as sum of $k$ terms, each term a  $d$-th power of a  linear form in 
the $n$ variables. 
It is the  variety associated with the class of diagonal 
depth three circuits (cf. Section~\ref{sterminology}) on $n$ variables with degree $d$ and top-fan-in $k$, and
is known in algebraic geometry as the $k$-th secant variety of the Veronese  variety  \cite{landsberg}. 
It is explicit, the defining polynomial being the polynomial computed by the generic,
homogeneous,  diagonal depth three circuit (with 
indeterminate constants) on
$n$ variables with degree $d$ and top fan-in $k$. Specifically, this defining polynomial
is $\sum_{i=1}^k (\sum_{j=1}^n y_{i,j} x_j)^d$, where $x_j$'s are the variables in the 
circuit, and  $y_{i,j}$'s are the indeterminate constants.

Let $X(d,k,n) \subseteq P(S^d_n)$ be
the projective closure of the set of polynomials that can be expressed as sum of $k$ terms, each term a product of $d$ linear forms
in the $n$ variables.
It is the  variety associated with the class of  depth three circuits on $n$ variables with degree $d$
and  top-fan-in $k$, and is known in algebraic geometry as the $k$-th secant variety of the Chow variety  \cite{landsberg}. 
It is explicit, the defining polynomial being the polynomial computed by the generic,
homogeneous,  depth three circuit (with 
indeterminate constants) on $n$ variables with degree $d$ and top fan-in $k$. 
Specifically, this defining polynomial 
is $\sum_{i=1}^k \prod_{r=1}^d (\sum_{j=1}^n y_{i,r,j} x_j)$, 
where $x_j$'s are the variables in the 
circuit, and  $y_{i,r,j}$'s are the indeterminate constants. 

\subsubsection{The explicit variety associated with the universal  circuit} \label{shy}
Following \cite{GCT1}, we  now define  an explicit variety without any degree restrictions, which 
plays the same role  in the study of  general polynomial identity testing
that $\Delta[\det,m]$ plays in the study of symbolic determinant identity testing.

First, we  define a universal  circuit over $K$ of depth $k$ and
width $m$. Let $S_i$, $0 \le i \le k$, denote the set of nodes in  this circuit
with level $i$. We assume  that $S_0$ contains just one node, called the {\em root},
and for all $i>0$, $|S_i|=m$.  
For all levels $0\leq i\leq k-1$, we introduce
indeterminates $y_{v,w}^u$'s for each $u\in S_i $ and 
 distinct $v,w\in S_{i+1}$. For the $k$-th level, we introduce  
indeterminates  
$y^u$'s, $u\in S_k $. Let $Y$ be the tuple of all these indeterminates together.
Beginning at the level $k$, for each element 
$u$ in $S_i $, we recursively  define  
the form $h(u)$ in the indeterminates $Y$ as follows.
For  $u\in S_k $, let $h(u)=y^u$. For $u\in S_i $,
with $i<k$, let $h(u) =\sum_{v,w} y^u_{v,w} h(v)h(w) $, where 
the sum ranges over all distinct $v,w \in S_{i+1} $. 
The form $H(Y)=H_{k,m}(Y)$ computed by this universal circuit 
is the form $h(u)$,  where $u\in S_0$ is the root. 

Any circuit over $K$ of size $s$ can 
can be obtained by specializing this  universal 
circuit with $k=O(s)$ and $m=O(s)$; cf. \cite{GCT1}. 

Let ${\cal X}$  be the space of homogeneous forms in $Y$ of total degree $d:=\deg(H(Y))$ over the field $K$. 
Let $l$ denote the number of variables in $Y$. Then ${\cal X}$ has a natural action of $G=GL_l(K)$, similar to the action  in Section~\ref{sNNLdetintro}. 
Let $\Delta[H(Y),k,m] \subseteq P({\cal X})$ denote the closure of the $G$-orbit of $H(Y)$ in $P({\cal X})$.
This  is an explicit variety 
without any degree restrictions (cf. Definition~\ref{dexpvariety} (c)). 

We also define  an explicit variety (with the usual low-degree restrictions), which 
plays the same role  in the study of  low-degree  polynomial identity testing
that $\Delta[\det,m]$ plays in the study of symbolic determinant identity testing.

Given any positive integer $c$, let $H(Y)_c$ denote the homogeneous degree $c$ part of $H(Y)$.
If $c=\poly(k,m)$, then $H(Y)_c$ can be computed by a circuit of $\poly(k,m)$ size,
and furthermore, the family $\{H(Y)_m\}$, with $k=c=m$, is VP-complete; cf. Section 5.6 in \cite{burgbook}.
Now let ${\cal X}$  be the space of homogeneous forms in $Y$ of total degree $m$ over the field $K$.
Define $\Delta[H(Y)_m,k,m]$ just as we defined $\Delta[H(Y),k,m]$ above, with $H(Y)_m$ in place of $H(Y)$.
This  is an explicit variety,
with the usual low-degree restrictions (cf. Definition~\ref{dexpvariety}).

\subsubsection{The categorical quotients} \label{scategorical}
Let $V$ be a representation of  $G=SL_m(K)$ of dimension $n$.
Then the invariant ring $K[V]^G$ is finitely generated \cite{hilbert}. So we can consider the
variety  $V/G=\spec(K[V]^G)$, called the {\em categorical quotient}  \cite{mumford}. 
It can be constructed  concretely as follows.

Fix any set $F=\{f_1,\ldots,f_t\}$  of non-constant homogeneous generators of $K[V]^G$. 
Consider the morphism $\pi_{V/G}$ from $V$ to $K^t$ given by 
\begin{equation} \label{eqmorphismgeneral}
\pi_{V/G}: \quad v \rightarrow (f_1(v),\ldots,f_t(v)).
\end{equation}
Then $V/G$ can be identified with the closure of the image of this morphism. As we shall
see below, this image is  already closed (cf. Theorem~\ref{tmumfordnew}).
Let $z=(z_1,\ldots,z_t)$ be the coordinates of $K^t$,  $I$ 
the ideal of $V/G$ under this embedding, and  $K[V/G]$ its coordinate ring. 
Then $K[V/G]=K[z]/I$, and we have the comorphism $\pi_{V/G}^*: K[V/G] \rightarrow K[V]$ given by 
\begin{equation}\label{eqmorphism2}
 \pi_{V/G}^*(z_i)=f_i. 
\end{equation}
Since $f_i$'s are homogeneous, $K[V/G]$ is a graded ring, with the grading given by $\deg(z_i)=\deg(f_i)$.
Furthermore, $\pi_{V/G}^*$ gives an isomorphism between $K[V/G]$ and $K[V]^G$. Thus, we have
$\pi_{V/G}^*(K[V/G])=K[V]^G$.

The general definition of explicit varieties (Definition~\ref{dexpvariety}) 
specializes, when applied to the map $\pi_{V/G}$ in (\ref{eqmorphismgeneral}),  to the following definition.
Let $v_1,\ldots,v_n$ denote the standard monomial basis \cite{raghavan}  of $V$ as in Section~\ref{sgeneralintro}.

\begin{defn} \label{dexpffthilbert}
\noindent (a) The categorical quotient   $V/G$ is  called {\em explicit}  if,
given the specification $\langle V, G \rangle$ of $V$ and $G$ as in Section~\ref{sgeneralintro},
one can compute in $\poly(n,m)$ time a set of  circuits
$C=C[V,m,c]$'s, $1 \le c \le q=\poly(n,m)$,   over $\Q$ of $\poly(n,m)$ bit size, including the bit-sizes of the constants, and over 
the variables $x=(x_1,\ldots,x_l)$, $l=\poly(n,m)$,  and $v=(v_1,\ldots,v_n)$, such that 
the polynomials $C[V,m,c](x,v)$'s  computed by  $C[V,m,c]$'s are of $\poly(n,m)$ degree and can be expressed  in the form 
\begin{equation} \label{eqfj}  C[V,m,c](x,v)=\sum_j f_{j,c}(v) g_{j,c}(x), \end{equation}
with homogeneous $f_{j,c}$'s $\in K[V]^G$ and $g_{j,c}$'s, 
so that $K[V]^G$ is  generated by  $f_{j,c}(v)$'s,  and $g_{j,c}(x)$'s are linearly independent.

\noindent (b) It is called {\em explicit without any degree restrictions}
if the degree requirement on $C[V,m,c](x,v)$'s is dropped.

\noindent (c) It  is called {\em strongly explicit} if, in addition to all the properties in (a), 
 the circuits $C[V,m,c]$'s are weakly skew (cf. Section~\ref{sterminology}).

\noindent (d) If $V/G$ is explicit, we  say that an {\em explicit First Fundamental Theorem} holds for 
$K[V]^G$, with the circuits $C[V,m,c]$'s constituting an
explicit (polynomial-time-computable)  encoding of a set of  generators for $K[V]^G$.

If $V/G$ is strongly explicit, we  say that a strongly explicit First Fundamental Theorem holds for
$K[V]^G$. 

\noindent (e) The notions in (a)--(d) are defined in the {\em relaxed sense} 
by requiring that  $f_{j,c}(v)$'s in (\ref{eqfj})
only  form a set of separating 
invariants \cite{derksenbook} (cf. also  Section~\ref{sstit} here) of $K[V]^G$, rather than
a set of generators.
\end{defn}

We are abusing the terminology a bit here.
Formally, instead of saying that $V/G$ is explicit, we should really be saying that the family 
$\{W_{\langle V, G \rangle}\}$, indexed by the specification $\langle V, G \rangle$,
where $W_{\langle V, G \rangle}:=V/G$, is explicit.

\begin{conj} \label{cexplicitcategorical}
The categorical quotient  $V/G$  is   explicit,
without any degree restrictions in general.
\end{conj}

It may be conjectured that $V/G$ is  explicit (with the usual low-degree 
restrictions), if $K[V]^G$ has a set of 
generators of $\poly(n,m)$ degree.

For all the applications in this article and in geometric
complexity theory (cf. Remark 2 after Theorem~\ref{tmarbitrary}),
a weaker form of this conjecture stipulating  explicitness of
$V/G$  only in the relaxed sense (cf. Definition~\ref{dexpffthilbert} (e)) suffices.

Conjecture~\ref{cexplicitcategorical}
is proved  in this article for
$V=M_m(K)^r$ with the adjoint action of $G$ (cf. Theorem~\ref{texplicitmatrix}), and 
for arbitrary $V$ when $m$ is constant
(cf. Theorem~\ref{texplicitconst0}).  The relaxed form 
of the  conjecture in positive characteristic 
is also proved  for the ring of matrix invariants
(cf. Theorem~\ref{texprelaxedmatrix}).

For constant $m$, we shall construct  a $C[V,m,c]$ with
depth four; cf. Theorem~\ref{texplicitconst0}.
The degrees of the generators encoded by 
$C[V,m,c]$ are at most exponential in its depth.
Comparing this bound with the degree bound 
in Derksen \cite{derksen} (cf. Theorem~\ref{tderksen}), one may  expect $C[V,m,c]$ in 
Conjecture~\ref{cexplicitcategorical} to have  $O(\poly(m,\log n))$  depth in general.

The simplest instance of the conjecture that the reader can check  is the following. 
Let $G=SL_m(K)$, and $V=K^m \oplus \cdots K^m$ ($r$ times), with the action of $G$ from the left. 
The coordinate ring $K[V]$ can  be identified with  the ring $K[U]$ generated by the entries of an $m \times r$ variable matrix $U$.
By the First Fundamental Theorem  of invariant theory 
\cite{fultonrepr,weyl}, 
the invariant ring $K[V]^G$ in this case  is generated by the $r \times r$ minors of $U$. 
The corresponding  map (\ref{eqmorphismgeneral})  in this case is the well-known
 Pl\"ucker map \cite{fultonrepr}
$U \rightarrow (\ldots, m_\alpha(U), \ldots)$, where $m_\alpha(U)$ ranges over all $r \times r$
minors of $U$. 
The categorical quotient $V/G$ in this case 
is the  Grassmanian.  It can be checked that the Grassmanian is strongly explicit, with the defining map being
the Pl\"ucker map. 

For explicit varieties in general, the image of the map $\psi_n$ in (\ref{eqpsiexp}) need not be closed. In contrast, for
categorical quotients we have:

\begin{theorem}[Mumford, Fogarty, and Kirwan \cite{mumford}] 
(cf. Theorem 1.1 in \cite{mumford} and Theorem 4.6 and 4.7 in \cite{popovbook}) 
\label{tmumfordnew}

\noindent (a) 
The image of $\pi_{V/G}$ in (\ref{eqmorphismgeneral}) is  closed. Hence, the map $\pi_{V/G}: V \rightarrow V/G$ is surjective.

\noindent (b) For any $x \in V /G$, $\pi_{V/G}^{-1}(x)$ contains a unique closed $G$-orbit.

\noindent (c) For any $G$-invariant (closed) subvariety $W \subseteq V$, $\pi_{V/G}(W)$ is a closed subvariety of $V/G$.

\noindent (d) Given $v,w \in V$, the closures of the $G$-orbits of $v$ and $w$ intersect iff $r(v)=r(w)$ for all $r \in K[V]^G$.
\end{theorem}

These additional properties of $V/G$ play a crucial role in this article; cf. Remark 3 after 
Theorem~\ref{tclosedimage}.

\subsubsection{Explicit variety associated with  $p$-computable polynomials} \label{scircuitexp}
Let $\{p_n(v,x)\}$, $v=(v_1,\ldots,v_r)$, $r=r_n= \poly(n)$,  $x=(x_1,\ldots,x_n)$, 
be a uniform $p$-computable \cite{valiant2} family of  polynomials, homogeneous in $v$, with
rational constants.
This means $p_n(v,x)$ has $\poly(n)$ degree, has a circuit over $\Q$ of $\poly(n)$ bit-size,
and, given $n$,  the specification of this  circuit  can be computed in $\poly(n)$ time.

Let $p_n(v,x)=\sum_\mu f_\mu(v) \mu(x)$, where $\mu$ ranges over all monomials
in $x$ of  degree $\le \deg(p_n)=\poly(n)$. Let $m=m_n$ be the number of 
such monomials. Let $\psi=\psi_n$ 
be the map
\[ \psi: v \in K^r \rightarrow (\ldots,f_\mu(v),\ldots) \in K^m.\] 
Then $\{W_n=\overline{Im(\psi_n)}\}$ is an explicit family of 
varieties, with the defining map $\psi_n$ and the defining polynomial $p_n$.

\subsubsection{Explicit toric variety}
\label{storic}
Let $\{p_n(x)\}$, $x=(x_1,\ldots,x_n)$, be a uniform $p$-computable family of homogeneous
polynomials over $x$ and $K$. Let 
$p_n(x)=\sum_\mu a_\mu \mu(x)$, where $a_\mu \in K$, and 
$\mu$ ranges over all monomials
in $x$ of total degree $=\deg(p_n)=\poly(n)$. Let $m=m_n$ be the number of 
such monomials. Consider the monomial map $\psi_n$:

\[ \psi_n: v=(v_1,\ldots,v_n) \in K^n \rightarrow  (\ldots, a_\mu \mu(v),\ldots) \in  K^m.\] 
Let $W_n=\overline{Im(\psi_n)}$, and $P(W_n)$ its projectivization. Then 
$\{P(W_n)\}$ is  an explicit family of  toric varieties, with the defining polynomial 
\[ F_n(v,x)=\sum_\mu a_\mu \mu(v) \mu(x).\] 
This polynomial  is $p$-computable and uniform, since a circuit for computing it can be
obtained from the one for $p_n$ by replacing each  $x_i$ with $v_i x_i$. 

The main difference between the explicit toric variety here and the 
more general explicit variety in Section~\ref{scircuitexp} is that $\mu(v)$ here 
is a monomial, whereas $f_\mu(v)$ in
Section~\ref{scircuitexp} can be any homogeneous polynomial.

\subsection{Unconditional upper bound for the problem of constructing an h.s.o.p.} \label{shsopgenexp}
We now study the problem of constructing an h.s.o.p. for an explicit variety.
The following  generalization of  Theorem~\ref{tnnldetexpspace} gives the currently best upper bound for this problem.

\begin{theorem} \label{tnnlexplicithsop}
The problem of constructing an h.s.o.p. for an explicit variety $W_n$ (cf. Definition~\ref{dexpvariety}) 
belongs to EXPSPACE. (This means it can be solved in work-space that is exponential in $n$.)

Assuming the Generalized Riemann Hypothesis, it belongs to EXPH (the exponential hierarchy), if 
$W_n$ has  defining equations
that can be computed in time that is exponential in $n$.
\end{theorem}

\proof The proof is similar to that of Theorem~\ref{tnnldetexpspace},
with $W_n$ in place of $\Delta[\det,m]$.
\qed

\subsection{The problem NNL  for explicit varieties}  \label{simpliexp}
If we insist on an h.s.o.p., then Theorem~\ref{tnnlexplicithsop} is the best that we can do at present.
However, if we are  willing to settle for a small homogeneous set $S \subseteq K[W_n]$ of $\poly(n)$ size, but not necessarily of the
optimal size, such that $K[W_n]$ is integral over the subring generated by $S$, then Theorem~\ref{tnnlexpfull}
proved  below says that we can do much better.
Relaxing the optimality constraint on  cardinality, 
but insisting on succinctness of specification in exchange, we are thus led  to the following notion of an s.s.o.p.
It   generalizes  the notion  of an s.s.o.p for $\Delta[\det,m]$ (cf. 
Section~\ref{sNNLdetintro}) to arbitrary explicit varieties.

\begin{defn} \label{desopexp}
Let $\{W_n\}$ be an explicit family of varieties  as in
Definition~\ref{dexpvariety}, 
$z_1,\ldots,z_{m_n}$  the coordinates of the ambient 
space $K^{m_n}$ containing $W_n$, and 
$\psi_n^*$ the comorphism of $\psi_n: K^{r_n} \rightarrow K^{m_n}$ in (\ref{eqpsiexp}).
Note  that $K[W_n]$ is graded, with  $\deg(z_j)=\deg(f_j)$.

\noindent (a) We say that  $s \in K[W_n]$ has  a {\em short specification} if 
$\psi^*_n(s)$ has a  circuit
over $\Q$ and  $v_1,\ldots,v_{r_n}$  of  $O(\poly(n))$  bit-length (not just size),
which   computes
the polynomial function on $K^{r_n}$ corresponding to $\psi_n^*(s)$.

\noindent (b) We say that a set $S \subseteq K[W_n]$ is a {\em small  system of
parameters (s.s.o.p.)} for $K[W_n]$ (and $W_n$) if 
(1) each element $s \in S$ has a short specification as in  (a) and 
is homogeneous of $\poly(n)$ degree, 
(2) $K[W_n]$ is integral over its subring  generated by $S$, and
(3) the size of $S$ is $\poly(n)$.

We say that $S$ is an {\em  explicit system of parameters (e.s.o.p.)} if, in addition, 
the specification of $S$, consisting of a circuit
for  $\psi_n^*(s)$ for each $s \in S$ as in (a), can be computed in $\poly(n)$ time.

If $W_n$ is strongly explicit then, by convention, we assume that the short specification as in (a) for
each element of $s \in S$ is a weakly skew circuit (cf. Section~\ref{sterminology}).

\noindent (c) S.s.o.p. and  e.s.o.p. 
without any degree restrictions are defined by dropping the degree requirement in (b) (1).
Quasi-e.s.o.p. and quasi-s.s.o.p. are defined by replacing $\poly(n)$ by $2^{\polylog(n)}$.

\noindent (d) We call $S$ 
{\em separating} if, for any two distinct points $u,v \in W_n$, 
there exists an $s \in S$ such that 
$s(u)\not = s(v)$. 

\end{defn} 

Let $F_n$ and $C_n$ be as in Definition~\ref{dexpvariety}.
For any $0 \le c \le \deg(F_n)$, let $C_{n,c}$ be the circuit computing the degree $c$-component (in $v$) of $F_n$. 
If $\deg(F_n)$ is $\poly(n)$, then, given $C_n$ and $c$, 
we can compute $C_{n,c}$ in $\poly(n)$ time using the Vandermonde interpolation technique 
as per   Strassen \cite{strassen}; cf.  also the survey by Shpilka and Yehudayoff \cite
{yehu}.
Alternatively, the explicit variety $W_n$ can  be specified
by giving the circuits $C_{n,c}$'s, $1 \le c \le \deg(F_n)$, instead of the circuit $C_n$.
For any  $b\in \N^n$ of $\poly(n)$ bit-length, let $C_{n,c,b}$ be the instantiation of $C_{n,c}$ at $x=b$. 

\begin{defn} \label{dstrict}
We say that  $s \in K[W_n]$ is {\em strict} if, for some $b\in \N^n$ 
of $\poly(n)$ bit-length
and $0<c \le \deg(F_n)$,   $\psi_n^{*}(s)(v)=C_{n,c,b}(v)$. This means $s  = \sum_j z_j g_j(b)$,
where  $j$ ranges over all indices such that $\deg(f_j)=c$.
Such a strict $s$ can be specified succinctly by the triple  $(b,c, C_n)$, or the pair $(b,C_{n,c})$, or just $b$ if 
$C_{n,c}$ is implicit (as in the case of the explicit variety $\Delta[\det,m]$).

We say that  an s.s.o.p. or an e.s.o.p. $S$ is {\em strict} if each $s\in S$ is strict.
It can then be specified by the set of pairs $(b,C_{n,c})$'s, or just by the set of $b$'s if $C_{n,c}$'s are implicit.
Strict quasi-s.s.o.p. and quasi-e.s.o.p. are 
defined by replacing $\poly(n)$ by $2^{\polylog(n)}$.
\end{defn}

Thus a strict  s.s.o.p. 
has a short specification (cf. Definition~\ref{desopexp} (a)) 
based on the circuit $C_n$ defining the variety $W_n$ itself.  As such
strictness  is a natural way to ensure succinctness. 
We shall prove  later (cf. Corollary~\ref{corssopexistsgen}) that a strict s.s.o.p. exists.

By {\em NNL for $W_n$}, we mean the problem of constructing an s.s.o.p. for $K[W_n]$.
By {\em NNL in a strong form  for $W_n$}, we mean the problem of constructing a separating  s.s.o.p. for $K[W_n]$.
By {\em NNL in a strict and strong form  for $W_n$}, we mean the problem of constructing a strict, separating  s.s.o.p. for $K[W_n]$.
We say that  NNL for $W_n$ has an explicit solution, if $K[W_n]$ has an e.s.o.p.

As the reader can check, an s.s.o.p. for $\Delta[\det,m]$ defined in Section~\ref{sNNLdetintro} is a specialization of the 
general definition of a strict s.s.o.p. given above.
The variety $W_n$ here is the variety $\Delta[\det,m]$ there, the map $\psi_n$ here is the map $\phi: M_{m^2}(K) \rightarrow {\cal X}$ in Section~\ref{sorbitclosuredet},
and a strict s.s.o.p. $S$  specified by a set of $b$'s here is  $S({\cal B})$ specified by
a set ${\cal B}$ of $m\times m$ matrices in Section~\ref{sNNLdetintro}.

Strictness is used in the proof of  Theorem~\ref{tequisditdet} to derive
a lower bound from  NNL; cf. the proof of Lemma~\ref{lnnldetblack}.
It is open  if similar lower bounds can be derived from non-strict NNLs.
The s.s.o.p.'s constructed in all the main results of this article 
stated in Section~\ref{sintro} are strict.

For simplicity, in what follows, 
we often keep $n$ implicit and  denote $W_n$ by $W$, $m_n$ by $m$, $r_n$ by $r$,
$\psi_n$ by $\psi$, and so on.

\subsection{Monte Carlo algorithm}  \label{smontecarloexp}
The following generalization of Theorem~\ref{tmontedet}
proves a stronger form of Theorem~\ref{tmonteintroexplicitnew}.

\begin{theorem} \label{tmonteexplicit}
Let $\{W_n\}$ be an explicit family of varieties. Then
there  is a $\poly(n)$-time Monte Carlo algorithm to construct a separating, strict  s.s.o.p. for $K[W_n]$, which is correct with
a high probability.
\end{theorem}
\proof 
Let  $W=W_n \subseteq K^m$, $m=m_n$,  be an explicit variety as in Definition~\ref{dexpvariety}, 
and $C_n$  the circuit computing $F_n(v,x)$, $v=(v_1,\ldots,v_r)$, $r=r_n$,  and $x=(x_1,\ldots,x_n)$,  as there. 
Let $s=\poly(n)$ be its size, and $d=\poly(n)$ its degree. Let $u= 2 s (d+1)^2$.

Choose  $T \subseteq [u]^n$ of size $6 (s+1 + n)^2$ randomly. By Theorem~\ref{theintz}, it is 
a hitting set  with a high probability against  all nonzero polynomials  $h(x)$ of degree $\le d$ that can be 
approximated infinitesimally closely  by  circuits over $K$ and $x$ of size $\le s$. 
More strongly, replacing $s$ by $2 s +1$, we can also assume  that, given
any two distinct polynomials $h_1(x)$ and $h_2(x)$ of degree $\le d$ that 
can be 
approximated infinitesimally closely  by  circuits over $K$ and $x$ of size $\le s$, 
there exists $b \in T$ such that $h_1(b) \not =  h_2(b)$.

This probabilistic construction of $T$  takes $\poly(s)=\poly(n)$ time.
In what follows, we assume that $T$ is such a hitting set.

For each $b \in T$ and $0< c \le \deg(F_n)$, define
$h_{b,c}(z):=\sum_j z_j g_j(b) \in K[W_n]$, 
where $j$ ranges over all indices such that $\deg(f_j)=c$, and $z=(z_1,\ldots,z_m)$ denote the coordinates of the ambient
space $K^m$ containing $W=W_n$.
Then $\deg(h_{b,c})=c$, since $\deg(z_j)=\deg(f_j)$, as per the grading on $K[W_n]$. Let 
\begin{equation} \label{eqS}
 S=\{ h_{b,c}(z) \ | \ b \in T, 0< c \le \deg(F_n)\} \subseteq K[W_n].
\end{equation}

It now follows from  Lemma~\ref{lintegral} (c) and (d) below
that $S$ is  a separating, strict s.s.o.p.  \qed 

\begin{lemma} \label{lintegral} 
Suppose  $W=W_n$ is an  explicit variety.
Let $S$ and $T$ be as in (\ref{eqS}). Then:

\noindent (a) $W \cap Z(S) = \{0\}$, where $Z(S)\subseteq K^m$ is the zero set of $S$, and $0$ denotes the origin in $K^m$. 

\noindent (b) The coordinate ring  $K[W]$ is integral over the  subring generated by $S$.

\noindent (c) The set $S$ is a strict  s.s.o.p.

\noindent (d) The set $S$ is also separating.

\end{lemma} 

\proof Let $\psi=\psi_n$, $f_j$, $g_j$, and  $F=F_n(v,x)$ be as in Definition~\ref{dexpvariety}. 

\noindent (a) By Hilbert's Nullstellensatz, we can assume that $K=\C$, since $F_n$, and hence $W$, and $Z(S)$ are
defined over $\Q$.
Consider any nonzero point $w=(w_1,\ldots,w_m) \in W \subseteq K^m$. We have to show that $h_{b,c}(w)\not = 0$ for some 
$b \in T$ and $0< c \le \deg(F_n)$. Let $F_w(x)=\sum_j w_j g_j(x)$.  Since $g_j(x)$'s are linearly independent, 
$F_w(x)$ is not identically zero as a polynomial in $x$.
Recall that  $K[W]$ is graded, with $\deg(z_j)=\deg(f_j)$. 
Let $F_w(x)_c = \sum_j w_j g_j(x)$, where $j$ ranges over all indices such that $\deg(f_j)=c$. Then $F_w(b)_c= h_{b,c}(w)$. 
So we have to show that $F_w(b)_c \not = 0$ for some $b \in T$ and $0< c \le \deg(F_n)$.

Since $W=\overline{\mbox{Im}(\psi)}$, and the closure in the Zariski topology coincides with the closure in the complex topology
(cf. Theorem 2.33 in \cite{mumfordalg}), there exists, for any $\delta>0$,
 $p_\delta \in K^r$, $r=r_n$, such that $||\psi(p_\delta)-w||_2 \le \delta/(m A)$,
where $A=\max\{||g_j||_2\}$ and, for any polynomial $e$,  $||e||_2$ denotes 
the $L_2$-norm of the coefficient vector of $e$. Since $w\not = 0$, 
taking $\delta$ to be small enough, we can assume
that $\psi(p_\delta)\not = 0$. Since $\psi(p_\delta)=(f_1(p_\delta),\ldots,f_m(p_\delta))$,
 and $g_j(x)$'s are linearly independent, 
$F_n(p_\delta,x)=\sum_j f_j(p_\delta) g_j(x)$ is not an identically zero
polynomial in $x$.  Let $C_n$ be the circuit computing $F_n(v,x)$ as in Definition~\ref{dexpvariety}. 
Let $C_{n,\delta}$ be   the circuit obtained from $C_n$ by specializing $v$ to $p_\delta$.
Then the size of $C_{n,\delta}$ is $s=\mbox{size}(C_n)=\poly(n)$, and the degree is $d=\mbox{deg}(C_n)=\poly(n)$. Furthermore, 
\[ ||C_{n,\delta}(x)-F_w(x)||_2 = || \sum_j (f_j(p_\delta)-w_j) g_j(x)||_2 \le m A ||\psi(p_\delta)-w||_2 \le \delta.\] 
Since $\delta$ can be made arbitrarily small,
it follows that $F_w(x)$ can be approximated infinitesimally closely 
by circuits of  degree $\le d$ and size $\le s$. Since $T$ is a hitting set, and $F_w(x)$ is not identically zero as a polynomial
in $x$, there exists $b\in T$ such that $F_w(b)\not = 0$. Hence $F_w(b)_c \not = 0$ for some $c \le \deg(F_n)$. This proves (a).

\noindent (b) By (a) and Hilbert's Nullstellensatz, 
it follows that, given any  $t\in K[W]$, $t^l$ belongs to the ideal $(S)$
in $K[W]$ generated by $S$, for some large enough positive integer $l$. 
Since $K[W]$ is graded, it now follows from the  graded Noether's normalization lemma
(Lemma~\ref{lnoether20}) that $K[W]$ is integral over its subring generated by $S$. This proves (b).

\noindent (c) $S$ is clearly strict by its definition. 
So it remains to verify the properties (1)-(3) in Definition~\ref{desopexp} (b).

\noindent (1)  We have to show that each $h_{b,c}(z) \in S$ has a short specification.
We have $\psi^*(h_{b,c})(v)=F_n(v,b)_c$, the degree $c$ component of $F_n(v,b)$. 
Since $W$ is explicit, cf.  Definition~\ref{dexpvariety}, we 
can compute the description of the circuit $C_n$ over $\Q$ computing $F_n$ in $\poly(n)$ time. Hence the total size of $C_n$,
including the bit-lengths of the constants in it, is $\poly(n)$. 
Using Vandermonde interpolation as in \cite{strassen,yehu}, we can construct, using $C_n$, in $\poly(n)$ time
a circuit $C_{n,c}$, for every $0 < c \le \deg(F_n)$, that  computes the degree $c$-component (in $v$) of $F_n$.
The circuit $C_{n,c,b}$ for computing $\psi^*(h_{b,c})(v)=F_n(v,b)_c$  is obtained by instantiating the circuit $C_{n,c}$
at $x=b$. Its total size (including the bit-lengths of the constants) is $\poly(n)$, and its degree is $\poly(n)$.
This shows that each $h_{b,c}(z)$ (or rather $\psi^*(h_{b,c})$) has a short specification.

\noindent (2) By (b), $K[W]$ is integral over the subring generated by $S$.

\noindent (3) Since the size of $T$ is $\poly(s)=\poly(n)$, and $\deg(F_n)$ is $\poly(n)$, the size of $S$ is  $\poly(n)$.

This shows that $S$ is a strict  s.s.o.p.  

\noindent (d) 
Consider any two distinct  points $w=(w_1,\ldots,w_m), w'=(w_1',\ldots,w'_m) \in W \subseteq K^m$.
Let $F_w(x)=\sum_j w_j g_j(x)$, and  $F_{w'}(x)=\sum_j w'_j g_j(x)$. 
These are distinct polynomials, since $w$ and $w'$ are distinct and $g_j$'s are linearly 
independent.
It also follows as in the proof of (a) that $F_w(x)$ and $F_{w'}(x)$ can be 
approximated infinitesimally closely 
by circuits of  degree $\le d$ and size $\le s$. Hence, by the stronger assumed property of the
hitting set $T$,  there exists $b \in T$ such that $F_w(b) \not = F_{w'}(b)$. 
Hence, $F_w(b)_c \not = F_{w'}(b)_c$, for some $c \le \deg(F_n)$. This means
$h_{b,c}(w) \not = h_{b,c}(w')$. Hence $S$ is separating.
\qed

The following is a corollary of Theorem~\ref{tmonteexplicit}.

\begin{cor}  \label{corssopexistsgen}
A  separating, strict  s.s.o.p. exists for the coordinate ring $K[W_n]$ of any explicit variety $W_n$.
\end{cor}

\subsection{Conditional derandomization} \label{sremove}
We now  derandomize the 
Monte Carlo algorithm in Theorem~\ref{tmonteexplicit}  using an appropriate 
black-box-derandomization or hardness hypothesis.

The following result proves the analogues of   Theorems~\ref{tnnldetblack} and \ref{tnnldetlower0} 
for any explicit variety.

\begin{theorem} \label{tnnlexpfull}
Let  $\{W_n\}$ be 
an explicit family of varieties as in Definition~\ref{dexpvariety}. Then:

\noindent (a) The variety $W_n$ has a  separating, strict e.s.o.p.,
assuming the strengthened black-box derandomization hypothesis for polynomial identity testing for small degree circuits  over $K$. 

\noindent (b) The variety $W_n$ has a  separating, strict quasi-e.s.o.p.,  assuming that 
there exists a family $\{h_n(x_1,\ldots,x_n)\}$ 
of  exponential-time-computable, multi-linear, integral polynomials
such that $h_n$ cannot be approximated infinitesimally closely by  circuits over $K$ of $O(2^{n^\epsilon})$ size,  for some 
constant $\epsilon>0$, as $n \rightarrow \infty$.
\end{theorem}

\proof

\noindent (a)  We have to show that 
a separating, strict s.s.o.p. for an explicit variety $W_n$
can be constructed in $\poly(n)$ time,
assuming the strengthened black-box derandomization hypothesis for polynomial identity testing for small degree circuits  over $K$. 

This follows if,
instead of the randomly chosen hitting set $T$ in the proof of Theorem~\ref{tmonteexplicit},
we use an explicit ($\poly(n)$-time computable) hitting set $T$  provided by the  strengthened black-box derandomization hypothesis. 

\noindent (b) This follows from  (a) and Theorem~\ref{trussell2}. 
\qed

\noindent {\em Remark 1:} If the multi-linear polynomial in (b)  is the permanent, then,
as for $\Delta[\det,m]$ (cf. Theorem~\ref{tnnldetlower}),
a separating, strict s.s.o.p. for  $W_n$ can be constructed  fast in parallel.

\noindent {\em Remark 2:} 
If in (b)  we assume instead that  there exists a family $\{h_n(x_1,\ldots,x_n)\}$ 
of  exponential-time-computable, multi-linear, integral polynomials such that $h_n$
cannot be  approximated infinitesimally closely by 
 circuits over $K$ of  $O(n^a)$ size, for any constant $a>0$, as
$n\rightarrow \infty$, then  it can be proved similarly that the strict 
form of  NNL for
$W_n$ can be solved in $O(2^{n^\epsilon})$-time
(assuming that $\poly(n)$ is replaced by $2^{n^\epsilon}$ in 
Definition~\ref{desopexp}), for any constant $\epsilon >0$. 

\noindent {\em Remark 3:} The statement  (b), 
in conjunction with \cite{kayalnew}, 
implies that an explicit  $W_n$ has a separating, strict quasi-e.s.o.p.,
assuming that 
there exists a family $\{h_n(x_1,\ldots,x_n)\}$ of 
exponential-time-computable, multi-linear, integral polynomials
such that $h_n$ cannot be approximated infinitesimally closely by  depth three 
circuits over $K$ of $O(2^{n^{\f {1}{2} + \epsilon}})$ size,  for some constant $\epsilon>0$, as
$n \rightarrow \infty$. A similar result also holds for homogeneous depth four 
circuits. The known $\Omega(2^{n^{1/2} \log n})$ lower bounds for the restricted versions
of these  circuits 
\cite{kayalsurvey} do not  imply
any nontrivial result for NNL for explicit varieties. Thus there is a sharp phase transition 
in the difficulty of the lower bound problem in this model  at the exponent $1/2$.

\noindent {\em Remark 4:} The statement (a) also holds for 
explicit varieties without any degree restrictions,  with the 
general (strengthened) polynomial identity testing without any degree restrictions in place of the (strengthened) polynomial identity testing for small degree circuits.

\noindent {\em Remark 5:} If $W_n$ is strongly explicit (cf. Definition~\ref{dexpvariety}), then the
(strengthened) polynomial identity testing for small degree circuits in the statement (a)
can be replaced with (strengthened) symbolic determinant identity testing. In this case it can also be shown that
NNL for $W_n$ belongs to DET, assuming that the strengthened black-box derandomization problem  for symbolic determinant identity testing belongs to DET (as may be
conjectured).

\noindent {\em Remark 6:} If $W_n$ is strongly explicit, then it can be assumed that
s.s.o.p.'s and e.s.o.p.'s in Theorems~\ref{tmonteexplicit}, \ref{tnnlexpfull} and Corollary~\ref{corssopexistsgen}
consist of weakly skew circuits (cf. Definition~\ref{desopexp}). 

\noindent {\em Remark 7:} The derandomization  hypothesis in the statement (a) is  only needed 
for the class of circuits used in the definition of $W_n$ (cf. Definition~\ref{dexpvariety}).

\subsection{An unconditional  EXPSPACE-algorithm} \label{sexpunconddet}
The following result gives the current best, unconditional, deterministic upper bound for NNL for explicit varieties.

\begin{theorem} \label{tssopexplicit}
Let $\{W_n\}$ be an explicit family of varieties. Then
the problem of constructing or verifying  a strict  s.s.o.p. for $K[W_n]$ belongs to EXPSPACE. This means it can be solved in 
$O(2^{\poly(n)})$ work-space.  It belongs to EXPH, assuming  the Generalized Riemann Hypothesis.
\end{theorem}

\proof For construction, this follows from 
Theorem~\ref{tnnlexpfull} (a)  and Theorem~\ref{tstrongexp}. The proof for verification
is implicit in the proof for construction.
\qed

\subsection{NNL for explicit varieties with closed defining maps} \label{sclosedmap}
Theorem~\ref{tnnlexpfull} (a) can be improved as  follows if the 
image of the defining map $\psi_n$ in (\ref{eqpsiexp}) is closed.

\begin{theorem} \label{tclosedimage}
Let $\{W_n\}$ be an explicit family of varieties such 
that the image of the  map $\psi_n$  in (\ref{eqpsiexp}) is closed.
Then the coordinate ring $K[W_n]$  of $W_n$ has a  separating,  strict  e.s.o.p.,
assuming the standard 
(instead of the strengthened) black-box derandomization
hypothesis for polynomial identity testing for small degree circuits over $K$. 
\end{theorem}

\proof 
The only reason we needed the strengthened black-box derandomization hypothesis in the proof of Theorem~\ref{tnnlexpfull} (a)
is because
the image of $\psi_n$ need not be closed, in general. If it is closed, then we can use the standard black-box derandomization 
hypothesis instead. \qed

\noindent {\em Remark 1:} 
Theorem~\ref{tclosedimage}, 
in conjunction with the PSPACE-bound for the standard black-box derandomization 
(Proposition~\ref{pstandard}), 
implies that NNL for $W_n$ belongs to PSPACE unconditionally if the image of
$\psi_n$ is closed.

\noindent {\em Remark 2:} 
Theorem~\ref{tclosedimage}, in conjunction with Theorem~\ref{trussell},
implies that $W_n$ has a strict quasi-e.s.o.p., assuming 
that there exists a family $\{h_n(x_1,\ldots,x_n)\}$ 
of  exponential-time-computable, integral, multi-linear polynomials such that $h_n$  cannot be
computed by circuits over $K$ of size sub-exponential in $n$. This 
improves Theorem~\ref{tnnlexpfull} (b) when the image of $\psi_n$ is closed.

\noindent {\em Remark 3:} 
If $W$ is strongly explicit (cf. Definition~\ref{dexpvariety}), then the low-degree polynomial identity testing 
in Theorem~\ref{tclosedimage}   can be replaced by symbolic determinant identity testing.
This result, in conjunction with Theorem~\ref{tmumfordnew} (a),  implies that if $W_n$ is a strongly explicit categorical quotient,
then an e.s.o.p. for $W_n$ exists, assuming the standard (instead of the strengthened)
black-box derandomization hypothesis for symbolic determinant 
identity testing.
This fact will play a crucial role in the proofs of Theorems~\ref{tmatrixNew} and \ref{thilbertnew}.

\noindent {\em Remark 4}: We only need the derandomization
hypothesis in Theorem~\ref{tclosedimage} 
 for the class of circuits used in the definition  of  $W_n$ (cf. Definition~\ref{dexpvariety}).

\subsection{Equivalence} \label{sequivfull}
The following is the analogue of Theorem~\ref{tequisditdet} for general polynomial identity testing.

\begin{theorem}  \label{tequisdit}

\noindent (a) 
The strengthened black-box derandomization hypothesis for general  polynomial identity testing  over $K$,
without any degree restrictions,  holds iff 
the orbit closure $\Delta[H(Y),k,m]$ (cf. Section~\ref{shy}), with $k=m$, has a strict e.s.o.p. 

\noindent (b) 
The strengthened black-box derandomization hypothesis for  low-degree  polynomial identity testing  over $K$  holds
iff  the orbit closure $\Delta[H(Y)_m,k,m]$ (cf. Section~\ref{shy}), with $k=m$, has a strict e.s.o.p. 

\noindent (c) Ignoring a quasi prefix, 
a sub-exponential lower bound for a family of
exponential-time-computable, integral, multi-linear polynomials as in 
Theorem~\ref{trussell2} holds iff 
the orbit closure $\Delta[H(Y)_m,k,m]$ (cf. Section~\ref{shy}), with $k=m$,  has a strict e.s.o.p.
\end{theorem}

\proof 

\noindent (a) The polynomial  $H(Y)$ corresponds to a universal circuit
(cf. Section~\ref{shy}), just as the determinant corresponds  \cite{malod} to a universal
weakly skew circuit. The polynomials corresponding to the points in  $\Delta[H(Y),k,m]$ 
can be approximated infinitesimally closely by circuits over $K$ of size $\poly(k,m)$ and degree 
$= \deg(H(Y))$. 
Though $\deg(H(Y))$ is exponential in $k$, Theorem~\ref{theintz} still implies existence of a small 
hitting set, with  $O(\poly(k,m))$ bit-length of specification, 
against such  polynomials.
The rest of the proof is similar to that of 
Theorem~\ref{tequisditdet} (a), with 
$\Delta[H(Y),k,m]$, with $k=m$,  in place of $\Delta[\det,m]$.

\noindent (b) The proof is similar to that of (a), with $\Delta[H(Y)_m,k,m]$, with $k=m$,  in place of $\Delta[H(Y),k,m]$.

\noindent (c) This follows from (b), Theorem~\ref{trussell2}, and Proposition~\ref{pconverserussell}.
\qed

\noindent {\em Remark:} 
It can be shown similarly that 
the strengthened black-box derandomization hypothesis for  polynomial identity 
testing  for depth three circuits over $K$ and $n$ variables with  degree $\le d$ 
and top fan-in $\le k$ holds iff the $k$-th secant variety $X(d,k,n)$ of the Chow variety (cf. 
Section~\ref{lexporbitclosuresthree}) has a strict e.s.o.p.

\subsection{The NNL for the orbit closure of the permanent}
Analogue of Theorem~\ref{tnnldetlower0} also holds for the orbit closure $\Delta[\perm,n,m]$  of the permanent
defined in Section~\ref{snnldetquasi}, though this variety  is not explicit.
So let us call it {\em weakly explicit}.

Specifically, given any set ${\cal B}=\{B_1,\ldots,B_k\} \subseteq M_m(\N)$, define 
$\psi_{{\cal B}}: {\cal X} \rightarrow K^k$ as in Section~\ref{sNNLdetintro},
replacing $\det(X)$  by  $z^{m-n} \perm(Y)$ in that definition, where $Y$ is any $n\times n$ sub-matrix of $X$, and
$z$ is any entry in $X$ outside $Y$. We say that ${\cal B}$ specifies a 
{\em strict s.s.o.p.} 
for $\Delta[\perm,n,m]$ if 
(1) the total bit-length of $B_i$'s is $\poly(m)$, and (2)
$\psi_{{\cal B}}$ does not vanish on any non-zero point in the affine cone  $\hat \Delta[\perm,n,m] \subseteq {\cal X}$ of
$\Delta[\perm,n,m]$.
We say that ${\cal B}$ specifies a {\em strict   e.s.o.p.} if, in addition,  it is $\poly(m)$-time-computable. 
A strict quasi-e.s.o.p. is defined by replacing $\poly(m)$  by $2^{\polylog(m)}$.

\begin{theorem} 
The variety  $\Delta[\perm,n,m]$ has a strict quasi-e.s.o.p., assuming that there exists 
a family $\{p_k(x_1,\ldots,x_k)\}$ of  exponential-time-computable, 
multi-linear, integral polynomials  such that $p_k$
cannot be approximated infinitesimally closely by permanents 
of symbolic matrices  over $K$ of  
$O(2^{k^\epsilon})$ size, for some constant
$\epsilon >0$, as $k\rightarrow \infty$.
\end{theorem}

\proof 
By inserting the oracle for the permanent in 
appropriate places in the proof of Theorem~\ref{trussell2}, it follows that
strengthened polynomial identity testing for small degree 
circuits  over $K$ of size $\le s$, with oracle gates for the permanent,
has $O(2^{\polylog(s)})$-time-computable hitting set (defined in the obvious way), assuming 
that there exists  a family $\{p_k(x_1,\ldots,x_k)\}$ 
of  exponential-time-computable,  multi-linear, integral
polynomials such   that $p_k$ 
cannot be approximated infinitesimally closely  by   circuits  over $K$, with  oracle gates
for the permanent,  of  $O(2^{k^\epsilon})$ size, for some   constant
$\epsilon >0$, as $k\rightarrow \infty$.  It easily follows from 
Valiant \cite{valiant2} that 
 a low-degree circuit of size $s$,  with oracle gates for the permanent, can be simulated by the permanent
of a symbolic matrix of $\poly(s)$ size. Hence, the same conclusion holds assuming instead that
$p_k$ cannot be approximated infinitesimally closely  by permanents of symbolic matrices 
 over $K$  of  $O(2^{k^\epsilon})$ size, for some   constant
$\epsilon >0$, as $k\rightarrow \infty$. The proof is now similar
to that of Theorem~\ref{tnnldetlower0}, using this fact in place of Theorem~\ref{trussell2}. \qed

\section{Explicitness of $V/G$ for  the ring of matrix invariants} \label{sexpmatrixring}
In this section we prove Theorem~\ref{tintroexplicitmatrixnew}, assuming that the base field $K$ has characteristic zero.

Let $V=M_m(K)^r$, with the adjoint action of $G=SL_m(K)$, be as in Section~\ref{snnlmatrixnew}.
Given $\sigma \in G$, the adjoint action maps $(A_1,\ldots,A_r) \in V$ to 
$(\sigma A_1 \sigma^{-1}, \ldots, \sigma A_r \sigma^{-1})$.
Let $n=\dim(V)=r m^2$. Let $U_1,\ldots,U_r$ be variable $m\times m$ matrices, and let
$U=(U_1,\ldots,U_r)$. Identify
the coordinate ring $K[V]$ of $V$ with 
the ring $K[U_1,\ldots,U_r]$ generated by the variable entries of $U_i$'s. 
Let $K[V]^G \subseteq K[V]$ be the ring of invariants with respect to the adjoint action of $G$.
Let $V/G=\spec(K[V]^G)$.

Call two words in $[r]^*$
 {\em equivalent} if one can be obtained from the other by a circular rotation. Recall that 
(cf. Section~\ref{sstrongblack}) $[r]$ denotes $\{1,\ldots,r\}$.

The following is a restatement of Theorem~\ref{tintroexplicitmatrixnew} in characteristic zero for convenience.

\begin{theorem} \label{texplicitmatrix}
The categorical quotient  $V/G$ is strongly explicit (cf. Definition~\ref{dexpffthilbert}), or in other words,
a strongly explicit First Fundamental Theorem holds for $K[V]^G$.

Specifically, let $X=(X_1,\ldots,X_r)$ be an $r$-tuple of  $k\times k$ variable matrices, where $k=m^2$,
and $m$ is the dimension of $U_i$'s as above.
Then there exist $\poly(n)$-time-computable weakly skew (Section~\ref{sterminology}) 
circuits $C_l$'s, $l \le m^2$,  
over $\Q$ and the variable entries of $X_i$'s and $U_i$'s,
such that
(1) the polynomial functions $C_l(X,U)$'s computed by $C_l$'s are of $\poly(n)$ degree,  homogeneous in $X$ and $U$,
and can be written as 

\begin{equation} \label{eqclxu}
C_l(X,U)=\sum_{[\alpha]}  f_{[\alpha],l} (U) g_{[\alpha],l} (X) , 
\end{equation}
where $[\alpha]=[\alpha_1 \cdots \alpha_l]$ ranges over the equivalence classes of all words of length $l$ with each $\alpha_j \in [r]$,
(2) $g_{[\alpha],l}(X)$'s are linearly independent homogeneous polynomials in the entries of $X_i$'s, 
and (3) $f_{[\alpha],l}(U)$'s are homogeneous invariants that generate $K[V]^G$.
\end{theorem}

\subsection{Geometric invariant theory}
Before proving Theorem~\ref{texplicitmatrix}, we recall some results in  geometric invariant theory
that are needed for its proof.

\begin{theorem}[Procesi-Razmyslov-Formanek] \cite{procesimatrix,razmyslov,formanek1} (The First Fundamental Theorem for  matrix
invariants; cf. Theorems 6 and 10 in \cite{formanek1}) \label{tprf}
The ring $K[V]^G$ is generated by the traces of the form $\trace(U_{i_1}\cdots U_{i_l})$, $l \le m^2$, 
$i_1,\ldots,i_l \in [r]$. 
\end{theorem}

Let $K[S_r]$ be the group algebra of the symmetric group $S_r$ on $r$ letters. Write any $\sigma \in S_r$ as a product of
disjoint cycles:
\[ \sigma = (a_1 \cdots a_{k_1}) (b_1 \cdots b_{k_2}) ...,\] 
where $1$-cycles are included, so that each of the numbers $1,\ldots,r$ occurs exactly once. 
Define
\begin{equation} \label{eqtsigma}
 T_\sigma(U_1,\ldots,U_r) = \trace(U_{a_1} \cdots U_{a_{k_1}}) \trace(U_{b_1} \cdots U_{b_{k_2}}) \cdots.
\end{equation}

The following result is a consequence of the Second Fundamental Theorem for 
matrix invariants due to Procesi and Razmyslov \cite{procesimatrix,razmyslov}.

\begin{theorem} (cf. Theorem 1 in \cite{formanek1}) \label{tprocesisecond} 
Define the $K$-linear map 
$\phi:K[S_r] \rightarrow K[V]^G$ by 
\[ \phi(\sum a_\sigma \sigma)= \sum a_\sigma T_\sigma(U_1,\ldots,U_r).\] 
Then $\mbox{Ker} (\phi)=\{0\}$ if $r \le m$. 
\end{theorem}

Let $X_1,\ldots,X_r$ be $k \times k$ variable matrices. 
For any word $\alpha=i_1,\ldots,i_l$, $i_j \in [r]$, 
let 
\begin{equation}  \label{eqtracemono} 
T_\alpha(X)=\trace(X_{i_1}\cdots X_{i_l}),
\end{equation} 
where $X=(X_1,\ldots,X_r)$.  Let  $T_{[\alpha]}(X)=T_\alpha(X)$, where
$[\alpha]$ denotes the equivalence class of words equivalent to $\alpha$ under circular rotation. 
The choice of $\alpha$ in $[\alpha]$ does not matter. 

\begin{cor} \label{cprocesi} 
The traces $\{T_{[\alpha]}(X)\}$, where  $[\alpha]$ ranges over all equivalence classes of 
words of length  $l \le k$,  are linearly independent.
\end{cor} 

\proof 
Suppose to the contrary that there is a linear dependence 
\begin{equation} \label{eqdepend}
 \sum_{[\alpha]} b_{[\alpha]} T_{[\alpha]}(X)  = 0, \quad b_{[\alpha]} \in K.
\end{equation} 
Without loss of generality, we can assume that this relation is homogeneous 
in every $X_i$. We can also assume that it is multi-linear
in $X_i$'s. 
Otherwise, we can multi-linearize it by (1) substituting
\[ X_i=\sum_{j=1}^{d_i} t_{i,j} X_{i,j}\] 
in the l.h.s. of (\ref{eqdepend}),
where $d_i$ is the (homogeneous) degree of $X_i$ in the relation, $t_{i,j}$'s  are new variables, and $X_{i,j}$'s 
are new variable $k\times k$ matrices, and then (2)
equating  the coefficient of $\prod_i \prod_{j=1}^{d_i} t_{i,j}$ to zero.

So assume that the dependence (\ref{eqdepend})  is multi-linear and homogeneous. Without of loss of generality,
assume that the variables occurring in this dependence are $X_1,\ldots,X_l$, $l \le k$. Then each  $[\alpha]$
in (\ref{eqdepend})  corresponds to a cyclic 
permutation $(i_1,\ldots,i_l) \in S_l$, which we denote by $\hat \alpha$.
Hence, the l.h.s. of (\ref{eqdepend}) equals $\phi(\sum_{\hat \alpha} b_{[\alpha]} \hat \alpha)$,
where $\phi: K[S_l]\rightarrow K[M_k(K)^l]^{SL_k(K)}$ is the map (cf. Theorem~\ref{tprocesisecond}) that takes 
$\sum_\sigma a_\sigma \sigma \in K[S_l]$ to $\sum_\sigma a_\sigma T_\sigma(X_1,\ldots,X_l)$.  
Since $l \le k$, it  follows from  (\ref{eqdepend}) and  Theorem~\ref{tprocesisecond} that 
all $b_{[\alpha]}$'s are  zero.
\qed 

\noindent {\em Remark:} The proof above also shows that 
the monomials in $T_{[\alpha]}(X)$'s of total 
degree $l \le k$ in $X_i$'s are linearly independent.

\subsection{Proof of Theorem~\ref{texplicitmatrix}} \label{sproofexpmatrix}
For any word $\alpha=i_1,\ldots,i_l$, $l \le m^2$, $i_j \in [r]$, cf. (\ref{eqtracemono}), let
\begin{equation} \label{eqtraceu}
T_{[\alpha]}(U)=T_\alpha(U)=\trace(U_{i_1}\cdots U_{i_l}),
\end{equation}
where $U=(U_1,\ldots,U_r)$. 
Let 
\begin{equation} \label{eqFmatrix}
F=\{T_{[\alpha]}(U)\},
\end{equation} 
where $[\alpha]$ ranges over the equivalence classes (for circular rotation) of  all words in $1,\ldots,r$ of length $\le m^2$.
Then   $F$ generates $K[V]^G$ by Theorem~\ref{tprf}.  

Consider the map $\pi_{V/G}$ from  $M_m(K)^r$ to $K^t$, $t=|F|$, defined as
\begin{equation}  \label{eqprocesiembed}
\pi_{V/G}: \quad A=(A_1,\ldots,A_r) \rightarrow (\ldots, T_{[\alpha]}(A),\ldots),
\end{equation} 
where $A_i \in M_m(K)$ for all $i$.
By Theorem~\ref{tmumfordnew} (a), its image  is  closed, and 
can be identified with  $V/G$.

For any $l\le k=m^2$, let

\begin{equation} \label{eqgenerixmatrix}
 T_l(X,U)= \trace((X_1\otimes U_1+ \cdots + X_r \otimes U_r)^l),
\end{equation}
where $X_i$'s are new $k \times k$ variable matrices, $X=(X_1,\ldots,X_r)$,   $U=(U_1,\ldots,U_r)$, and $\otimes$ denotes 
the Kronecker product of matrices. Thus each $X_i \otimes U_i$ is an $m' \times m'$ matrix, where $m'= k m=m^3$.
We have

\begin{equation} \label{eqexpfftmatrix}
T_l(X,U) = \sum_\alpha T_\alpha(X) T_\alpha(U) =\sum_{[\alpha]} |[\alpha]| T_{[\alpha]}(X) T_{[\alpha]}(U),
\end{equation}
where $[\alpha]=[\alpha_1 \cdots \alpha_l]$ ranges over the equivalence classes of all words of length $l$ with each $\alpha_j \in [r]$,
$|[\alpha]|$ denotes the cardinality of the equivalence class $[\alpha]$ of the word $\alpha$, 
and $T_\alpha(U)$ and $T_{\alpha}(X)$ are as  in (\ref{eqtraceu}) 
and (\ref{eqtracemono}).

Clearly $T_l(X,U)$, cf.  (\ref{eqgenerixmatrix}),  can be computed by an explicit ($\poly(n)$-time computable)
weakly skew circuit (Section~\ref{sterminology}). 
Fix such an explicit circuit $C_l$ computing $T_l(X,U)$.  Then 
$C_l(X,U)=T_l(X,U)$. Let   $g_{[\alpha],l}(X)= |[\alpha]| T_{[\alpha]}(X)$, and $f_{[\alpha],l}(U)=T_{[\alpha]}(U)$.
Then (\ref{eqclxu}) holds by (\ref{eqexpfftmatrix}).
Furthermore,  $g_{[\alpha],l}(X)$'s are linearly independent by Corollary~\ref{cprocesi}, and 
$f_{[\alpha],l}(U)$'s generate $K[V]^G$ by Theorem~\ref{tprf}.

This proves Theorem~\ref{texplicitmatrix}. 

\section{NNL for the ring of matrix invariants} \label{sstit} 
In this section, Theorem~\ref{tmatrixNew} is proved, assuming that the base field $K$ has characteristic zero.

Let $V=M_m(K)^r$, $n=\dim(V)=r m^2$,  $G=SL_m(K)$, $K[V]^G$, and $V/G$  be as in Section~\ref{sexpmatrixring}. 
By Theorem~\ref{texplicitmatrix}, $V/G$ is strongly  explicit. 
Hence, we can specify $V/G$ succinctly, as per the general definition of an explicit variety 
(cf. Definition~\ref{dexpvariety}), by  the circuits $C_l(X,U)$'s in 
Theorem~\ref{texplicitmatrix}. Instead,
we shall  specify $V/G$ and $K[V]^G$  succinctly by just giving the pair $(m,r)$ in unary. This is sufficient and also equivalent,
since, given $(m,r)$, we can compute the circuits $C_l(X,U)$'s in Theorem~\ref{texplicitmatrix} in $\poly(m,r)$ time. 

An s.s.o.p. or an e.s.o.p. for $K[V]^G$ 
is defined as in Section~\ref{snnlmatrixnew}. The symbolic determinants that
were used in the definition of an s.s.o.p. in Section~\ref{snnlmatrixnew} are equivalent to weakly skew circuits 
(cf. Section~\ref{sterminology}). Quasi-s.s.o.p. and quasi-e.s.o.p. are   defined, as before,
 by replacing $\poly(n)$ by $2^{\polylog(n)}$.

Following Derksen and Kemper \cite{derksenbook},
we call $S \subseteq K[V]^G$ {\em separating} if,
for any two distinct $v,w \in V$ such that $r(v)\not = r(w)$ for some $r \in K[V]^G$, there exists an $s \in S$
such that $s(v) \not = s(w)$. This is a general notion that applies to
any finite dimensional representation  of a reductive group.

By the  problem NNL for $K[V]^G$, we mean, as in Section~\ref{snnlmatrixnew},
the problem of constructing an
s.s.o.p., given $(m,r)$ in unary.
By the strong form of NNL, we mean the problem of constructing a separating
s.s.o.p.

The reader  should check that these definitions are specializations of the general 
Definition~\ref{desopexp} for strongly explicit varieties.
We can also define strict s.s.o.p. and e.s.o.p. for $V/G$ (cf. Definition~\ref{dstrict}) 
using the circuits $C_l(X,U)$'s in Theorem~\ref{texplicitmatrix}.
All s.s.o.p.'s constructed in this section are strict. However, strictness is not as important for $V/G$ as it is for $\Delta[\det,m]$,
since existence of a strict e.s.o.p. for  $V/G$ does not imply any lower bound. 
Hence, we shall not worry about strictness in this section.

We   prove in this section the following stronger form of Theorem~\ref{tmatrixNew} in characteristic zero.

\begin{theorem}[Cf. \cite{GCT5focs}, \cite{fs2,fs3}, and Remark 1 in Section~\ref{sprooftech}] \label{tmatrixstrongform}
The ring  $K[V]^G$ has a separating e.s.o.p,
assuming the standard  black-box derandomization hypothesis for symbolic determinant identity testing. 
It has a separating quasi-e.s.o.p. unconditionally.
\end{theorem}

The following will turn out to be a corollary of the proof of this result.

\begin{theorem} \label{tintroorbitclosureNew}
The problem of deciding if the $G$-orbit-closures of two rational points in $V$ intersect belongs to DET $\subseteq$ NC.
\end{theorem}

\subsection{Construction of an h.s.o.p.}
Before we prove Theorem~\ref{tmatrixstrongform}, 
we   study  the problem of constructing an h.s.o.p. (homogeneous system of parameters) 
for $K[V]^G$ (cf. Definition~\ref{dhsop}). 
The following result gives the currently best upper bound for this problem.

\begin{theorem} \label{thsop}
The problem of constructing an h.s.o.p. for  $K[V]^G$ belongs to EXPH,
assuming the Generalized Riemann Hypothesis.
\end{theorem}

For the proof, we need the following result.

Recall that the trace function $T_\sigma$  defined  in (\ref{eqtsigma}) satisfies 
\cite{procesimatrix}  the fundamental trace identity 
\[ F(U_1,\ldots,U_{m+1})= \Sigma_{\sigma \in S_{m+1}} \mbox{sign}(\sigma) T_\sigma(U_1,\ldots,U_{m+1}) = 0.\] 

\begin{theorem}[Procesi-Razmyslov](The Second Fundamental Theorem for matrix invariants) 
(cf. Theorem 4.5 in \cite{procesimatrix})
\label{tprocesi3}
The ideal of all relations among the trace monomial generators of $K[V]^G$ given by Theorem~\ref{tprf} 
is generated by the elements of the form $F(M_1,\ldots,M_{m+1})$, where $M_i$'s range over all possible monomials in $U_j$'s so that
the total length of $M_i$'s is $\le m^2$.
\end{theorem}

This follows from the proof of Theorem 4.5 in \cite{procesimatrix}.

\noindent {\em Proof of Theorem~\ref{thsop}:} 
The defining equations for $V/G$ given in Theorem~\ref{tprocesi3} can clearly 
be computed in time  exponential in $n$.
Hence the result follows from Theorem~\ref{tnnlexplicithsop}.
\qed 

If we insist on an h.s.o.p., then Theorem~\ref{thsop} is the best that we can do at present.
But if we only require a small homogeneous  $S$  of $\poly(n)$ 
cardinality such that $K[V]^G$ is integral over the subring generated by $S$, and do not insist 
on optimality of $|S|$, then Theorem~\ref{tmatrixstrongform} 
says  that the double exponential time bound in Theorem~\ref{thsop} 
can be brought down to  quasi-polynomial. (Theorem~\ref{thsop} only 
implies a double-exponential time
bound for the problem of constructing an h.s.o.p.,  since conjecturally 
EXPH $\not \subseteq$ EXP.)

We now turn towards the proof of Theorem~\ref{tmatrixstrongform}.

\subsection{A Monte Carlo algorithm} 
The first step is an efficient  Monte Carlo algorithm to construct an s.s.o.p.

\begin{theorem} \label{tmontecarlomatrix}
A separating  s.s.o.p. for $K[V]^G$ can be constructed by a $\poly(n)$-time Monte Carlo algorithm that is correct with a high probability.

In particular, a separating  s.s.o.p. for $K[V]^G$ exists.
\end{theorem} 
\proof By Theorem~\ref{texplicitmatrix}, $V/G$ is explicit. Hence the result follows from Theorem~\ref{tmonteexplicit}.
\qed

\subsection{Reduction of  NNL to black-box symbolic determinant identity testing} 
\label{ssditcond}
The next step is to derandomize the Monte Carlo algorithm in Theorem~\ref{tmontecarlomatrix}
assuming a suitable derandomization hypothesis. The first statement in Theorem~\ref{tmatrixstrongform} following from
the following result.

\begin{theorem} \label{tmatrixmain}
Assume that the standard 
black-box derandomization hypothesis for  symbolic determinant identity testing over $K$ holds. 
Then $K[V]^G$ has a separating   e.s.o.p.
\end{theorem}

\proof 
Since $V/G$ is strongly explicit (cf. Theorem~\ref{texplicitmatrix}), and the image of $\pi_{V/G}$ in (\ref{eqprocesiembed}) is closed
by Theorem~\ref{tmumfordnew} (a), 
it follows from Theorem~\ref{tclosedimage} and Remark 3 thereafter  that the Monte Carlo algorithm in Theorem~\ref{tmontecarlomatrix} 
can be derandomized assuming the standard  black-box derandomization hypothesis for symbolic determinant identity testing.  \qed 

We now give a second more refined proof of this  result, since it  is needed for the proof of the  second unconditional statement in
Theorem~\ref{tmatrixstrongform}.
For this proof,  we need the following  result from 
geometric invariant theory. 
We state in a more general form than what is needed here, since it will 
be needed in such generality in Sections~\ref{sexpconstant}  and \ref{snnlconstant}.

\begin{theorem}[Derksen and  Kemper] (cf. Theorem 2.3.12 in \cite{derksenbook}) \label{tsepderksen}
Let $W$ be a finite dimensional representation of any 
algebraic reductive group $H$ over 
$K$. Let $S \subseteq K[W]^H$ be a finite separating set (cf. Section 2.3.2 in   \cite{derksenbook}, and the beginning 
of this section) of homogeneous invariants.
Then $K[W]^H$ is integral over the subring generated by $S$.
\end{theorem}

In this section, we shall use this result with $W=V$ and $H=G$. 

\noindent {\em A refined proof of Theorem~\ref{tmatrixmain}:}

We follow  the same notation as in Section~\ref{sproofexpmatrix}.

Let $T_l(X,U)$ be as in (\ref{eqgenerixmatrix}).
Let $U'=(U_1',\ldots,U_r')$ be another tuple of variable $m\times m$ matrices, in addition to $U$. 
For each $l \le k=m^2$, define  the symbolic trace difference

\begin{equation} \label{eqtildet}
  \tilde T_l(X,U,U')= T_l(X,U) - T_l(X,U'). 
\end{equation}

Clearly $\tilde T_l(X,U,U')$ has a weakly skew circuit (Section~\ref{sterminology})
over $X,U$, and $U'$   of $\poly(n)$ size. 
Since symbolic determinants are polynomially equivalent to  weakly skew circuits 
(cf. Section~\ref{sterminology} and \cite{malod}), it follows that 
each $\tilde T_l(X,U,U')$ can be expressed as $\det(N_l(X,U,U'))$ for some symbolic matrix $N_l(X,U,U')$ 
of size $q=\poly(n)$  over $X$, $U$, and $U'$.

By our black-box derandomization hypothesis for symbolic determinant identity testing, there exists an explicit ($\poly(n)$-time computable) 
hitting set $B = B_{s,q} \subseteq \N^s$ for symbolic determinant identity testing  for   $q\times q$ matrices whose entries are 
linear functions of the $s=r k^2$ variable entries of $X_i$'s with coefficients in $K$.
It has to be stressed here that the hitting set $B$ is against non-zero
symbolic determinants of size $q$ over $X$, not over $X$, $U$, and $U'$. The reason will become clear in a moment.
Fix such an explicit $B$. We think of each $b\in B$ as an $r$-tuple $b=(b_1,\ldots,b_r)$ of $k\times k$ integral matrices.

Let 
\begin{equation} \label{eqSmatrix}
S = \{ T_l(b,U) \ | \ b \in B, 1 \le l \le k\} \subseteq K[V]^G. 
\end{equation}

Suppose $A,A' \in V=M_m(K)^r$ are two $r$-tuples such that,
for some invariant $h \in K[V]^G$, $h(A) \not = h(A')$. 
By Theorem~\ref{tprf}, it  follows that some generator $T_{[\alpha]}(U)$, cf. (\ref{eqFmatrix}), 
assumes different values at $A$ and $A'$. 
By (\ref{eqexpfftmatrix}) and Corollary~\ref{cprocesi}, 
this implies that  $\tilde T_l(X,A,A')=T_l(X,A)-T_l(X,A')$
is not identically zero, as a polynomial in $X$, for some $l \le m^2$.

Since $\tilde T_l(X,U,U')$ can be expressed as a symbolic determinant 
of size $q=\poly(n)$  over $X$, $U$, and $U'$, $\tilde T_l(X,A,A')$ is a symbolic determinant of size $q=\poly(n)$ over $X$.
Since $B$ is a hitting set against such symbolic determinants over $X$, 
and $\tilde T_l(X,A,A')$ is not identically zero as a polynomial in $X$,
there exists $b \in B$ such that $\tilde T_l(b,A,A') \not = 0$, i.e.,
$T_l(b,A) \not = T_l(b,A')$. It follows that $S$ is separating. 

Every element of $S$ is clearly homogeneous of $\poly(n)$ degree.
By Theorem~\ref{tsepderksen}, it follows that
$K[V]^G$ is integral over the subring generated by $S$.

Since the hitting set $B$ is explicit,
and matrix powering, Kronecker product,  and trace  
have  explicit weakly-skew circuits  (cf. Section~\ref{sterminology} and \cite{malod}), 
it follows from (\ref{eqgenerixmatrix}) that the specification of $S$ consisting of a weakly 
skew  circuit
for its every element can be computed in $\poly(n)$ time. Hence $S$ is  a separating  e.s.o.p.  

This proves Theorem~\ref{tmatrixmain}. \qed 

\noindent {\em Remark 1:} The e.s.o.p. constructed in Theorem~\ref{tmatrixmain} is also 
strict (cf. Definition~\ref{dstrict}) with respect to the defining 
polynomials $C_l(X,U)$'s in Theorem~\ref{texplicitmatrix} for $V/G$.

\noindent {\em Remark 2:}
Assuming a stronger  parallel black-box derandomization hypothesis for symbolic determinant identity testing over $K$, 
the problem of constructing 
a separating   s.s.o.p. for $K[V]^G$ can be shown to belong to $\mbox{DET} \subseteq \mbox{NC}^2 \subseteq \mbox{P}$.
This hypothesis is that the problem
of constructing, given $m$ in unary, 
a hitting set against non-zero symbolic determinants of size $m$ over (say) $m^2$   variables 
belongs to DET.

\subsection{Deciding if two orbit-closures intersect}
The following is a consequence of the above proof in conjunction with  the standard geometric invariant theory.

\begin{theorem} \label{torbitclosure}
The problem of deciding if the closures of the $G$-orbits of two rational points in $V$ intersect, and finding some invariant in $K[V]^G$ that separates the two
if they do not,  belongs to co-RDET $\subseteq$ co-RNC.
\end{theorem}

The complexity class co-RDET here is the randomized version of co-DET (the complement of DET) \cite{cook}.

\proof By Theorem~\ref{tmumfordnew} (d) and the refined proof of Theorem~\ref{tmatrixmain},
the closures of the $G$-orbits of $A, A' \in V$ intersect iff 
the symbolic trace difference $\tilde T_l(X,A,A')=T_l(X,A)-T_l(X,A')$ 
is identically zero for every  $l \le k=m^2$. For rational $A$ and $A'$, this can  be tested by 
a co-RDET algorithm \cite{ibarra}:
just substitute large enough random integer values for the entries of $X$ and test if all the differences vanish.
If the symbolic trace difference is not identically zero for some $l$, then this test 
returns a matrix $C$ such that the test fails for that  $l$ when $X=C$.
The symbolic trace $T_l(C,U)$ is an invariant that separates $A$ and $A'$ in that case.
\qed

\subsection{Replacing symbolic determinants by read-once oblivious algebraic branching programs} 
\label{sforbes}
In this section we describe how the symbolic determinant identity testing
in Theorem~\ref{tmatrixmain} can be replaced by polynomial identity testing
for  read-once oblivious algebraic branching programs (cf. Section~\ref{sterminology}), 
as pointed out by Forbes and Shpilka \cite{fs3}. In conjunction with their
earlier quasi-derandomization of polynomial identity testing for
such programs in \cite{fs2}, 
this implies  existence of  a quasi-e.s.o.p.
for $K[V]^G$, as stated in Theorem~\ref{tmatrixstrongform},  unconditionally.

\begin{lemma}[Forbes and  Shpilka] (cf. Lemmas 2.3 and 3.4 in \cite{fs3}) \label{lforbes}
For any positive integer $l$, there exists a  read-once oblivious algebraic branching program
$P_l(Y,U,U')$
over $\Z$,  the variable entries of $U$, $U'$ (thought as indeterminate constants), and the tuple  $Y=(y_1,\ldots,y_l)$ of 
auxiliary variables,  with the 
specification of $\poly(l,m,r)$ bit-size, such that 

\begin{equation} \label{eqforbes1}
P_l(Y,U,U')= \sum_\alpha Y_\alpha (T_\alpha(U)-T_\alpha(U')),
\end{equation}

where $\alpha=\alpha_1 \alpha_2 \cdots$ ranges over  all words of length $l$, with each $\alpha_j \in [r]$, and $Y_\alpha=\prod_j y_j^{\alpha_j}$.
\end{lemma}
\proof The r.h.s. of (\ref{eqforbes1}) equals $(\trace(\prod_{j=1}^l (\sum_{i=1}^r y_j^i U_i)))-(\trace(\prod_{j=1}^l (\sum_{i=1}^r y_j^i U'_i)))$, which 
can clearly be computed by a read-once oblivious algebraic 
branching program with the specification of   $\poly(l,m,r)$ bit-size. \qed

Since the monomials $Y_\alpha$'s are linearly independent, we can replace $T_l(X,U,U')$ by $P_l(Y,U,U')$ in the refined proof of
 Theorem~\ref{tmatrixmain}.
This implies  that Theorem~\ref{tmatrixmain} also holds after replacing
the symbolic determinant identity testing in its statement
by polynomial identity testing  for read-once oblivious algebraic branching programs
(cf.  Section~\ref{sterminology}).
The existence of a separating quasi-e.s.o.p. as in Theorem~\ref{tmatrixstrongform}
(and even a quasi-NC algorithm for the strong form of NNL in this case)
follows in view of the  quasi-NC black-box algorithm for polynomial identity testing 
for read-once oblivious algebraic branching programs  in \cite{fs2}.
This replacement also derandomizes 
the co-RDET-algorithm in  Theorem~\ref{torbitclosure} in view of 
the  white-box (cf. Section~\ref{sblackstd}) 
 DET-algorithm  for polynomial identity testing  for read-once oblivious
algebraic branching programs  in Raz and Shpilka and Arvind et al. \cite{raz,arvind}. 
This proves Theorem~\ref{tintroorbitclosureNew}.
(Unlike the co-RDET algorithm in Theorem~\ref{torbitclosure},
this algorithm does not return a separating invariant if 
the two orbit closures do not intersect.)

\section{Explicitness of $V/G$ when $G$ has constant dimension}\label{sexpconstant}
In this section we prove Theorem~\ref{tintroexplicitconstantnew}.

Let $V$ be a rational  representation of $G=SL_m(K)$ of dimension $n$.
The following is a restatement of Theorem~\ref{tintroexplicitconstantnew} for convenience.

\begin{theorem} \label{texplicitconst0}
The categorical quotient $V/G=\spec(K[V]^G)$ is strongly explicit (Definition~\ref{dexpffthilbert}), i.e.,
a strongly explicit First Fundamental Theorem holds for $K[V]^G$,   if $m$ is constant. 
\end{theorem} 

We begin by recalling some results from invariant theory and standard monomial theory 
\cite{raghavan,doubillet} that are needed to prove this result, and then we prove some 
complexity-theoretic lemmas.

\subsection{A degree bound for the ring of invariants} \label{sdegreebound}
First, we recall from Derksen \cite{derksen} a degree bound for a set of generators for $K[V]^G$.

Since  $G$ is reductive \cite{fultonrepr}, 
$V$ can be decomposed  as a direct sum of irreducibles:
\begin{equation} \label{eqweyl1}
V=\oplus_{\lambda} m(\lambda) V_\lambda(G),
\end{equation} 
where $\lambda:\lambda_1 \ge \cdots \lambda_r>0$, $r <m$, is a partition, i.e., a non-increasing sequence 
of positive integers,  $V_\lambda(G)$ is the irreducible Weyl module \cite{fultonrepr} of $G$ labelled by $\lambda$,
and $m(\lambda)$ is its multiplicity. 
We assume that $V$ and $G$  are specified 
by the tuple
\begin{equation} \label{eqtuplenew}
 \langle V, G \rangle := (n,m; (\lambda^1,m(\lambda^1));\ldots; (\lambda^s,m(\lambda^s))),
\end{equation}
which   gives $n$ and $m$ in unary, and the 
multiplicity $m(\lambda^j)$ in unary for  each Weyl module $V_{\lambda^j}(G)$ that occurs in the 
decomposition  (\ref{eqweyl1}) with nonzero multiplicity.
The bit-length of this specification  is $O(n+m)$.

The {\em degree}  $d$ of $V$ is defined to be the maximum of $|\lambda|=\sum_i \lambda_i$ over  the $\lambda$'s that occur in this decomposition with nonzero 
multiplicity. 
For each copy of $V_\lambda(G)$ that occurs in this decomposition,
fix  the   standard monomial basis of $V_\lambda(G)$ as defined in \cite{raghavan}. It will be 
reviewed in Section~\ref{sstandard} below. 
This yields a basis $B(V)$ of $V$, which we call  the {\em standard monomial basis} of $V$. 
Let $v_1,\ldots,v_n$ be the coordinates of $V$ in this basis. In what follows, we use these concrete coordinates of $V$ throughout.
So the elements of $K[V]$ are regarded as polynomials in $v_1,\ldots,v_n$.

\begin{theorem}[Derksen](cf. Theorem 1.1, Proposition 1.2 and Example 2.1   in \cite{derksen}) \label{tderksen}
The invariant ring $K[V]^G$ is generated by homogeneous invariants of degree $\le l= n m^2 d^{2 m^2}$.
\end{theorem} 

This bound is $\poly(n)$, when $m$ is constant, since $d \le n$ by the following result.

\begin{lemma} \label{lbound}
Let $V$ be as in (\ref{eqweyl1}). Then (a) $\dim(V)=n \ge d$, and 
(b) $n=\Omega(2^{\Omega(m)})$, if $d = \Omega(m^2)$. 
\end{lemma} 

This can be  shown using 
the fact that the dimension of $V_\lambda(G)$ is equal \cite{fultonrepr} 
to the number of semi-standard tableau of shape $\lambda$. See the preliminary version \cite{GCT5focs}  for the details. The fact (b) will be needed later 
for  the proof of Theorem~\ref{tquasiexplicit}.

Theorem~\ref{tderksen}  allows the following concrete realization of $V/G$.

Let $l$ be as in Theorem~\ref{tderksen}.
Let $K[V]^G_l \subseteq K[V]^G$ be the subspace
of homogeneous invariants of degree $l$, and $K[V]^G_{\le l}$ the subspace of non-constant invariants of degree $\le l$.  The spaces $K[V]_l$ and $K[V]_{\le l}$ are defined similarly.
The dimension $t$ of $K[V]^G_{\le l}$  is bounded by $\dim(K[V]_{\le l})=\sum_{c\le l}  {c+n-1 \choose n-1}$. This  bound is exponential in $n$, even
when $m$ is constant. This worst case upper bound on $t$ is not tight. But we cannot expect a significantly better bound, since
the function $h(l)=\dim(K[V]^G_l)$ is a quasi-polynomial \footnote{This means there exist polynomials $h_1(l), \ldots, h_k(l)$
such that $h(l)=h_j(l)$ if $l=j$ (mod $k$). The {\em degree} of $h$ is the maximum degree of $h_j$'s.}
of degree $\dim(V/G)\ge \dim(V)-\dim(G)=n-m^2$. This follows from \cite{flenner}  since  the singularities of $V/G$ 
are rational \cite{boutot}. To prove
Theorem~\ref{texplicitconst0}, we have  to show 
that some spanning set of $K[V]^G_{\le l}$  of  cardinality exponential in $n$ can still be encoded by a small uniform circuit.

Let $F=\{f_1,\ldots,f_t\}$ be a set of  non-constant  homogeneous invariants that span
$K[V]^G_{\le l}$.
By Theorem~\ref{tderksen},  $F$ generates  $K[V]^G$.
Consider the morphism $\pi_{V/G}$ from $V$ to $K^t$ given by 

\begin{equation} \label{eqmorphismhilbert}
\pi_{V/G}: \quad  v \rightarrow (f_1(v),\ldots,f_t(v)).
\end{equation}
By Theorem~\ref{tmumfordnew} (a), the image of this morphism  is closed, and 
$V/G$ can be identified with this closed image.
Let $z=(z_1,\ldots,z_t)$ be the coordinates of $K^t$,  $I$ 
the ideal of $V/G$ under this embedding, and $K[V/G]$ its coordinate ring. 
Then $K[V/G]=K[z]/I$, and we have the comorphism $\pi_{V/G}^*: K[V/G] \rightarrow K[V]$ given by 

\begin{equation}\label{eqmorphism2new}
 \pi_{V/G}^*(z_i)=f_i. 
\end{equation}

Since $f_i$'s are homogeneous, $K[V/G]$ is a graded ring, with the grading given by $\deg(z_i)=\deg(f_i)$.
Furthermore, $\pi_{V/G}^*$ gives the isomorphism between $K[V/G]$ and $K[V]^G$:

\[\pi_{V/G}^*(K[V/G])=K[V]^G. \]

\subsection{The standard monomial  basis of $V$} \label{sstandard}
We now define the standard monomial basis of $V$ mentioned above   following 
\cite{raghavan}, and prove some lemmas concerning its complexity-theoretic properties.

Let $\bar G=GL_m(K)$. 
Let $Z$ be an  $m\times m$ variable matrix. Let $K[Z]$ be the ring generated by the variable entries of
$Z$. Let $K[Z]_d$ denote the degree $d$ part of $K[Z]$. 
It has commuting left and right actions of $\bar G$, where $(\sigma,\sigma') \in \bar G \times \bar G$
maps $h(Z) \in K[Z]_d$ to $h(\sigma^{t} Z \sigma')$.
For each partition $\lambda: \lambda_1 \ge \cdots \lambda_q >0$, $q\le m$, the Weyl module $V_\lambda(\bar G)$
labelled by $\lambda$ can be embedded in $K[Z]_d$, $d=|\lambda|=\sum_i \lambda_i$, as follows.

Let $(A,B)$ be a bi-tableau of shape $\lambda$. This means both $A$ and $B$ are Young tableau
\cite{fultonrepr} of
shape $\lambda$ such that (1) each box of $A$ or $B$ contains a number in $[m]=\{1,\ldots,m\}$, (2)
all columns of $A$ and $B$ are strictly increasing, and (2) all rows are non-decreasing.  
Let $A_i$ and $B_i$ denote the $i$-th column of $A$ and $B$, respectively. With any pair $(A_i,B_i)$ of
columns, we associate the minor $Z(A_i,B_i)$ of $Z$ indexed by the row numbers occurring in $A_i$ and the column numbers
occurring in $B_i$. With each bi-tableau $(A,B)$, we associate the monomial in the minors of $Z$ defined 
by $Z(A,B):=Z(A_1,B_1) Z(A_2,B_2) Z(A_3, B_3)  \cdots $. We call such a monomial {\em standard} of shape $\lambda$ and 
degree $d=|\lambda|$. We call a monomial in the minors of $Z$ {\em non-standard} if it is not standard.
It is shown in Doubillet, Rota, and Stein \cite{doubillet} that the standard monomials of degree $d$ form a basis of $K[Z]_d$. We denote this basis of $K[Z]_d$  by {\em $B(Z)_d$}.

A standard monomial $Z(A,B)$ is called 
{\em canonical} if the column $B_i$, for each $i$, just consists of the entries $1,2,3,\ldots$ in the increasing
order.
It is known \cite{raghavan}  that, for each partition $\lambda$,  the subspace of $K[Z]$ spanned 
by the canonical monomials of shape $\lambda$ is a representation of $G$ under its left action  on $K[Z]$.
It is also  known \cite{raghavan} that this representation is isomorphic to the Weyl module $V_\lambda(\bar G)$ of $\bar G$, and
that the set of canonical monomials of shape $\lambda$ form its basis. We refer to it as the 
{\em standard monomial basis} of $V_\lambda(\bar G)$, and denote it by $B_\lambda=B_\lambda(\bar G)$.
Each Weyl module $V_\lambda(G)$ of  $G=SL_m(K)$ is also a Weyl module of $\bar G$ in a natural way. 
Hence this also specifies the standard monomial basis $B_\lambda$  of $V_\lambda(G)$.

Fix the standard monomial basis $B_\lambda$ in  each copy of $V_\lambda(G)$ in the complete decomposition
of $V$  as in (\ref{eqweyl1}). This yields a basis $B(V)$ of $V$, which we call its {\em standard monomial basis}.
It depends on the choice of the decomposition of $V$ (if the multiplicities are greater than one). 
But this choice does not matter in what follows.

\begin{lemma} \label{ldrsstraight}

\noindent (a) Given  any   nonstandard monomial $\mu$  of degree $d$ in the minors of $Z$,
the  coefficients  of $\mu$  in the basis $B(Z)_d$ can be computed 
in $\poly(d^{m^2})$ time. More strongly, they can be computed 
by a uniform $\mbox{AC}^0$-circuit of   $\poly(d^{m^2})$ bit-size with oracle access to DET (the determinant function).

\noindent (b)  Consider $K[Z]_d$ as a left $\bar G$-module, 
where $g \in \bar G$ maps $h(Z)$ to $(g \cdot h)(Z)=h(g^{t} Z)$. Then,
given the specifying label (a bi-tableau)  of any basis element $b \in B(Z)_d$  and 
$g \in GL_m(\Q)$, the coefficients  of $g \cdot b$ in the basis $B(Z)_d$ 
can be computed in $\poly(d^{m^2},\bitlength{g})$ time, where $\bitlength{g}$ denotes the bit-length of the specification of $g$. 
More strongly, they  can be computed 
by a uniform $\mbox{AC}^0$-circuit of   $\poly(d^{m^2},\bitlength{g})$ bit-size with oracle access to DET.

\noindent (c) Let $V_\lambda(\bar G)$ be a Weyl module of degree $d$, and
$B_\lambda$ its standard monomial basis as above. 
For any  basis element $b \in B_\lambda$ specified by a tableau and $g \in GL_m(\Q)$, 
the coefficients  of $g \cdot b$ in the basis $B_\lambda$ 
can be  computed in $\poly(d^{m^2},\bitlength{g})$ time.
More strongly, they  can be computed 
by a uniform $\mbox{AC}^0$-circuit of   $\poly(d^{m^2},\bitlength{g})$ bit-size with oracle access to DET.
\end{lemma}

When $m$ is constant, the $\poly(d^{m^2})$  bound becomes $\poly(d)=O(\poly(n))$. 

\proof 

\noindent (a) 
Let $B'(Z)_d$ denote the usual monomial 
basis of $K[Z]_d$ consisting of the monomials in the entries  $z_{i j}$ of $Z$ of total degree $d$.
The cardinality of $B'(Z)_d$ is equal to 
the number of monomials  of degree $d$ in the $m^2$ variables $z_{i j}$'s. This number 
is  ${d + m^2 -1 \choose m^2 -1} = O(\poly(d^{m^2}))$.
The cardinality  of $B(Z)_d$ is the same.
Let ${\cal A}_d$ be the matrix for the change of basis so that:

\begin{equation} 
B(Z)_d = {\cal A}_d B'(Z)_d, \quad \mbox{and} \quad B'(Z)_d = {\cal A}_d^{-1} B(Z)_d.
\end{equation}

The matrix ${\cal A}_d$ can be computed in $\poly(d^{m^2})$ time. 
For this, observe that each row of ${\cal A}_d$ 
corresponds to the expansion of a standard monomial $b\in B(Z)_d$ 
in the usual monomial basis $B'(Z)_d$.  Since the number of monomials of degree $\le d$ in the $m^2$ variable entries of 
$Z$ is $O(\poly(d^{m^2}))$ and the degree of $b$ is $d$, this expansion can be  computed 
by a uniform weakly skew (Section~\ref{sterminology}) circuit
of  $\poly(d^{m^2})$ bit-size (constructed by induction on $d$). 
It follows \cite{malod} 
that it can also be computed  fast in parallel by a uniform  $\mbox{AC}^0$-circuit of $\poly(d^{m^2})$ bit-size with oracle access to DET.
This yields the representation of $b$ in the basis $B'(Z)_d$.  Thus ${\cal A}_d$ can be computed by a uniform $\mbox{AC}^0$-circuit 
of $\poly(d^{m^2})$ bit-size with oracle access to DET.

Once ${\cal A}_d$ has been computed, ${\cal A}_d^{-1}$  can also be computed fast in parallel \cite{malod} by a uniform  $\mbox{AC}^0$-circuit of 
$\poly(d^{m^2})$ bit-size with oracle access to DET.

The standard representation in the basis $B(Z)_d$ of any nonstandard monomial 
$\mu \in K[Z]_d$ in the minors of $Z$ can now be
computed fast in parallel as follows. Let $b(\mu)$ and $b'(\mu)$ be the row vectors of the coefficients of $\mu$ in
the bases  $B(Z)_d$ and $B'(Z)_d$, respectively. Clearly $b(\mu)=b'(\mu){\cal A}_d^{-1}$.
Expand $\mu$ fast in parallel (as we expanded $b$  above) to get its representation $b'(\mu)$. 
Multiply this on the right by ${\cal A}_d^{-1}$ fast in parallel to get $b(\mu)$.

\noindent (b) First, we expand $g \cdot b$ fast in parallel (as above) to get its
representation in the usual monomial basis $B'(Z)_d$.
The representation in $B(Z)_d$ can now be computed fast in parallel
by multiplication on the right by ${\cal A}^{-1}_d$.

\noindent (c) 
This follows from (b),  using  the concrete realization of $V_\lambda(G)$ described before,
as the $G$-submodule of  $K[Z]_d$, $d=|\lambda|$,  spanned by the canonical 
 monomials of shape $\lambda$.
\qed

\subsection{Encoding generators of $K[V]^G$ by a depth four  circuit} \label{sencode}
We shall deduce Theorem~\ref{texplicitconst0} from a stronger result (Theorem~\ref{texplicitconst}) described below, which  shows how 
to encode a set of generators of $K[V]^G$ by a depth four   circuit. To state it we need a 
few definitions.

Let $v=(v_1,\ldots,v_n)$ be the  coordinates of  $V$ in the standard monomial basis $B(V)$ of $V$ as above.
Let $x = (x_1,\ldots,x_n)$ be new variables.
Let 
\begin{equation} \label{eqX}
X=  \sum_i x_i v_i \in K[V;  x]
\end{equation}
be a generic affine combination of $v_i$'s.
Here $K[V;x]$ denotes the ring obtained by adjoining $x_1,\ldots,x_n$ to $K[V]=K[v_1,\ldots,v_n]$.
Then, for any $c>0$,

\begin{equation} \label{eqgeneric}
 X^c= \sum_{a_1,\ldots,a_n \ge 0: \sum a_i = c} {c \choose a_1,\ldots,a_n} (\prod_{i\ge 1} x_i^{a_i}) 
(\prod_{i \ge 1} v_i^{a_i}).
\end{equation}
Here ${c \choose a_1,\ldots,a_n}$ denotes the multinomial coefficient, and the monomials 
$(\prod_{i \ge 1} v_i^{a_i})$ occurring in this expression form a basis of
the subspace $K[V]_c \subseteq K[V]$ of polynomials on $V$ of degree $c$.

Let $R=R_G:K[V]\rightarrow K[V]^G$ denote the Reynolds' operator for $G$ (cf. Section 2.2.1 in \cite{derksenbook}).
We denote the induced map from $K[V; x]$ to $K[V]^G[x]$ by $R$ as well.
Here $K[V]^G[x]$ denotes the ring obtained by adjoining $x_1,\ldots,x_n$ to $K[V]^G$.
Now consider a generic invariant 
\begin{equation}  \label{eqgeneric2}
R(X^c)(v,x) =
\sum_{a_1,\ldots,a_n \ge 0: \sum a_i =c} {c \choose a_1,\ldots,a_n} 
R(\prod_{i \ge 1} v_i^{a_i}) (\prod_{i\ge 1} x_i^{a_i}) \in K[V]^G[x].
\end{equation}

Since the  monomials $(\prod_{i \ge 1} v_i^{a_i})$ in (\ref{eqgeneric}) form a basis of $K[V]_c$, 
it follows from the properties of the Reynold's operator that the elements 
$R(\prod_{i \ge 1} v_i^{a_i}) \in K[V]^G$ occurring in (\ref{eqgeneric2}) span the subspace 
$K[V]^G_c \subseteq K[V]^G$ of invariants of degree $c$. 
By Theorem~\ref{tderksen},  the invariants of degree $\le l=n m^2 d^{2 m^2}$ 
generate $K[V]^G$. Hence,
the set 
\begin{equation} \label{eqF}
 F=\{R(\prod_{i \ge 1} v_i^{a_i}) \ | \ \sum_i a_i =c, 0<c \le l \}
\end{equation}
generates $K[V]^G$.

Let $\Delta_3[n,l,k]$ denote the class of diagonal depth three circuits  
(cf. Section~\ref{sterminology}) over 
$K$ and  the variables $x_1,\ldots,x_n$ with total degree $\le l$ and top fan-in $\le k$.
The size of any such circuit  is $O(k n l)$.

Theorem~\ref{texplicitconst0} follows from the following stronger result.

\begin{theorem}  \label{texplicitconst}
Let $N=n^{m^2} d^{m^4}$, and let $l=n m^2 d^{2 m^2}$ as in Theorem~\ref{tderksen}.
Given $n,m$, $0<c\le l$,  and the specification $\langle V, G \rangle$ of $V$ and $G$ as in (\ref{eqtuplenew}),
one can compute in $\poly(N)$ time the specification of a depth four  
circuit $C=C[V,m,c]$ 
over $\Q$ such that (1) $C$  computes  the polynomial $R(X^c)(v,x)$ in 
$x=(x_1,\ldots,x_n)$ and $v=(v_1,\ldots,v_n)$, and (2)  for  any fixed 
$h \in V$, the circuit $C_h$, obtained by specializing the variables $v_i$'s  in $C$ 
to the coordinates of $h$ in the standard monomial basis $B(V)$ of $V$,  is 
a diagonal depth three circuit in the class $\Delta_3[n,c,k]$, with $k=O(\poly(N))$.

More strongly,  $C$   can be computed by a uniform $\mbox{AC}^0$-circuit of
$\poly(N)$ bit-size with oracle access to DET.
\end{theorem} 

\noindent {\em Proof strategy:} 
The proof  proceeds in four  steps:
(1) Show that the computation of 
the Reynolds operator on $K[V;x]$ can be reduced to (a) the computation of the Reynolds 
operator on the coordinate ring $K[G]$  of $G$, and (b) the computation of 
a certain  comorphism  $\psi^*$ on $K[V;x]$  (defined below) 
associated with the representation $V$ (cf. Lemma~\ref{lreynold1}). 
(2) Give an efficient
algorithm for the computation of  the Reynolds operator on $K[G]$, as needed in (1)(a), 
for constant $m$ 
(cf. Lemma~\ref{lreynolds}). 
(3) Show that  the computation of $\psi^*(X^c)$
can be encoded by a small  circuit of constant depth, for constant
$m$ (cf. Lemma~\ref{lC'}).
(4) Put  (1), (2), and (3) together  to  construct efficiently a small circuit of  depth four
for computing $R(X^c)(v,x)$, for constant $m$.

The following  lemma concerning the  computation of the Reynolds operator $R=R_G$, $G=SL_m(K)$,
addresses the first step in this proof strategy.

Consider the representation morphism  $\psi: V \times G \rightarrow V$
given by: $(v,\sigma) \rightarrow \sigma^{-1}  v$.
Let $\psi^*: K[V] \rightarrow K[V \times G] \cong K[V] \otimes K[G]$ denote the corresponding
comorphism.  This is defined so that, for any $f \in K[V]$ and $t \in V \times G$,

\begin{equation} \label{eqpsistar}
\psi^*(f)(t)= f(\psi(t)).
\end{equation} 

By extending the base from $K$ to 
$K[x]=K[x_1,\ldots,x_n]$, we get the morphism $\psi^*$ from 
$K[V;x]$ to $K[V;x] \otimes K[G]$. 
Given 
$f \in K[V;x]$, let  $\psi^*(f)=\sum_i g_i \otimes h_i$, where 
$g_i \in K[V;x]$ and $h_i \in K[G]$. 

\begin{lemma}\label{lreynold1} (cf. Proposition 4.5.9 
and Remark 4.5.29 in  \cite{derksenbook})
\[ R_G(f)= \sum_i g_i R_G(h_i).\] 
\end{lemma} 

This reduces the computation of $R_G$ on $K[V;x]$ 
to (a) the computation of $R_G$ on $K[G]$, and (b) the computation of $\psi^*$.

Now we address the step (a) above. 
Since $G$, as an affine variety, has just one $G$-orbit (with respect 
to the left-action of $G$ on itself), it follows that $K[G]^G=K$. Hence,
$R_G$ maps $K[G]$ to $K[G]^G=K$. Let $\Q[G]$ denote the coordinate ring of $G$ over $\Q$.
Then $R_G$ similarly  maps $\Q[G]$ to $\Q[G]^G=\Q$.  
For the proof of Theorem~\ref{texplicitconst},
we   only need to compute  $R_G$ on $\Q[G]$. 

Let $Z$ be an $m\times m$ variable matrix.
Then   $K[G]=K[Z]/J$, where $J$ is the principal ideal  generated by $\det(Z)-1$.
Furthermore, by the First Fundamental Theorem of invariant theory \cite{fultonrepr},
$K[Z]^G=K[\det(Z)]$, where $K[Z]$ is considered as a left $G$-module 
as in Section~\ref{sstandard}.

The following lemma addresses the second step in the proof strategy, namely, the computation of $R_G$ on $K[G]$. 

\begin{lemma} \label{lreynolds}
Given $g \in \Q[G] \subseteq K[G]$, represented as a polynomial  $f \in \Q[Z]$, 
$R_G(g) \in \Q$ can be  computed in  $\poly(\deg(f)^{m^2},\bitlength{f})$ time,
where $\bitlength{f}$ denotes the total bit-length of the coefficients of $f$. 
 
More strongly, $R_G(g)$ can be computed by  a uniform  $\mbox{AC}^0$-circuit of 
$\poly(\deg(f)^{m^2},\bitlength{f})$ bit-size with oracle access to DET.
\end{lemma}

The computation of $R_G$ on $K[G]$ can be reduced to the computation of  $R_G$ on $K[Z]$, considered as a left $G$-module as in
Section~\ref{sstandard},  where $R_G$ maps $K[Z]$ to $K[Z]^G=K[\det(Z)]$. Indeed, 
if $g \in K[G]$ is represented by $f \in K[Z]$,  then
$R_G(g)=R_G(f)$ (mod $J$).

Hence, to prove Lemma~\ref{lreynolds}, it suffices to show how $R_G(f)$, $f \in \Q[Z]$,
can be computed in the stated running time. 

Towards this end,  we first recall 
how  $R_G$ on $K[Z]$ can be computed 
using Cayley's $\Omega$ process  \cite{hilbert}. 
Here $\Omega$ is a differential operator on $K[Z]$ defined as follows. Let $z_{i,j}$'s denote the variable entries of $Z$.
Then, for any $h(Z) \in K[Z]$, 

\[ \Omega(h(Z)):= \sum_{\pi \in S_m} \mbox{sign}(\pi) 
\f{\partial^m h}{\partial z_{1,\pi_1} \partial z_{2,\pi_2} \cdots \partial z_{m,\pi_m}},\] 
where $S_m$ is the symmetric group on $m$ letters.

\begin{lemma} (cf. Proposition 4.5.27 in \cite{derksenbook}) \label{lderksenbook2}
Suppose $f \in K[Z]$ is homogeneous. If the  degree of $f$ is $m r$, then
\[R_G(f)= \det(Z)^{r} \f {\Omega^r f}{c_{r,m}},\] 
where $c_{r,m}=\Omega^r(\det(Z)^r) \in \Z$ (it is non-zero).  If the degree of $f$ is not divisible by $m$,
then $R_G(f)=0$.

If $g \in K[G]$ is represented by $f \in K[Z]$, then $R_G(g)=\f {\Omega^r f}{c_{r,m}}$, if 
the degree of $f$ is $m r$,  and $R_G(g)=0$, if the degree of $f$ is not divisible by $m$.
\end{lemma} 

\noindent {\em Proof of Lemma~\ref{lreynolds}:} 
By Lemma~\ref{lderksenbook2}, it suffices to compute  $\Omega^r(f)$ and 
$c_{r,m}=\Omega^r(\det(Z)^r)$ within the stated running time, when $\deg(f)=mr$.

Write $\det(Z)^r=\sum_\alpha a_\alpha \alpha(z_{1,1},\ldots,z_{m,m})$, where $\alpha$ ranges over 
the monomials in $z_{i,j}$'s of degree $m r$, and $a_\alpha \in \Z$. Then 
\[
\Omega^r=\sum_\alpha a_\alpha \alpha(\f{\partial}{\partial  z_{1,1}},\ldots, \f{\partial} {\partial z_{m,m}}).
\] 

The number of $\alpha$'s  here is ${mr + m^2 -1 \choose m^2 -1}= O(\poly(\deg(f)^{m^2}))$, when $\deg(f)=mr$,
and the bit-length of each $a_\alpha$ is $\poly(m,r)=\poly(\deg(f))$. 
Hence  $\f {\Omega^r f}{c_{r,m}} \in \Q$, for $f \in \Q[Z]$ of degree $m r$,
can be  computed in $\poly(\deg(f)^{m^2},\bitlength{f})$ time.

The coefficients $a_\alpha$'s can also  be computed fast in parallel 
by  a uniform  $\mbox{AC}^0$-circuit of 
$\poly(\deg(f)^{m^2})$ bit-size with oracle access to DET,
using  multi-variate Vandermonde interpolation \cite{strassen,yehu}; 
cf. also the proof of Lemma~\ref{lC'} below for the use of this technique. 
Hence  $\f {\Omega^r f}{c_{r,m}}$  can also be computed  fast in parallel.  \qed 

Next  we address the third step in the proof strategy, namely, the computation 
of $\psi^*(X^c)$.

Let $\bar G=GL_m(K)$. Then  $V$ as in (\ref{eqweyl1})
is also a polynomial $\bar G$-representation in a natural way so that, as a $\bar G$-module:
\begin{equation} \label{eqweyl2}
V=\oplus_{\lambda} m(\lambda) V_\lambda(\bar G).
\end{equation} 

Let $u \in \bar G$ be a generic (variable)  matrix. Let $0<c\le l=O(\poly(n,d^{m^2}))$ and $N=n^{m^2} d^{m^4}$ be as in  Theorem~\ref{texplicitconst}. Let 
$u^{-1}= {\mbox Adj}(u)/\det(u)$, where $\mbox{Adj}(u)$ denotes the adjoint of $u$.
Let $u_{i, j}$ denote the $(i,j)$-th entry of $u$.

For any $f \in K[V;x]$, 
let $u \cdot f \in K[V;x]$ denote the result of applying $u \in \bar G$ to $f$, thinking of 
$K[V;x]$ as a $\bar G$-module in the natural way. Formally, 
\begin{equation} \label{eqactionkvx}
(u \cdot f) (w) = f(u^{-1} \cdot w),
\end{equation}
for all $w \in V$, thinking of $f$ as a polynomial function on $V$ with coefficients in $K(x)$.
Here $u^{-1} \cdot w$ denotes the result of applying $u^{-1}$ to $w$.
Let

\begin{equation} \label{eqactionkvx2}
(u \diamond f) (w) := f(Adj(u) \cdot  w),
\end{equation}
for all $w \in V$. If $u$ were a generic matrix of $G$, instead of $\bar G$,
then $u \diamond f$ and $u \cdot f$ would coincide.

For $X \in K[V;x]$ as in (\ref{eqX}),
  $u \diamond  X$ can be expressed as:
\begin{equation}  \label{eqlinform1}
u \diamond  X = \sum_i x_i (u \diamond v_i) = \sum_i  e_i(x, u) v_i,
\end{equation}
where $e_i \in \Q[ x, u]$ is a polynomial in $x_j$'s and the entries $u_{i,j}$'s
of $u$, which is determined by the action of $\bar G$ on $V$. It 
is  linear in $x_j$'s, and  has total degree $\le d (m-1) \le d m$ 
in the entries of $u$. The latter fact follows
from (\ref{eqactionkvx2}), 
since $V$ is a representation of $\bar G$ of degree $d$, and $Adj(u)$ has degree $m-1$ in $u_{i,j}$'s.

By (\ref{eqactionkvx2}), $(u \diamond  X^c)(w)= X^c(Adj(u) \cdot  w)$, for all $w \in V$. Hence, it follows
from (\ref{eqlinform1}) that

\begin{equation}  \label{eqform2}
(u \diamond  X^c) = (u \diamond  X)^c = \sum_{\mu}   \mu \beta_\mu(v, x), 
\end{equation}
where $\mu$ ranges over the  monomials  in $u_{i,j}$'s 
of total degree at most 
$d m c \le d m l = O(\poly(n,d^{m^2}))$,  and 
$\beta_\mu(v, x)$ is a polynomial of degree $c$ in $v=(v_1,\ldots,v_n)$ as well as  $x=(x_1,\ldots,x_n)$.
The number of $\mu$'s here  is  $\le {d m c + m^2 -1 \choose m^2 -1}= O(\poly(N))$.

Thinking of  $\mu$'s 
as elements of $K[G]$,  $\psi^*(X^c)$, by the definition of $\psi^*$, cf. (\ref{eqpsistar}), 
equals  the r.h.s. of (\ref{eqform2}). This is because
$u \cdot X^c$ and $u \diamond X^c$ coincide  if we think of  $u$ as a generic
element of $G$ (rather than $\bar G$), whence $\det(u)=1$.

Thus:

\begin{equation}  \label{eqform3}
\psi^*(X^c) =  \sum_{\mu}   \mu \beta_\mu(v, x), 
\end{equation}
where $\mu$ and $\beta_\mu$ are as in (\ref{eqform2}).

Hence, to encode $\psi^*(X^c)$ efficiently 
by a circuit, it suffices to encode $\beta_\mu(v,x)$'s efficiently 
by a circuit. This is done in the following result.

\begin{lemma} \label{lC'}
Let $N=n^{m^2} d^{m^4}$. Then, given $n,m,d,c$ as above, and the specification $\langle V, G \rangle$ of $V$ and $G$  as in 
(\ref{eqtuplenew}), 
one can compute in $\poly(N)$ time, and more strongly, by a uniform $\mbox{AC}^0$-circuit 
of $\poly(N)$ bit-size with oracle access to DET, the
specification of a circuit $C'$ over $\Q$ of $\poly(N)$ bit-size
on the  input variables $v_1,\ldots,v_n$ and $x_1,\ldots,x_n$,  and with multiple outputs 
that compute  the polynomials $\beta_\mu(v, x)$'s in (\ref{eqform2}).
The top (output) gates of $C'$ are all addition gates. Furthermore, for any fixed $h \in V$, 
the circuit $C'_h$ obtained from $C'$ by specializing the variables $v_i$'s to 
the coordinates  of $h$ (in the standard monomial basis of $V$) is a diagonal depth three circuit with multiple outputs in the class 
$\Delta_3[n,c,e]$, $e=\poly(N)$. By this, we mean that the sub-circuit of $C'$ 
below  each output gate is in $\Delta_3[n,c,e]$.
\end{lemma}
\proof 
We cannot compute $\beta_\mu(v, x)$ in (\ref{eqform2}) by expanding $(u \diamond  X)^c$ 
as a polynomial in $x$, $u$, and $v$, 
since the number of terms in this expansion is exponential in $n$. 
But we can compute it  by a constant depth  circuit,
by evaluating $(u \diamond  X)^c$ at several values of $u$ and then performing 
multivariate Vandermonde interpolation in the spirit of Strassen \cite{strassen,yehu},  as follows.

First,  we show how to construct, for any 
fixed $g \in M_m(\Q)$,
a constant depth  circuit $A_g$  that computes the polynomial in $v$ and $x$ given by 
\begin{equation} 
g \diamond  X^c = (g \diamond  X)^c  = (\sum_i  e_i( x, g) v_i)^c,
\end{equation}\
where $e_i( x,g)$ is a linear form in $ x$ that is obtained by evaluating
$e_i( x, u)$ in (\ref{eqlinform1}) at $u=g$. By the definition of $\diamond$, 
cf. (\ref{eqactionkvx2}),
this is well defined at any element of $M_m(\Q)$ (not just $GL_m(\Q)$). 

Towards this end,  we first construct a depth two circuit $A_g'$, 
with an addition gate at the top,
that computes the quadratic 
polynomial in $v$ and $x$
\begin{equation} 
g \diamond  X = \sum_i  e_i( x, g) v_i,
\end{equation}
obtained by instantiating (\ref{eqlinform1}) at $u=g$. 
Recall that  $v_1,\ldots,v_n$ are the  coordinates of $V$ corresponding to the 
standard monomial basis $B(V)$ of $V$ compatible with the decomposition (\ref{eqweyl2}).
Hence, using Lemma~\ref{ldrsstraight} (c), the coefficients  of the 
linear form $e_i(x,g)$, for given $g \in M_m(\Q)$,
can be computed in $\poly(n,d^{m^2},\bitlength{g})$ time, and more strongly, by a uniform $\mbox{AC}^0$-circuit of
$\poly(n,d^{m^2},\bitlength{g})$ bit-size with oracle access to DET.
After this, the specification
of $A_g'$ can  be computed in  $\poly(n,d^{m^2},\bitlength{g})$ time, and more strongly, by a uniform 
$\mbox{AC}^0$-circuit of  $\poly(n,d^{m^2},\bitlength{g})$ bit-size with oracle access to DET.

Next,  we construct $A_g$, with a single multiplication (powering) gate of fan-in $c$ at its top,
that  computes the $c$-th power of $g \diamond X$ computed by the output node of $A_g'$. 
The polynomial $A_g(v, x)$ computed by $A_g$ is 
$(g \diamond  X^c)(v, x)$. Furthermore, for any fixed $h \in V$,
the circuit obtained by instantiating $A_g$ at $v=h$ is a depth two circuit with a multiplication
(powering) gate at the top. 

Next, we show how to efficiently construct 
a  circuit $C'$  for computing the polynomials $\beta_\mu$'s,
using $A_g$'s for several $g$'s of $\poly(N)$ bit-length.

Let $u_{i,j}$ denote the $(i,j)$-th entry of  $u$ as before.
Let $e$ be the  number of monomials $\mu$'s in $u_{i,j}$'s with the  degree in each $u_{i,j}$ at most $d':=d m c$.
Then $e=O(( d m c)^{m^2})=O(\poly(N))$, since $c \le l=\poly(n,d^{m^2})$.
Order these monomials lexicographically. 
For $r\le e$, let $\mu_r$ denote the $r$-th monomial in this order. 
Choose $m\times m$ non-negative integer  matrices $g_1,\cdots,g_e$ such that 
(1) the $e\times e$ matrix $B=[\mu_r(g_s)]$, whose $(s,r)$-th entry, for  $s,r \le e$, is
 $\mu_r(g_s)$,   is non-singular, and 
(2) every entry of each $g_s$ is $\le d'$. We can choose such $g_s$'s explicitly so that $B$ is a 
multivariate Vandermonde matrix as described in  Section 3.9 in \cite{pan}. Specifically, 
let $E=\{0,\ldots,d'\}^{m^2}$ be the set of $e$ integral points in $\Z^{m^2}$.
Order $E$  lexicographically. 
Let $g_s$ be the $s$-th point in $E$, interpreted as an $m \times m$ matrix.
Then $B$ is a non-singular multivariate Vandermonde matrix 
(cf. Sections 3.9 and 3.11 in \cite{pan}). It can be computed in $\poly(N)$ time, and
more strongly, by a uniform $\mbox{AC}^0$-circuit of $\poly(N)$ bit-size.
Its inverse $B^{-1}$ can  be computed by a uniform $\mbox{AC}^0$-circuit of 
$\poly(N)$ bit-size with oracle access to DET.

Let $\bar \beta$ denote the column-vector of length  $e$ whose $r$-th entry, for $r\le e$,
is $\beta_{\mu_r}(v, x)$ (which we define to be zero if the total degree of $\mu_r$ exceeds $d'=d m c $). 
Let $\bar  A$ denote the column vector of length $e$  whose $s$-the entry, for $s\le e$, is 
$A_{g_s}(v, x)=(g_s \diamond  X^c)(v,x)$.
Then, by (\ref{eqform2}),
\[ \bar  A= B \bar \beta, \quad \mbox{and} \quad \bar \beta = B^{-1} \bar  A.\]
Using the second  equation here,
we can construct a constant depth circuit $C'$ (with multiple outputs) 
for computing the entries of $\bar \beta$, using the constant depth circuits $A_{g_s}$'s constructed 
above. 
Each output  gate of $C'$ is an addition gate with fan-in $e=\poly(N)$. 
Each gate at the second level from the top is a  powering gate with fan-in $c$, 
because the top gate of each  $A_{g_s}$ is the powering  gate with fan-in $c$. 
For a fixed $h \in V$, the circuit $C'_h$  obtained by instantiating $C'$ at $v=h$ 
is thus a  diagonal depth three circuit with multiple outputs in the class $\Delta_3[n,c,e]$.

Since $A_{g_s}$, for every $g_s \in E$, and $B^{-1}$ can be constructed in $\poly(N)$ time,
the construction of $C'$ takes $\poly(N)$ time. 
More strongly, it can be computed by a uniform $\mbox{AC}^0$-circuit of   $\poly(N)$ bit-size with oracle access to DET. \qed

In the final step, we put everything together
to construct the circuit $C=C[V,m,c]$ 
for computing $R(X^c)$, as required in Theorem~\ref{texplicitconst}, 
given $n,d,m,c$, and the specification $\langle V, G \rangle$ of $V$ and $G$
as in (\ref{eqtuplenew}).

By Lemma~\ref{lreynold1} and (\ref{eqform3}),
\[R(X^c)(v, x)= \sum_\mu R_G(\mu) \beta_\mu(v, x).\]
Here $R_G(\mu)$ is a rational number that can be computed in $\poly(N)$ time using 
Lemma~\ref{lreynolds}, since the  degree of  $\mu$ is  $\poly(n,d^{m^2})$. 
Let $C'$ be the circuit for computing $\beta_\mu$'s as in Lemma~\ref{lC'}.
The circuit $C$ is obtained by adding a single  addition gate that
performs linear combinations of the various output nodes 
of $C'$ computing $\beta_\mu$'s, the coefficients in the linear combination being the $\poly(N)$-time-computable rational numbers  
$R_G(\mu)$'s.
Since the top gates of $C'$ are addition gates with fan-in $e$, 
we can ensure, by merging the  addition gates in the top two levels,
that the depth of $C$ is the same as that of $C'$. 
The top gate of $C$ after this merge is an addition gate with fan-in $k=e^2=O(\poly(N))$.

Given $n,d,m,c$, and $\langle V, G \rangle$ as in (\ref{eqtuplenew}), 
the specification of $C'$ can be computed in $\poly(N)$ time by  Lemma~\ref{lC'}.
After this, the specification of the circuit $C$  as above 
can also be computed in $\poly(N)$ time. More strongly, it can be computed by a uniform $\mbox{AC}^0$-circuit of 
$\poly(N)$ bit-size with oracle access to DET.

For  any fixed 
$h \in V$, the circuit $C_h$, obtained by specializing the variables $v_i$'s  in $C$ 
to the coordinates of $h$, is 
a diagonal depth three circuit in the class $\Delta_3[n,c,k]$, with $k=e^2=O(\poly(N))$.
This is because, by Lemma~\ref{lC'}, 
$C'_h$  is a diagonal depth three circuit with multiple outputs 
in the class $\Delta_3[n,c,e]$, $e=O(\poly(N))$.

This completes the proof of Theorem~\ref{texplicitconst}. 

More generally:

\begin{theorem} \label{tquasiexplicit}
The categorical quotient $V/G$ is quasi-explicit (cf. Definition~\ref{dexpvariety} (d))
when $m=O(\sqrt{d})$.
\end{theorem}
\proof 
By  Lemma~\ref{lbound} (b),
$l$ and $N$ in Theorems~\ref{tderksen} and \ref{texplicitconst} are  $O(2^{\polylog(n)})$,
if $m=O(\sqrt{d})$. The result follows from Theorem~\ref{texplicitconst}, in conjunction with 
this fact.\qed

\section{NNL for the general ring of invariants} \label{snnlconstant}
In this section we prove Theorem~\ref{thilbertnew}.

Let $V$ be a finite dimensional representation of $G=SL_m(K)$.
Let $K[V]^G \subseteq K[V]$ be the ring of invariants, and $V/G:=spec(K[V]^G)$, the categorical quotient \cite{mumford}.
We assume that $V/G$ and $K[V]^G$ are specified succinctly by  the tuple 
$\langle V,G\rangle$ in  (\ref{eqtuplenew}). The bit-length of this succinct specification is $O(n+m)$.

Since $V/G$ is explicit when $m$ is constant (cf. Theorem~\ref{texplicitconst0}), 
we can also specify it succinctly in this case,
 as per the general definition of an explicit variety
(Definition~\ref{dexpvariety}), by the circuits $C[v,m,c]$'s in Theorem~\ref{texplicitconst}. 
This specification is equivalent when $m$ is constant, because, given 
$\langle V, G \rangle$,  one can compute the circuits $C[V,m,c]$'s in Theorem~\ref{texplicitconst} in $\poly(n,m)$ time.

If $V/G$ is explicit (cf. Conjecture~\ref{cexplicitcategorical}),
the general definition of an s.s.o.p.  for explicit varieties  (Definition~\ref{desopexp}) 
specializes to the following  concrete definition.

\begin{defn} \label{dexplicit}

\noindent (a) 
A set  $S \subseteq K[V]^G$ is an {\em s.s.o.p. (small system of parameters)} for   $K[V]^G$ if
(1) $K[V]^G$ is integral over the subring generated by $S$,
(2) the cardinality  of $S$ is $\poly(n,m)$, 
(3) every invariant in $S$ is homogeneous of $\poly(n,m)$ degree, and (4) every $s \in S$  has a small specification in the form of 
a  circuit (Section~\ref{sterminology})   of $\poly(n,m)$ bit-length over $\Q$ and the coordinates $v_1,\ldots,v_n$ of $V$
in the standard monomial basis.

\noindent (b) 
A set $S \subseteq K[V]^G$ is  an {\em e.s.o.p. (explicit system of parameters)}
for $K[V]^G$ if (1) $S$ is an s.s.o.p. for $K[V]^G$, and
(2) the specification of $S$, consisting of a 
circuit  as above for each $s \in S$, can be computed in $\poly(n,m)$ time, 
given the  specification $\langle V, G\rangle$ as in (\ref{eqtuplenew}).

If $V/G$ is strongly explicit then, by convention, we assume that a small  specification as in (a) (4) for
each element of $s \in S$ is a weakly skew circuit (cf. Section~\ref{sterminology}).

\noindent (c)  Quasi-s.s.o.p. and quasi-e.s.o.p. are defined by  replacing $\poly(n,m)$  by $2^{\polylog(n,m)}$.

\noindent (d) S.s.o.p., e.s.o.p., and the related notions without degree restrictions are defined by dropping 
the degree requirement in (a) (3).

\noindent (e) We call an s.s.o.p. or  an e.s.o.p. separating if $S$ in (a) is 
separating \cite{derksenbook}  (cf.  Section~\ref{sstit}).
\end{defn}

By {\em the problem NNL for $K[V]^G$}, we mean  the problem of constructing an s.s.o.p. for $K[V]^G$, given $\langle V, G\rangle$.
By {\em the strong form of NNL}, we mean the problem of constructing a
separating s.s.o.p.

For constant $m$,  define a {\em separating near-e.s.o.p.} 
for $K[V]^G$ by replacing $\poly(n,m)$ in the definition of a separating e.s.o.p. above
by $O(n^{O(\log \log n)})$. 

We prove the following stronger form of Theorem~\ref{thilbertnew} in this section.

\begin{theorem} \label{thilbertmainlower}
There exists 
a separating near-e.s.o.p. for $K[V]^G$, if $m$ is constant, and a separating quasi-e.s.o.p., if $m=O(\sqrt{d})$.
\end{theorem}

\subsection{An EXPSPACE-algorithm for constructing an h.s.o.p.} \label{sexpspacehsop}
Before  we turn to this goal, we begin with the following result for the construction of an h.s.o.p.

\begin{prop} \label{pfftandsfthilbert}
The problem of constructing an h.s.o.p. (cf. Definition~\ref{dhsop})
for  $K[V]^G$ belongs to EXPSPACE for any $m$, not necessarily constant.
\end{prop}

When $m$ is constant, this result follows from Theorem~\ref{tnnlexplicithsop}, 
since $V/G$ is then explicit (Theorem~\ref{texplicitconst0}).
For general $m$, it cannot be deduced from Theorem~\ref{tnnlexplicithsop}, since $V/G$ in general is not yet known be explicit,
though it is conjectured to be so; cf. Conjecture~\ref{cexplicitcategorical}.

\proof
Let $F=\{f_1,\ldots,f_t\}$ be the set of generators of $K[V]^G$ as in (\ref{eqmorphismhilbert}), and 
$\pi_{V/G}$  the morphism from $V$ to $K^t$ based on $F$ as there. Here $t$  is  the dimension of 
$K[V]^G_{\le l}$, with $l$ as in Theorem~\ref{tderksen}. This can be exponential in $n$ even when $m$ is 
constant; cf. the discussion before (\ref{eqmorphismhilbert}).

Using this embedding $\pi_{V/G}$  of $V/G$ and Gr\"obner basis theory,
we can compute the equations of $V/G \subseteq K^t$ 
in work-space that is exponential in $\dim(V/G) \le n$, polynomial in the dimension $t$ of the
ambient space, and poly-logarithmic in the maximum degree of the elements in $F$; cf. Theorem 1 in \cite{mayr2}. 
This work-space requirement is  single exponential in $n$ and $m$ (since $d \le n$
by Lemma~\ref{lbound}). 

Applying Gr\"obner basis theory \cite{mayr2} again  to these equations of $V/G$,
we compute  an h.s.o.p. for $K[V]^G$.
The work-space requirement of this algorithm is also exponential in $n$ and $m$; cf. the proof of Theorem~\ref{tnnldetexpspace}.
\qed

If we insist on an h.s.o.p., then Proposition~\ref{pfftandsfthilbert} is the best that we can do at present. 
But if we are willing to settle for an s.s.o.p. (which need not have the optimal cardinality) instead of an h.s.o.p.,  then
Theorem~\ref{thilbertmainlower} shows   that a near-s.s.o.p. for $K[V]^G$ can be constructed in
near-$\poly(n)$ time if $m$ is constant, and  more generally, 
a quasi-s.s.o.p. for $K[V]^G$ can be constructed in
quasi-$\poly(n)$ time if $m=O(\sqrt{d})$.

We now turn to the  proof of Theorem~\ref{thilbertmainlower}.

\subsection{A Monte Carlo algorithm} 
We begin with the following result, which 
gives an efficient  Monte Carlo algorithm for constructing
a separating s.s.o.p.,  when $m$ is constant.

\begin{theorem} \label{tmontecarloconstant}
Suppose $m$ is constant. 
Then a separating   s.s.o.p. 
for $K[V]^G$ can be constructed by a $\poly(n)$-time Monte Carlo algorithm that is correct with a high probability.
In particular, a separating  s.s.o.p. for $K[V]^G$ exists.
\end{theorem} 
\proof By Theorem~\ref{texplicitconst0}, 
$V/G$ is explicit when $m$  is constant. Hence the result follows from  Theorem~\ref{tmonteexplicit}.
\qed

\subsection{Reduction of  NNL to the  standard black-box identity testing for diagonal depth three circuits}
The goal now  is to derandomize this algorithm. 

Since $V/G$ is strongly explicit (cf. Theorem~\ref{texplicitconst0})  when $m$  is constant, 
and the image of the map $\pi_{V/G}$ in (\ref{eqmorphismhilbert}) is closed (cf. Theorem~\ref{tmumfordnew} (a)),
it follows from Theorem~\ref{tclosedimage} and Remark 3 thereafter  that 
the algorithm in Theorem~\ref{tmontecarloconstant} can be derandomized, assuming the 
standard  black-box derandomization hypothesis for symbolic determinant identity testing.
The following  result shows that this derandomization is, in fact, possible assuming a much weaker hypothesis,
namely, the standard  black-box derandomization hypothesis for polynomial identity testing for diagonal depth three circuits (cf. Section~\ref{sterminology}).
This  hypothesis is that a hitting set 
against  diagonal depth three circuits on $n$ variables with degree $\le e$ and top fan-in $\le k$ can be computed in $\poly(s)$ time,
where $s=O(n e k)$ is the size of such circuits.
The  parallel black-box derandomization hypothesis in this context 
is that such a hitting set can be computed by a uniform  $\mbox{AC}^0$-circuit of 
$\poly(s)$ bit-size with oracle access to DET. It is known that such a hitting set can be computed by a uniform $\mbox{AC}^0$-circuit of 
quasi-$\poly(s)$ bit-size \cite{manindra}.

\begin{theorem} \label{thilbertmainblack}
Suppose  the standard black-box derandomization hypothesis for   polynomial identity testing for  diagonal depth three circuits over $K$ 
holds. Then $K[V]^G$ has a separating  e.s.o.p. if $m$ is constant.

Specifically, there  then exists a set $S \subseteq K[V]^G$ of $\poly(N)$ homogeneous invariants,
$N=n^{m^2} d^{m^4}$, such that  (1) $S$ is separating, and hence (cf. Theorem~\ref{tsepderksen}) 
$K[V]^G$ is integral over its subring generated by $S$, (2) every invariant in $S$ has   $\poly(N)$ degree, (3) 
every $s \in S$ has a weakly skew (Section~\ref{sterminology}) circuit 
over $\Q$ and the coordinates $v_1,\ldots,v_n$ of $V$ 
of $\poly(N)$ bit-length, and (4)
the specification of $S$, consisting of such a  weakly-skew  circuit for  every
invariant in $S$, can be computed in $\poly(N)$ time. 

Assuming the parallel black-box derandomization hypothesis for  polynomial identity testing for  diagonal  depth three circuits, the 
specification of  $S$ can  be computed by a uniform  $\mbox{AC}^0$-circuit of $\poly(N)$ bit-size with oracle access to DET.
\end{theorem} 
\proof
Let $N$ be as above. Let $k=O(\poly(N))$ and $l$ be  as in Theorem~\ref{texplicitconst}.
Consider  the class $\Delta_3[n,l,2 k]$ (cf. Section~\ref{sencode}) 
of diagonal depth three circuits over $n$ variables, with total degree $\le l$, and top fan-in
$\le 2 k$.

By our black-box derandomization hypothesis  for  diagonal depth three circuits over $K$,
there exists a  hitting set  $T$ against $\Delta_3[n,l,2 k]$ that can be computed  in $\poly(n,k,l)=\poly(N)$ time.
Assuming the  parallel black-box derandomization hypothesis, $T$ can  be computed by a uniform $\mbox{AC}^0$-circuit of 
$\poly(N)$ bit-size.

Fix such a $T$. By the definition of a hitting set, for any circuit $D \in \Delta_3[n,l,2 k]$ such that $D(x)$, $x=(x_1,\ldots,x_n)$,
is not an identically zero polynomial,
there exists $b \in T$ such that $D(b) \not = 0$.

For any $b \in T$ and $0<c\le l$, define the invariant 
\[ r_{b,c}:=R(X^c)(v,b) \in K[V]^G,\] 
where $R(X^c)$ is as in (\ref{eqgeneric2}).
Let 
\begin{equation} \label{eqShilbert}
 S= \{ r_{b,c}  \ | \  b \in T, 0 < c \le l\}  \subseteq K[V]^G.
\end{equation}
The elements of $S$ are homogeneous polynomials in $v$ of  degree $\le l$, which is $\poly(n)$ if $m$ is constant.

\begin{claim} The set  $S$ is separating. 
\end{claim}
\proof Let $w_1,\ldots,w_n$ be auxiliary variables. For every $c \le l$, define the symbolic difference
\[ \tilde R^c(x,v,w)= R(X^c)(v,x) - R(X^c)(w,x), \] 
where $R(X^c)(w,x)$ is defined just like $R(X^c)(v,x)$, substituting $w$ for $v$.
Suppose $e,f  \in V$ are two points such that $r(e) \not = r(f)$ for some 
$r \in K[V]^G$. It follows that some generator in the set $F$ in (\ref{eqF}) assumes different values at $e$ and $f$.
From (\ref{eqgeneric2}), it follows that, for some $c \le l$, $\tilde R^c(x,e,f)$ is not an identically zero polynomial in
$x$. By Theorem~\ref{texplicitconst}, $R(X^c)(e,x)$ is computed by a diagonal depth three circuit in the class 
$\Delta_3[n,l,k]$. Hence $\tilde R^c(x,e,f)$ is computed by a diagonal depth three circuit in the class 
$\Delta_3[n,l,2 k]$. Since $T$ is a hitting set against such circuits,
it follows that, for some $b \in T$, $\tilde R^c(b,e,f) \not = 0$. That is,
\[ r_{b,c}(e)=R(X^c)(e,b) \not = R(X^c)(f,b)=r_{b,c}(f). \]  Thus $S$ is separating. This proves the claim.

It follows from the claim and Theorem~\ref{tsepderksen} that $K[V]^G$ is integral over the subring generated by $S$.

For any $b\in T$ and $0<c\le l$, let $D_{b,c}$ be the circuit obtained by specializing the circuit 
$C[V,m,c]$ in Theorem~\ref{texplicitconst} at $x=b$. Then $D_{b,c}$ computes $r_{b,c}=R(X^c)(v,b)$ as a polynomial in $v$.
We  specify $S$ by giving, for every invariant $r_{b,c} \in S$,
the specification of  $D_{b,c}$. By Theorem~\ref{texplicitconst}, the circuit $D_{b,c}$  has constant depth and $\poly(N)$ bit-size.
Hence, it can also be  specified by  a weakly skew circuit  of $\poly(N)$ bit-size.

By our black-box derandomization hypothesis, 
the specification of $T$ can be computed in $\poly(N)$ time.
Once $T$ is computed, 
using  Theorem~\ref{texplicitconst}, we can compute in $\poly(N)$ time,
for each $b\in T$ and $c \le l$,  the specification of the circuit $D_{b,c}$  computing 
the invariant $r_{b,c} \in S$. 
Thus the specification of $S$ in the form of a  circuit 
$D_{b,c}$ for  each $r_{b,c}$, or the corresponding weakly skew circuit, can be computed in $\poly(N)$ time.
Hence, $S$ is a separating e.s.o.p.

Assuming the  parallel black-box derandomization hypothesis,
$T$,  and hence $S$,  can be computed by a uniform  $\mbox{AC}^0$-circuit of  $\poly(N)$ bit-size with oracle access to DET.
\qed

\subsection{Proof of  Theorem~\ref{thilbertmainlower}}

\proof 
By Forbes, Saptharishi, and Shpilka  \cite{fs1}, 
a hitting set against diagonal depth three circuits of size $\le s$ can be computed in
$O(s^{O(\log \log s)})$ time.
The result  follows from the proof of Theorem~\ref{thilbertmainblack} in conjunction
with this fact; we also need Lemma~\ref{lbound}, if $m=O(\sqrt{d})$. \qed 

\subsection{General $m$} \label{sgeneralm}
The following is  the current best result  for general $m$.

\begin{theorem} \label{tmarbitrary}
Let  $V$  be a finite dimensional  representation of $G=SL_m(K)$.
Suppose $V/G$ is explicit (cf. Definition~\ref{dexpffthilbert}).
Then $K[V]^G$ has a separating  e.s.o.p.,
 assuming
the standard black-box derandomization hypothesis for low-degree  polynomial identity testing   over $K$
\end{theorem}

\proof  The proof is similar to that of Theorem~\ref{thilbertmainblack}, using the assumed 
explicitness of $V/G$ in place of Theorem~\ref{texplicitconst},
and the black-box derandomization hypothesis for low-degree 
polynomial identity testing in place of the black-box derandomization hypothesis for diagonal depth three circuits. 
\qed

\noindent {\em Remark 1:} If $V/G$ is strongly explicit
(cf. Definition~\ref{dexpffthilbert}), 
then it follows similarly that $K[V]^G$ has a separating e.s.o.p.,
assuming the standard 
black-box derandomization hypothesis for symbolic determinant identity testing. 
If $V/G$ is explicit without any degree restrictions, then one has to assume instead
the black-box derandomization hypothesis for polynomial  identity testing without
any degree restrictions. 

\noindent {\em Remark 2:} All these results (and Theorems~\ref{torbitclosuregen} 
 (a), and \ref{tsemisimple} (b)  below) also hold 
assuming explicitness of $V/G$ in the relaxed sense
(cf. Definition~\ref{dexpffthilbert} (e)). 

\noindent {\em Remark 3:} 
The derandomization hypothesis in Theorem~\ref{tmarbitrary}  can be traded, up to a quasi-prefix, 
 with the hardness hypothesis in Theorem~\ref{trussell}.

We also note down a consequence of the proof of Theorem~\ref{tmarbitrary}.

\begin{theorem} \label{torbitclosuregen}
Let $V$ be a finite dimensional representation of $G=SL_m(K)$. Then:

\noindent (a) 
The  problem of deciding if the closures of the $G$-orbits of two rational points in $V$ 
intersect, and finding an invariant separating the two if they do not,
 belongs to co-RNC,  if $V/G$ is explicit. 
It is in NC assuming, in addition, the white-box derandomization hypothesis \cite{russell}
for low degree arithmetic circuits over $\Q$ \footnote{This hypothesis is that,
given a low-degree  arithmetic circuit over $\Q$, 
one can decide  in polynomial time if the circuit
evaluates a non-zero polynomial, and if so, construct in the same time
an input on which the evaluation is non-zero.}.

\noindent (b) The problem   belongs to P, if $m$ is constant. 

\noindent (c) It belongs to DET $\subseteq$ NC, for constant $m$,  
if we do not ask for a separating invariant if
the closures do not intersect. 
\end{theorem}

\proof \noindent (a):
The proof of the first statement  is similar to that of Theorem~\ref{torbitclosure}, using the assumed explicitness of $V/G$ 
in place of Theorem~\ref{texplicitmatrix}. The second statement is implicit in the proof of the
first statement.

\noindent (b) and (c): Suppose $m$ is constant. Using Theorem~\ref{texplicitconst}
in place of Theorem~\ref{texplicitmatrix} in the proof of Theorem~\ref{torbitclosure}, we get a co-RNC-algorithm 
for the problem. This algorithm only uses white-box (cf. Section~\ref{sblackstd})
polynomial identity testing for diagonal depth three circuits.
It can be derandomized using 
the DET-algorithm for this test, which  follows from
Raz and Shpilka \cite{raz},  Arvind et al. \cite{arvind}, and Saxena \cite{saxena}. 
This yields  a DET-algorithm as stated in (c) that, however,   does not 
return a separating invariant if the orbit-closures do not intersect.
To get a separating invariant if the closures intersect, as needed
in (b), we use instead a polynomial time algorithm for 
white-box polynomial identity testing for diagonal depth three circuits
that returns a witness input  
if the polynomial computed by the circuit is not identically zero. Such an algorithm can be   obtained by combining
\cite{raz,arvind,saxena} with a proof technique in \cite{manindra} (cf. Appendix A
therein). Using this witness input, a separating invariant can be  constructed if the orbit-closures
do not intersect; cf. the proof of Theorem~\ref{torbitclosure}. 
\qed 

\noindent {\em Remark 3:} 
Theorem~\ref{torbitclosuregen} (c)
can also be proved without representation theory as follows.
The orbit closures of two points $v,w \in V$ intersect iff 
$\forall \epsilon \in \R \ \exists g,h \in SL_m(\C): ||g\cdot v - h \cdot w||_2 \le \epsilon^2,$ where $||\ ||_2$ denotes the $L_2$ norm on $V$. 
If $m$ is constant, this can be checked  in  polynomial time 
by the algorithm in \cite{basu} for quantifier elimination in the theory of reals (which can also be parallelized). 
This proof  does not 
return an invariant separating the two orbit closures if they do not intersect, as in 
Theorem~\ref{torbitclosuregen} (a) and (b).
This  is a serious limitation in the context of invariant theory.
Most importantly,  this proof  is based on an inherently white-box
technique, which does not work in the black-box setting of Theorem~\ref{thilbertnew}.

\noindent {\em Remark 4:} Theorem~\ref{torbitclosuregen} (a) implies that
the problem of deciding if a given rational point in $V$ belongs to the null cone \cite{mumford} 
of the $G$-action
belongs to co-RNC, if $V/G$ is explicit. It belongs to NC assuming, in addition, the white-box derandomization 
hypothesis for low degree arithmetic circuits over $\Q$.

\subsection{Generalization to 
reductive algebraic groups} \label{sgeneralize}
The preceding results for $SL_m$  can be generalized to other  reductive
algebraic groups  as follows.

Let $K$ be algebraically closed field of characteristic zero.
Let $G$ be  a connected, reductive, algebraic group over $K$,
specified by its root datum \cite{humphreys,milne}.
Let $V$ be a  finite dimensional  rational representation of $G$.
Given any highest weight $\lambda$ of $G$, let $V_\lambda(G)$ denote the 
associated irreducible representation of $G$ \cite{fultonrepr}. 
We specify $V$, as in (\ref{eqtuplenew}),
by giving (in unary)  $n=\dim(V)$ and  the multiplicities of $V_\lambda(G)$'s that occur with non-zero 
multiplicities in $V$.

\begin{theorem} \label{tsemisimple}
Let $V$ and $G$ be as above. Then:

\noindent (a) Analogues of Theorems~\ref{thilbertmainlower}, \ref{thilbertmainblack},  and 
\ref{torbitclosuregen} (b) hold,  when $\dim(G)$ is constant

\noindent (b) Analogues of Theorems~\ref{tmarbitrary} and \ref{torbitclosuregen}  (a)
hold, without any restriction on $\dim(G)$. In particular, if $V/G$ is explicit, 
$K[V]^G$ has a separating e.s.o.p.,  assuming the standard black-box derandomization hypothesis
for low-degree polynomial identity testing over $K$.
\end{theorem}

This can be proved by extending the proof for $SL_m$;
cf. the preliminary version \cite{GCT5focs} for details.

It may be conjectured that,
for any finite dimensional representation $V$ of a connected, reductive, algebraic
group $G$ in characteristic zero, with the specification of $V$ and $G$ as above,
$V/G$ is explicit, if $K[V]^G$ has a set of generators
of  $\poly(n,\dim(G))$ degree, and is explicit 
without any degree restrictions, in general (cf. Definition~\ref{dexpffthilbert}).

More generally, let $V$ be a finite dimensional representation
of any  reductive, possibly disconnected, algebraic group $G$ over an algebraically closed field 
of any characteristic \footnote{Here, we assume that $V$ and $G$ are specified as follows, since the direct-sum decomposition as in 
(\ref{eqweyl1}) need not hold
in positive characteristic.
Let $G_0 \subseteq G$ be the connected component of the identity, specified by its 
root datum \cite{humphreys,milne}, and $T$ its maximal torus.
We specify $V$ by 
giving, in its fixed basis, the representation matrices for: (1)  one-parameter multiplicative 
subgroups  generating $T$, (2) one-parameter additive  subgroups $U_\alpha$'s 
associated  with the roots $\alpha$'s of $G_0$ with respect $T$ 
(cf. Section 26.3  in \cite{humphreys}), and (3) 
a set of elements $a_1,\ldots,a_l \in G \setminus G_0$, which together with 
$T$ and  $U_\alpha$'s generate $G$. The entries of the representation
matrix of a one-parameter subgroup are assumed to be rational functions of the 
parameter with coefficients in a finite extension of $\Q$,
if the characteristic is zero, or $F_p$, if the characteristic  is $p>0$.}.
Then  it may be conjectured that $V/G$ is explicit in the relaxed sense, without any degree restrictions, in general; cf. Definition~\ref{dexpffthilbert} (e). 
It  may be conjectured to be explicit in the relaxed sense,
with the usual low degree restrictions, 
if there is an  upper bound on the degrees
of generators or separating invariants for $K[V]^G$ that is polynomial in the bit-length of the succinct specification of
$V$ and $G$. This  is the case, for example,  when $G$ is finite and
$V$ is its permutation representation \cite{goebel}.
The conjecture is proved  in the next section  in any characteristic 
for $G=SL_m(K)$ and $V=M_{m}(K)^r$, with the adjoint action of $G$ 
 (cf. Theorem~\ref{texprelaxedmatrix}). 
Analogues of Theorems~\ref{tmarbitrary} and \ref{torbitclosuregen}  (a) hold for any finite dimensional representation $V$ of
any reductive group $G$ in any characteristic,  assuming that  $V/G$ is explicit in the relaxed sense. 
If  $V/G$ is explicit in the relaxed sense without any degree restrictions (as conjectured in general), 
then analogues of Theorems~\ref{tmarbitrary} and  \ref{torbitclosuregen}  (a) still hold, 
replacing low degree polynomial identity testing with
general polynomial identity testing without any degree restrictions.
These results can be proved by replacing  Theorem~\ref{tmumfordnew} with its generalization in
 \cite{mumford}
for arbitrary reductive groups in any characteristic.

In view of the results and arguments above, 
the strong form of NNL for $K[V]^G$, for any 
finite dimensional representation $V$ of any reductive group $G$ in any characteristic,
may be conjectured to be in P, along with 
the $G$-orbit-closures-intersection and the null cone membership
problems (cf. Theorem~\ref{torbitclosuregen} and Remark 4 thereafter).

\section{Extensions} \label{sextensions}
In this section we extend the results in Sections~\ref{sexpmatrixring} and  \ref{sstit}
to arbitrary  characteristics (cf. Section~\ref{smatrixpos}), and to 
quivers (cf. Section~\ref{squiver}). We then
deduce their  implications for parametrization of closed orbits 
in representations of reductive groups (cf. Section~\ref{sclosed}),
and for parametrization of semi-simple representations of finitely generated
algebras (cf. Section~\ref{ssemisimple}). We also 
extend the results in Sections~\ref{snnldeter} and \ref{sexplicit} to large  enough positive characteristics
(cf. Section~\ref{sposcharequi}).
Henceforth, $K$ will denote an algebraically closed field of any characteristic $p$.

\subsection{Matrix invariants in arbitrary   characteristic} \label{smatrixpos}
First,  we  prove Theorem~\ref{tmatrixNew} in arbitrary characteristic.
It follows from the following  stronger result.

Let $V=M_m(K)^r$, with the adjoint action of $G=SL_m(K)$. 
Separating s.s.o.p. and e.s.o.p. for $K[V]^G$, and the black-box derandomization 
hypothesis for low-degree circuits are  defined as 
in characteristic zero (cf.  Sections~\ref{sstit} and  \ref{sblackstd}).
If the characteristic  $p$ is positive, 
the  constants in the circuits specifying  the s.s.o.p. 
and the entries of the elements of the hitting set against low-degree circuits 
are assumed to be in $F_{p^l}$, the finite field with $p^l$ elements, with $l=O(\log(m))$. 

\begin{theorem} \label{tmatrixarbi}
Let $V$ and $G$ be as above.

\noindent (a) Suppose  $p \not \in [2,\floor{m/2}]$.
Then a separating e.s.o.p. exists  for $K[V]^G$,
 assuming the standard  black-box derandomization hypothesis for
polynomial identity testing for read-once oblivious algebraic branching programs
(cf. Section~\ref{sterminology}).
A separating quasi-e.s.o.p.  exists  unconditionally.

\noindent (b) A separating e.s.o.p. exists for  $K[V]^G$ for  any  $p$,
assuming the standard black-box derandomization hypothesis for symbolic determinant
identity testing over $K$.

\noindent (c) The problem of deciding if the closures of the $G$-orbits of two rational
points in $V$ intersect belongs to 
co-RDET $\subseteq$ co-RNC for any $p$. 
It belongs to  NC if $p \not \in [2,\floor{m/2}]$. By a rational point in $V$, when $p>0$,
we mean a point whose coefficients  belong to  a finite extension of $F_p$.
\end{theorem} 

The known upper bound  \cite{domokos} on the degrees of the generators
in the  First Fundamental Theorem for matrix invariants in positive characteristic in  Donkin \cite{donkin} is exponential in $m$, 
unlike the polynomial bound in  the First Fundamental Theorem for matrix invariants in characteristic zero 
(cf. Theorem~\ref{tprf}).
Hence  the proof of Theorem~\ref{tmatrixmain} cannot be extended to arbitrary characteristic
using  Donkin \cite{donkin}  in place of Procesi and Razmyslov \cite{procesimatrix,razmyslov}.
But, as we shall see below, the  proof can be extended  using the following geometric alternative to  Theorem~\ref{tprf}
in arbitrary characteristic.

Let $U=(U_1,\ldots,U_r)$ denote an $r$-tuple of variable $m\times m$ matrices as in Section~\ref{sexpmatrixring}. Identify
$K[V]$ with the ring $K[U]=K[U_1,\ldots,U_r]$ generated by the variable entries of $U_i$'s.
Given any word $\alpha \in [r]^*$, define  $T_\alpha(U) \in K[V]^G$ as in (\ref{eqtraceu}).
For any $m\times m$ matrix $X$, let $c_i(X)$ denote the $i$-th coefficient of its characteristic polynomial, so that
\[ \det(\lambda I -X)= \lambda^m -c_1(X) \lambda^{m-1} + \cdots + (-1)^m c_m(X) I. \] 
Define a  separating set $S \subseteq K[V]^G$ in arbitrary characteristic $p$
just as it was defined in characteristic zero in Section~\ref{sstit}.

\begin{theorem}[Geometric First Fundamental Theorem  in arbitrary characteristic]   \label{tprocesigen}
\noindent (a) The set $\{c_i(U_j)\} \cup \{T_\alpha(U)\} \subseteq K[V]^G$, where 
$\floor{m/2}< i \le m$, $1 \le j \le r$, and 
$\alpha \in [r]^*$ ranges over all words of length $\le m^3$,
is separating, if $p \not \in [2,\floor{m/2}]$. 

\noindent (b) The set 
$\{c_{i,\alpha}(U) \ | \ 0 \le i \le m\} \subseteq K[V]^G$, where
$\alpha=i_1 i_2 \cdots \in [r]^*$ ranges over all words of length $\le m^2$,
and $c_{i,\alpha}(U)=c_i(U_{i_1} U_{i_2} \cdots)$, 
is  separating for any $p$.
\end{theorem}

In characteristic zero, this result follows  from  Theorem~\ref{tprf}, 
letting  $\alpha$ range over the words of length $\le m^2$.
For a proof in arbitrary characteristic, we  need the following two results.

Let $\hat R=K\langle U_1,\ldots,U_r\rangle$ be the free non-commutative algebra over $K$ generated by the $r$ 
matrix-variables $U_1,\ldots,U_r$ (not the $r m^2$ variable entries of $U_i$'s).
Given any $A=(A_1,\ldots,A_r) \in M_m(K)^r$, let $\rho_A: \hat R \rightarrow M_m(K)$ 
denote the $m$-dimensional representation of $\hat R$ given by $U_i \rightarrow A_i$. 
Clearly, two tuples $A, B \in M_m(K)^r$ belong to the same $G$-orbit iff $\rho_A$ and $\rho_B$ are isomorphic representations.
We say that $A \in M_m(K)^r$ is semi-simple if $\rho_A$ is a semi-simple representation of $\hat R$.

\begin{theorem}[cf. Theorem 4.1 in King \cite{king}]  
\label{tartin}

The $G$-orbit of $A \in M_m(K)^r$ is  closed iff $A$ is semi-simple.
\end{theorem}

Let $R$ be  any finite-dimensional algebra over $K$. Let 
$\rho: R \rightarrow M_m(K)$ be an   $m$-dimensional representation of $R$.
For any  $r \in R$, let $\chi_\rho(r)$ denote the characteristic polynomial of $\rho(r)$.
Let $Q \subseteq R$ be any subset that spans $R$ over $K$. 

\begin{theorem}[Brauer and  Nesbitt] (cf. \cite{brauer},  Theorem 5.7 in \cite{eggermont} and its proof) \label {tbrauer}
Two finite dimensional semi-simple representations $\rho$ and $\rho'$ of $R$ are isomorphic iff $\chi_\rho(q)=\chi_{\rho'}(q)$ 
for all $q \in Q$.
If $R$ is not a finite dimensional algebra, then the same statement also holds if $\rho(Q)$ and $\rho'(Q)$ span $\rho(R)$ and 
$\rho'(R)$, respectively.
\end{theorem}

This follows from the proof of Theorem 5.7 in \cite{eggermont}.

\noindent {\bf Proof of Theorem~\ref{tprocesigen}:} 

\noindent (a) By  the generalization of Theorem~\ref{tmumfordnew} (d) to
arbitrary characteristic 
(cf.  Theorem 1.1 in  \cite{mumford}),
it suffices to show that the set 
$\{c_i(U_j)\} \cup \{T_\alpha(U)\}$ of invariants in (a)
separates closed $G$-orbits in $M_m(K)^r$, i.e.,  given two distinct closed $G$-orbits, there exists an invariant in the
set that assumes different values on the orbits.

By Theorem~\ref{tartin}, $A \in M_m(K)^r$ has a closed $G$-orbit iff $A$ is semi-simple. By definition, this 
is so iff the $m$-dimensional representation
$\rho_A$ of $\hat R=K\langle U_1,\ldots,U_r\rangle$ given by $U_i \rightarrow A_i$ is semi-simple.

Furthermore, two semi-simple tuples $A, B \in M_m(K)^r$ are in the same (closed) $G$-orbit iff
the two representations $\rho_A$ and $\rho_B$ of $\hat R$ are isomorphic.
Let $S \subseteq [r]^*$ be the  subset of words  of length $\le m^3$. 
It suffices to show that, given any two semi-simple
$A, B \in M_m(K)^r$ with $\rho_A \not \cong \rho_B$, there exists an $\alpha \in S$ such that $T_\alpha(A) \not = T_\alpha(B)$, or there 
exist an $i$, with $\floor{m/2} < i \le m$,  and $j \le r$
such that $c_i(A_j) \not = c_i(B_j)$, where $A_j$ denotes the $j$-th matrix in $A$.

Let $K[A]$ denote the subalgebra of $M_m(K)$ generated by $A_i$'s, the subalgebra $K[B]$ being similar.
Clearly $\rho_A(\hat R)=K[A]$, and $\rho_B(\hat R)=K[B]$.
Since $\dim(K[A]) \le \dim(M_m(K))=m^2$, it follows (cf. Pappacena \cite{pappacena}) that the
words in $A_i$'s of length $\le m^2$ span $K[A]$.
Similarly, the words in $B_i$'s of length $\le m^2$ span $K[B]$.  Let $Q$ be the set of words in $U_i$'s of length $\le m^2$. 
It follows that $\rho_A(Q)$ and $\rho_B(Q)$ span $\rho_A(\hat R)=K[A]$ and $\rho_B(\hat R)=K[B]$, respectively.

Suppose to the contrary that $T_\alpha(A)=T_\alpha(B)$, for all $\alpha \in S$,
and $c_i(A_j)=c_i(B_j)$, for all $\floor{m/2} < i \le m$ and $j \le r$. 
Then we will show that $\chi_{\rho_A}(q)=\chi_{\rho_B}(q)$, for all $q \in Q$.
For this, we have to show that
$c_{k,\alpha}(A)=c_{k,\alpha}(B)$, for every $\alpha \in [r]^*$ of length $\le m^2$
and  $1 \le k \le m$, where $c_{k,\alpha}(A)=c_k(A_{i_1} A_{i_2} \cdots)$ if $\alpha=i_1 i_2 \cdots$.

Fix any word $\alpha=i_1 \cdots i_l$ of length $l \le m^2$.
It follows that $\alpha^j=\alpha \cdots \alpha$ ($j$ times), for any $j \le m$, 
belongs to $S$. Hence, by our assumption, it follows that $T_{\alpha^j}(A)=T_{\alpha^j}(B)$ for 
all $j \le m$, and $c_i(A_j)= c_i(B_j)$ for all $\floor{m/2} < i \le m$ and $j \le r$.  
By Lemma 2 in Domokos \cite{domokos}, 
$c_{k,\alpha}(A)$,  $1 \le k \le m$, is a polynomial in 
$c_{t,\alpha}(A)$'s, $t \le \floor{m/2}$, and $c_i(A_j)$'s, 
$\floor{m/2} < i \le m$ and $j \le r$.  
Since $p \not \in [2,\floor{m/2}]$, by Newton's identities, $c_{t,\alpha}(A)$,
for $t \le \floor{m/2}$, can be expressed as:

\[ c_{t,\alpha}(A) =  \f {1}{t} \sum_{t'=1}^{t} (-1)^{t'-1}  c_{t-t',\alpha}(A) 
T_{\alpha^{t'}}(A).\]

This shows that $c_{t,\alpha}(A)$, for $t \le \floor{m/2}$,  is a polynomial in 
$T_{\alpha^j}(A)$'s, $j\le \floor{m/2}$. 
The story for $B$ is similar.

It follows  that 
$c_{k,\alpha}(A)=c_{k,\alpha}(B)$, for every $\alpha \in [r]^*$ of length $\le m^2$
and  $1 \le k \le m$. That is, 
$\chi_{\rho_A}(q)=\chi_{\rho_B}(q)$ for all $q \in Q$.

The representations $\rho_A$ and $\rho_B$ are semi-simple, since $A$ and $B$ are semi-simple.
Furthermore,  $\rho_A(Q)$ and $\rho_B(Q)$  span $\rho_A(\hat R)=K[A]$ and $\rho_B(\hat R)=K[B]$, respectively.
Hence, it follows from Theorem~\ref{tbrauer},  applied to $\hat R$, 
that $\rho_A\cong \rho_B$; a contradiction. 

\noindent (b) The proof is similar to that of (a). It holds in arbitrary characteristic,
 since we do not need to use Newton's 
identities now. 
\qed

\subsubsection{Proof of Theorem~\ref{tmatrixarbi}}

\noindent (a):  Fix any  $p \not \in [2,\floor{m/2}]$. 
Let  $Y=(y_1,\ldots,y_{m^2})$ be a tuple of 
auxiliary variables. For any $l \le m^2$,  let
$P_l(Y,U) := \sum_\alpha Y_\alpha T_\alpha(U)$, 
where $\alpha=\alpha_1 \alpha_2 \cdots$ 
ranges over  all words of length $l$ with each $\alpha_j \in [r]$, 
and $Y_\alpha=\prod_{j=1}^l y_j^{\alpha_j}$.

By Theorem~\ref{tprocesigen} (a),
the coefficients $c_i(U_j)$'s of $\det(z I - U_j)$'s (considered as
polynomials in $z$ with coefficients in $K[U]$),  $1 \le j \le r$,
and the coefficients of $P_l(Y,U)$'s (considered as  polynomials in $Y$ with coefficients in $K[U]$), $l \le m^2$,  form
a separating set of invariants in $K[V]^G$.

Furthermore (cf. the proof of Lemma~\ref{lforbes}), each
$P_l(Y,U)$ can be
computed by a read-once oblivious algebraic branching program over $Y$ and $U$ of $\poly(l,m,r)$ size, thinking of the entries
of $U_i$'s as indeterminate constants.

Let $U'=(U_1', \ldots,U_r')$ be another tuple of $m\times m$ variable matrices.
Let $\tilde P_l(Y,U,U'):= P_l(Y,U) - P_l(Y,U')$. 
It follows that $\tilde P_l(Y,U,U')$ can also be
computed by a read-once oblivious algebraic branching program over $Y$, $U$, and $U'$ of 
size $q=O(\poly(l,m,r))$, thinking of the entries
of $U_i$'s and $U'_i$'s as indeterminate constants. By our assumption, there exists an
explicit $\poly(l,m,r)$-time-computable hitting set $B$ for
polynomial identity testing for read-once oblivious algebraic branching programs of size $q$ over $Y$. Fix such a $B$. In the definition of $B$,  we are 
considering programs over $Y$, and not over $Y$, $U$, and $U'$, for
the reasons that will become clear soon.
Fix also  $m+1$ distinct elements $a_0,\ldots,a_m \in F_{p^k}$, $k=O(\log m)$.

\begin{claim} \label{cSpos}
The set 

\begin{equation}
S :=\{P_l(b,U) \ | \  b \in B, 1 \le l \le m^2 \} \cup \{ \det(a_i I - U_j) \ |  \ 0 \le i \le m, 1 \le j \le r \}
\end{equation} 

is a separating set of invariants in $K[V]^G$, if $p \not \in [2,\floor{m/2}]$.
\end{claim} 

\noindent {\em Proof of the claim:} 
Let  $A=(A_1,\ldots,A_r)$ and $A'=(A_1',\ldots,A_r')$ be any  two $r$-tuples in
$V=M_m(K)^r$ such that,
for some invariant $h \in K[V]^G$, $h(A) \not = h(A')$. We have to show that
some element in $S$ assumes distinct values at $A$ and $A'$.

By Theorem~\ref{tprocesigen} (a), 
either (1) some
coefficient $c_i(U_j)$ 
of $\det(z I - U_j)$ (considered as a polynomial in $z$), for some $j \le r$, 
or (2) some coefficient of  $P_l(Y,U)$ (considered as a polynomial in $Y$), for some $l \le m^2$, 
assumes different values at $A$ and $A'$.

In the first case, since $\det(z I - U_j)$,
as a polynomial in $z$, has degree $m$, it follows that
$\det(a_i I - A_j) \not = \det(a_i I - A_j)$ for some $0 \le i \le m$. Hence 
$\det(a_i I - U_j) \in S$ assumes distinct values at  $A$ and $A'$ in this case.

In the second case,  $\tilde P_l(Y,A,A')=P_l(Y,A)-P_l(Y,A')$
is not identically zero as a polynomial in $Y$.
Since $\tilde P_l(Y,U,U')$ has a read-once oblivious algebraic branching program 
of size $q=O(\poly(m,r))$  over $Y$, $U$, and $U'$, thinking of the entries
of $U_i$'s and $U'_i$'s as indeterminate constants, $\tilde P_l(Y,A,A')$ 
has a read-once oblivious algebraic branching program  of size $q$ over $Y$.
Since $B$ is a hitting set against such programs over $Y$,
 and $\tilde P_l(Y,A,A')$ is not identically zero as a polynomial in $Y$,
there exists $b \in B$ such that $\tilde P_l(b,A,A') \not = 0$, i.e.,
$P_l(b,A) \not = P_l(b,A')$. Hence, $P_l(b,U) \in S$ assumes distinct values at
$A$ and $A'$ in this case.

It follows that  $S$ is a separating set of invariants in $K[V]^G$.
This proves the claim.

Every element of $S$ is clearly homogeneous of $\poly(m,r)$ degree.
By the generalization of Theorem~\ref{tsepderksen} to arbitrary characteristic (cf. Theorem 2.3.12
in \cite{derksenbook}), it follows that
$K[V]^G$ is integral over the subring generated by $S$.

The size of $S$ is   $\poly(m,r)$.
Since the hitting set $B$ is explicit,
and $P_l(Y,U_j)$'s and $\det(a_i I - U_j)$'s
have explicit   weakly skew circuits,
it follows that the specification of $S$, consisting of a 
weakly skew circuit  for its every element, can be computed in $\poly(m,r)$ time. 
Hence $S$ is  a separating  e.s.o.p. of $K[V]^G$.

This proves  the first statement in Theorem~\ref{tmatrixarbi} (a).

The second statement  follows from this proof of the first statement, 
inserting quasi-prefixes in appropriate places, in conjunction with 
the black-box quasi-derandomization of polynomial identity testing for
read-once oblivious algebraic branching programs in Forbes and Shpilka \cite{fs2}, 
which  holds in arbitrary characteristic.

\noindent (b) 
By Theorem~\ref{tprocesigen} (b), the set 
$\{c_{i,\alpha}(U) \ | \ 0 \le i \le m\} \subseteq K[V]^G$, where
$\alpha=i_1 i_2 \cdots \in [r]^*$ ranges over all words of length $\le m^2$
and $c_{i,\alpha}(U)=c_i(U_{i_1} U_{i_2} \cdots)$, 
is a  separating set of invariants in $K[V]^G$, for any  $p$.

Introduce new variables $y$ and $z_{j,s}$, $1 \le j \le m^2$, $0 \le s \le r$.
Let $z=(..,z_{j,s},..)$ denote the tuple of $z_{j,s}$'s. 
Let 
\begin{equation} \label{eqpuyz}
p(U,y,z):= \det(y I-\prod_{j=1}^{m^2} (z_{j,0} I + \sum_{s=1}^r z_{j,s} U_s)),
\end{equation}
where $I$ denotes the $m \times m$ identity matrix. This polynomial remains invariant under the
adjoint action of $G$ on the tuple $U=(U_1,\ldots,U_r)$.
Hence  the coefficients of $p(U,y,z)$, considered as a polynomial in $y$ and $z$ with coefficients
in $K[U]$, belong to  $K[U]^G=K[V]^G$.

For any $\alpha=i_1 i_2 \cdots \in [r]^*$ of length $\le m^2$, we can set each $z_{j,s}$ to either zero or one
so that $p(U,y,z)$ specializes to the characteristic polynomial of $U_{i_1} U_{i_2} \cdots$.
It follows that the coefficients of $p(U,y,z)$, considered as a polynomial in $y$ and $z$ with
coefficients in $K[U]^G$, form a separating set of invariants in  $K[V]^G$. 

The polynomial $p(U,y,z)$ in (\ref{eqpuyz}) has a weakly skew circuit (cf. Section~\ref{sterminology})
of  $O(\poly(m,r))$ size over $y,z$, and $U$. 
By the polynomial equivalence between weakly skew circuits
and symbolic determinants \cite{malod},  $p(U,y,z)$
can also be expressed as a symbolic determinant
of size $q=O(\poly(m,r))$ over $y,z$, and $U$. 
By our assumption, there exists an explicit
$\poly(m,r)$-time-computable hitting set $B$ against all symbolic determinants over $y$ and $z$ of size $q$. Note that  $B$ is defined by
considering  symbolic determinants over $y$ and $z$,
not over $y,z$, and $U$. Fix such as a $B$. 

\begin{claim} 
The set $S=\{ p(U,b_1,b_2) \ | \ (b_1,b_2) \in B \}$ is a set of separating invariants in 
$K[V]^G$, for  any  $p$. 
\end{claim}

The proof is similar to that of Claim~\ref{cSpos}, with Theorem~\ref{tprocesigen} (b) 
in place of Theorem~\ref{tprocesigen} (a).
The rest of the proof of Theorem~\ref{tmatrixarbi} (b) 
is similar to that of the first statement in Theorem~\ref{tmatrixarbi} (a).

\noindent (c): Given two rational points $A, A' \in V=M_m(K)^r$, 
we want to decide if the closures of the $G$-orbits of $A$ and $A'$ 
intersect. By the generalization of Theorem~\ref{tmumfordnew} (d) to
arbitrary characteristic (cf. Theorem
1 in \cite{mumford}), this is so  iff every invariant in $K[V]^G$ assumes the same value at $A$ and $A'$, or
equivalently, if every  invariant in any separating set of invariants in $K[V]^G$ 
assumes the same value  at $A$ and $A'$.  

As noted in the proof of (b), 
the coefficients of the symbolic determinant  $p(U,y,z)$ in (\ref{eqpuyz}), considered as a polynomial in $y$ and $z$ with
coefficients in $K[U]^G$, form a separating set of invariants in  $K[V]^G$. 
It follows that the closures of the $G$-orbits of $A$ and $A'$ intersect iff 
the polynomial $\tilde p(A,A',y,z):=p(A,y,z) - p(A',y,z)$ is  identically zero as a polynomial in $y$ and $z$. 
This can be tested by a co-RDET algorithm \cite{ibarra}: Just  substitute random values for $y$
and $z$, using a large enough extension of $F_p$, if $p>0$, and test if the resulting specialization 
of $\tilde p(A,A',y,z)$ is zero. 

If $p \not \in [2,\floor{m/2}]$, then we can give a deterministic NC-algorithm for the problem as follows. 

By Theorem~\ref{tprocesigen} (a), 
the coefficients
of $\det(z I - U_j)$'s, $1 \le j \le r$, considered as  polynomials in $z$ with coefficients in $K[U]$, 
and the coefficients of $P_l(Y,U)$'s, considered as  polynomials in $Y$ with coefficients in $K[U]$, form
a separating set of invariants in $K[V]^G$ in this case.
Hence it follows from the generalization of Theorem~\ref{tmumfordnew} (d) 
to  arbitrary characteristic \cite{mumford} 
that the closures of the $G$-orbits of $A$ and $A'$  intersect iff 
$f_j(z,A,A'):=\det(z I - A_j) -\det(z I - A_j')$, for every  $j \le r$, is identically zero, and
$\tilde P_l(Y,A,A') := P_l(Y,A) - P_l(Y,A')$, for every  $l \le m^2$, is identically zero. 
The problem of testing whether $f_j(z,A,A')$ is identically zero belongs to DET \cite{cook},
since $f_j(z,A,A')$ is the difference of two symbolic determinants in just one variable. 
The polynomial $\tilde P_l(Y,A,A')$ has a read-once oblivious algebraic branching program over $Y$ of 
$\poly(m,r)$ size. Hence, whether it is identically zero can be tested by an NC-algorithm,
using a straightforward parallelization of the white-box algorithm for polynomial identity testing
for read-once oblivious algebraic branching programs in Raz and Shpilka \cite{raz}. 
(The DET-algorithm for white box polynomial identity testing for read-once oblivious algebraic branching 
programs in  \cite{arvind} works only in characteristic zero.) 
This proves Theorem~\ref{tmatrixarbi} (c). 
\qed

We also note down the following consequence of the proof of Theorem~\ref{tmatrixarbi} (b). 
Define {\em strong explicitness of $V/G$ in the relaxed sense}  by extending 
Definition~\ref{dexpffthilbert} (e) to positive characteristic in 
the obvious way.

\begin{theorem} \label{texprelaxedmatrix}
The categorical quotient $V/G$ is strongly explicit  in the  relaxed sense in
any characteristic, 
with $p(U,y,z)$ in (\ref{eqpuyz}) as the defining polynomial.
\end{theorem}

\noindent {\em Remark (on matrix semi-invariants):}
We can also let $G= SL_m(K) \times SL_m(K)$, and $V=M_m(K)^r$, with the 
left-right action of $G$, which maps 
$(C_1, \ldots, C_r) \in V$, given $(A,B) \in G$, to $(A C_1 B^{-1}, \ldots, A C_r B^{-1})$.  In
characteristic zero, the recent polynomial degree bound 
in  \cite{makam,IQS},
in conjunction with  \cite{weyman,zubkov}, implies that
the categorical quotient $V/G$ is then strongly explicit. Hence, by 
Theorem~\ref{tsemisimple} (b) and 
Remark 1 after Theorem~\ref{tmarbitrary}, $K[V]^G$ has a separating e.s.o.p. in this case, 
 assuming the black-box derandomization hypothesis for symbolic determinant identity 
testing. It would be interesting to make this result unconditional.

\subsection{Generalization to quivers} \label{squiver}
Theorem~\ref{tmatrixarbi} can be generalized to arbitrary quivers as follows.

Let $Q$ be a quiver  (a directed graph allowing loops and multiple arrows) \cite{procesiquiver,weyman}, i.e., a four-tuple $(Q_0,Q_1,t,h)$, where $Q_0=\{1,\ldots,l\}$ is a set of vertices,
$Q_1$ is a finite set of arrows among these vertices, and the two maps $t,h: Q_1 \rightarrow Q_0$ assign to each arrow 
$\phi \in Q_1$ its tail $t(\phi)$ and head $h(\phi)$. 
A {\em representation} $W$ of the quiver  $Q$ over the   field $K$
is a family $\{W(i): i \in Q_0\}$ of finite dimensional vector spaces over $K$ 
together with a family of linear maps $W(\phi): W(t(\phi)) \rightarrow W(h(\phi))$,
$\phi \in Q_1$. The $l$-tuple  $(\dim(W(1)), \ldots, \dim(W(l)))$ 
of integers is called {\em the dimension vector} of $W$. 
For a fixed dimension vector $m=(m(1),\ldots, m(l)) \in \N^l$,
 the {\em representation space}
$V=V(Q,m)$ of the quiver $Q$ is  the set of all representations  of $Q$ with the dimension vector $m$.
Clearly,

\begin{equation} \label{eqquiver}
V=V(Q,m)= \oplus_{\phi \in Q_1} \mbox{Hom}_K (K^{m(t(\phi))}, K^{m(h(\phi))}) = \oplus_{\phi \in Q_1} M_\phi(K),
\end{equation}
where $M_\phi(K)$ denotes the space of $m(h(\phi)) \times m(t(\phi))$ matrices with entries in $K$. 
There is a canonical action of 
\begin{equation} \label{eqG}
 G=\prod_{i=1}^l GL_{m(i)} (K)
\end{equation}
on $V$,  defined by 
\[ (g \cdot W) (\phi) = g(h(\phi)) W(\phi) g (t(\phi))^{-1}, \] 
for any  $g=(g(1),\ldots,g(l)) \in G$ and $W \in V(Q,m)$.

Let $U=(\ldots,U_\phi,\ldots)$, $\phi \in Q_1$,  be a tuple of 
variable  matrices,  where $U_\phi$ is an $m(h(\phi)) \times m(t(\phi))$ variable matrix.
Then the  coordinate ring $K[V]$ of $V$ can be identified with the ring $K[U]$ over the variable entries of $U_\phi$'s.
Let  $K[V]^G \subseteq K[V]$  be the subring of $G$-invariants. 
When $Q$ consists of a single vertex with $r$ self-loops,
$K[V]^G$ for the dimension vector $(m)$, $m \in \N$, 
coincides with the invariant ring  in Theorem~\ref{tmatrixarbi}. 
If $Q$ has no directed cycles then $K[V]^G=K$.  So we are mainly interested in the
case when $Q$ has directed cycles. 

We specify $V$ and $G$ by giving  the graph of the quiver $Q$, and the
dimension vector $m=(m(1),\ldots,m(l)) \in \N^l$ (in unary).
Let  $|m|:=\sum_i m(i)$. The definitions of s.s.o.p. and e.s.o.p. for the ring of matrix invariants extend to this setting
in a natural way.

\begin{theorem} \label{tquiver}
The analogue of Theorem~\ref{tmatrixarbi}, after replacing $m$ there
with $|m|$ here,    holds for   $V$ and  $G$ as above.
\end{theorem}

For the proof, we recall some results concerning the path algebra of a quiver.

Let $R_Q$ be  the  path algebra (cf. Section 1.2 in \cite{brion}) of $Q$. This is the
associative algebra generated by the variables $e_i$, $i \in Q_0$,
and $e_\phi$, $\phi \in Q_1$, subject to the relations:

\begin{equation} 
e_i^2 = e_i, \quad e_i e_j = 0 \ (i\not = j), \quad 
e_{h(\phi)} e_\phi = e_\phi e_{t(\phi)} = e_\phi.
\end{equation}

Given two representations $W_1$ and $W_2$ of $Q$, a morphism $f: W_1 \rightarrow W_2$ between these 
two representations 
is  a family of linear morphisms $\{f(i): W_1(i) \rightarrow W_2(i)
\ | \ i \in Q_0\}$ such that, for all  $\phi \in Q_1$, $W_2(\phi) \circ f(t(\phi)) = f(h(\phi)) \circ W_1(\phi)$. Thus  the set of representations of $Q$ is a category, and two representations 
of $Q$  are  isomorphic iff they are in the same $G$-orbit.

\begin{prop} (cf. Proposition 1.2.2 in \cite{brion}) \label{pbrion}
The category of representations of $Q$ is equivalent to the category of 
left $R_Q$-modules.
\end{prop}

Let $W=(\{W(i)\},\{W_\phi\})$ be any representation of $Q$ with the dimension vector
$m=(m(1), m(2), \ldots)$. For any $i \in Q_0$, 
let $M^W_i$ denote 
the $|Q_0| \times |Q_0|$-block matrix  whose (1) $(i',j')$-th
block, for $1 \le i', j' \le |Q_0|$ with  $i'\not =j'$ or $i'=j'\not = i$,   is the  $m(i')\times m(j')$ zero-matrix, and 
(2) the $(i,i)$-th block is the $m(i)\times m(i)$ identity matrix.
For any $\phi \in Q_1$, let $M^W_\phi$ denote the 
$|Q_0| \times |Q_0|$-block matrix defined similarly,
whose $(h(\phi),t(\phi))$-th block
is $W_\phi$, and  all other blocks are zero. 

It can be checked that the  representation $W$ of $Q$ defines the  left $R_Q$-module
$\hat W := \bigoplus_i W(i)$,
on which the action of $e_i$, $i \in Q_i$, is given by the matrix $M^W_i$,
and  the action of $e_\phi$ is given by $M^W_\phi$.
The representation $\hat W$ is completely specified by the matrix tuple 
$M^W:=(\cdots, M^W_i, \cdots, M^W_\phi, \cdots)$, $i \in Q_0$, $\phi \in Q_1$, of
$|m| \times |m|$ matrices.
We think of this tuple  as an element of  
$\hat V := M_{|m|}(K)^{|Q_0| +|Q_1|}$.

\begin{theorem} (cf. Theorem 4.1 in King \cite{king}) \label{tkingquiver}
The $G$-orbit of a representation $W$ of $Q$ is closed iff
the $R_Q$-module $\hat W$ is semi-simple.
\end{theorem}

\noindent {\em Proof of Theorem~\ref{tquiver}:} 
We  only show how the analogue of Theorem~\ref{tmatrixarbi} (a)
for quivers can be deduced from Theorem~\ref{tmatrixarbi} (a) for
matrix invariants.
The story for the analogues of Theorem~\ref{tmatrixarbi} (b) and (c)
 is similar.

So assume that the characteristic
$p \not \in [2,\floor{|m|/2}]$, and that 
the  black-box derandomization hypothesis 
for polynomial identity testing for read-once oblivious algebraic branching 
programs  holds. 

Let $V$ be the representation space of $Q$ associated with the dimension vector $m$.
Consider the adjoint action of $\hat G=SL_{|m|}(K)$
on $\hat V = M_{|m|}(K)^{|Q_0| +|Q_1|}$. By Theorem~\ref{tmatrixarbi} (a)
and our black-box derandomization hypothesis, 
the invariant ring $K[\hat V]^{\hat G}$ has a separating 
e.s.o.p. $\hat S$ that  can be computed in $\poly(|m|,|Q_0|,|Q_1|)$ time.
Fix such an $\hat S$.

For the tuple  $U=(\ldots,U_\phi,\ldots)$, $\phi \in Q_1$, of 
variable  matrices as before, define
the matrices $M^U_i$, $i \in Q_0$, and $M^U_\phi$, $\phi \in Q_1$, just as
we defined $M^W_i$ and $M^W_\phi$, 
replacing $W_\phi$'s by $U_\phi$'s in the definition.

This defines a generic representation of $R_Q$ on $\oplus_i K^{m(i)}$ specified
by  the matrix tuple
$M^U=(\cdots, M^U_i, \cdots, M^U_\phi, \cdots)$, $i \in Q_0$, $\phi \in Q_1$, 
of $|m|\times |m|$ matrices. This tuple can be 
thought of as a generic point   in $\hat V$, and  we can evaluate each invariant in
$\hat S$ at $M^U$. 
It is easy to see  that, for each $\hat s \in \hat S$, $\hat s(M^U) \in K[V]^G$. 

Let $S=\{ \hat s(M^U) \ | \ \hat s \in \hat  S\}$. 

\begin{claim} \label{cquiverpos}
The set $S$ is a separating set of invariants in $K[V]^G$.
\end{claim}

\noindent {\em Proof of the claim:} 
By the generalization of Theorem~\ref{tmumfordnew} (d) to
arbitrary characteristic (cf. Theorem 1.1 in \cite{mumford}), it suffices to show that $S$ separates the closed $G$-orbits 
in $V$. 

So suppose $A,A' \in V$ are two representations of $Q$ whose $G$-orbits are closed and distinct.
We want to show that some invariant in $S$ assumes distinct values on these orbits. 

Since the $G$-orbits of $A$ and $A'$ are distinct, it follows that $A$ and $A'$ are not isomorphic
representations of $Q$.
Hence, by Proposition~\ref{pbrion}, the $R_Q$-modules $\hat A$ and $\hat A'$ are  not isomorphic.

Since the $G$-orbits of $A$ and $A'$ are closed, it follows from 
Theorem~\ref{tkingquiver} that  the $R_Q$-modules $\hat A$ and $\hat A'$ are semi-simple. 
Hence, by Theorem~\ref{tartin}, the $\hat G$-orbits of the matrix tuples $M^{\hat A}$ and 
$M^{\hat A'}$  are  closed.
Since $\hat A$ and $\hat A'$ are   not isomorphic, it follows that
the $\hat G$-orbits of $M^{\hat A}$ and $M^{\hat A'}$ are distinct and closed.
Since $\hat S$ is a separating set of invariants in $K[\hat V]^{\hat G}$, it follows,
by the generalization of Theorem~\ref{tmumfordnew} (d) to
arbitrary characteristic \cite{mumford}, 
that there exists an invariant $\hat s \in \hat S$ that assumes distinct values at $M^{\hat A}$
and $M^{\hat A'}$. 

But, $\hat s(M^{\hat A})= \hat s(M^U)(A)$, and similarly, 
$\hat s(M^{\hat A'})= \hat s(M^U)(A')$. It follows that 
the element $\hat s(M^U) \in S$ assumes distinct values at $A$ and $A'$. This proves the claim.

Since $\hat S$ is a separating e.s.o.p., a specification of $\hat S$, in the form of 
a weakly skew circuit for its every element, can computed in $\poly(|m|,|Q_0|, |Q_1|)$ time.
It follows that the specification of $S$,
in the form of a
weakly skew circuit for its every  element, can also computed in $\poly(|m|,|Q_0|, |Q_1|)$ time.
The size of $S$ is the same as the size of $\hat S$, which is $\poly(|m|,|Q_0|, |Q_1|)$.
Furthermore, each element of $S$ is homogeneous, since each element of 
$\hat S$ is homogeneous.
By Claim~\ref{cquiverpos}, $S$ is separating. 
Hence, by the generalization of Theorem~\ref{tsepderksen} to arbitrary characteristic
(cf. Theorem 2.3.12 in \cite{derksenbook}), 
$K[V]^G$ is integral over the subring generated by $S$.
It follows that $S$ is a separating e.s.o.p.
of $K[V]^G$. \qed 

Theorem~\ref{tquiver} has a simpler  proof in  characteristic zero.
This can be obtained
by extending the proof of Theorem~\ref{tmatrixmain},
using Proposition~\ref{pbrion},  and replacing Theorem~\ref{tprf} by its 
generalization for quivers
(cf. Theorem 1 in  \cite{procesiquiver}). This generalization states  that 
$K[V]^G$ is generated by the  trace-monomials associated with the oriented cycles in $Q$ of length $\le |m|^2$.

\subsection{Explicit  parametrization of closed orbits} \label{sclosed}
Now, let $V$ be a finite dimensional representation of any reductive \cite{milne} 
algebraic group $G$ over $K$.
The set of  $G$-orbits in $V$, in general,  cannot be given the structure of an
algebraic variety. The fundamental insight in \cite{mumford} is that the set of closed $G$-orbits
in $V$ can be given the structure of an algebraic variety. Indeed, by the generalization of Theorem~\ref{tmumfordnew} to
arbitrary characteristic \cite{mumford},
the points of $V/G$ are in
one-to-one correspondence with the closed $G$-orbits in $V$. But this algebraic structure is not
efficient from the complexity-theoretic perspective, since typically a  set of generators 
for $K[V]^G$, such as the one in Theorem~\ref{tprf},  has exponential cardinality.
So we ask if the set of closed
$G$-orbits in $V$ can be given the structure of a variety
that is efficient from the complexity-theoretic perspective.

For simplicity, we  confine
ourselves to the case when $V=M_m(K)^r$, with the adjoint action of 
$G=SL_m(K)$,
as in Section~\ref{smatrixpos}, and  we assume that $V$ and $G$ are 
specified by giving $m$ and $r$ in unary.
But the analogue of Theorem~\ref{texppara}  below  holds 
for any finite dimensional
representation  of any reductive algebraic group.

Given a set $S =\{s_1,\ldots,s_k\} \subseteq K[V]^G$, let $\psi_S: V \rightarrow K^k$ denote the 
map $v \rightarrow  (s_1(v), \ldots,s_k(v))$. Let $n=\dim(V)$.

\begin{defn} \label{dexppara}
We say that  the closed $G$-orbits in $V$ have an 
{\em explicit parametrization} if
there is exists a subset $S =\{s_1,\ldots,s_k\} \subseteq K[V]^G$, $k=O(\poly(n))$,
of homogeneous invariants of $\poly(n)$ degree such that 
(1) the image $\psi_S(V)$ of $\psi_S$  is closed,
(2)  for any $x \in \psi_S(V)$, $\psi_S^{-1}(x)$ contains a unique closed $G$-orbit in $V$, and
(3) given the specification of $V$ and $G$ as above, the specification of $S$, consisting of a circuit of $\poly(n)$  bit-size for every 
element in it, can be computed in $\poly(n)$ time. 
The constants in these circuits are rational, if the
characteristic $p$ of $K$ is zero. Otherwise, they are in a finite  extension field $F_{p^l}$, with $l=O(\log(m))$.
\end{defn}

In this case, the points of the variety $\psi_S(V)$ are in one-to-one correspondence
with the closed $G$-orbits in $V$, and given any $v \in V$, $\psi_S(V)$ can be
computed in $\poly(n)$ arithmetic operations over $K$. If 
the circuits specifying $S$ are   weakly skew, $\psi_S(V)$,
for a rational $v$ (cf. Theorem~\ref{tmatrixarbi}), can  be computed in time polynomial in $n$ and  the bit-length of $v$.

By the generalization of Theorem~\ref{tmumfordnew} to arbitrary characteristic \cite{mumford}, 
explicit parametrization of closed $G$-orbits in $V$ also yields explicit parametrization of 
the equivalence classes of $G$-orbits in $V$, where two $G$-orbits are considered equivalent iff their closures intersect.

\begin{theorem} \label{texppara}
The closed $G$-orbits in $V$ have an explicit parametrization
if $K[V]^G$ has a separating e.s.o.p.
\end{theorem}

For the proof, we need the following lemma.

\begin{lemma} \label{lderksenpara}
Let $S =\{s_1,\ldots,s_k\} \subseteq K[V]^G$ be a 
separating set of homogeneous invariants.
Then the image $\psi_S(V)$ of $\psi_S$ is a closed subvariety of $K^k$.
Furthermore, for any $x \in \psi_S(V)$, $\psi_S^{-1}(x)$ contains a unique closed $G$-orbit in $V$.
\end{lemma}
\proof The map $\psi_S$ can be factored as: 

\begin{equation} \label{eqfactorpi}
V \stackrel{\pi_{V/G}}{\longrightarrow} V/G  \stackrel {\psi'_S} {\longrightarrow}  K^k,
\end{equation}
where $\pi_{V/G}$ is defined as in (\ref{eqmorphismgeneral}).
By the generalization of Theorem~\ref{tmumfordnew} (a) to arbitrary 
characteristic \cite{mumford},
the first map is surjective.
Hence the image of $\psi_S$ coincides with the image of $\psi'_S$. 
Since $S$ is separating, by the generalization of Theorem~\ref{tsepderksen} to arbitrary
characteristic \cite{derksenbook}, 
the coordinate ring $K[V]^G$ of $V/G$ is integral over 
the subring generated by $S$. This means the map $\psi'_S$ is finite
(cf. Section 5.3 in \cite{shafa}), and hence its image is a closed 
subvariety of $K^k$. Thus  the image of $\psi_S$ is closed.

The map $\psi'_S$ is also one-to-one, since $S$ is separating. 
Hence, by the generalization of Theorem~\ref{tmumfordnew} (b) to
arbitrary characteristic \cite{mumford}, for 
any $x \in \psi_S(V)$, $\psi_S^{-1}(x)$ contains a unique closed $G$-orbit in $V$,
\qed 

\noindent {\em Proof of Theorem~\ref{texppara}:} 
Suppose $K[V]^G$ has a separating e.s.o.p. $S$. The properties (1) and (2) in Definition~\ref{dexppara} 
follow from Lemma~\ref{lderksenpara}. The  property (3) follows because $S$ is an e.s.o.p. 
\qed

Theorem~\ref{tmatrixarbi} (a), in conjunction with the proof of Theorem~\ref{texppara},  implies:

\begin{theorem} \label{tquasiexplicitorbits}
The closed $G$-orbits in $V$ have a quasi-explicit parametrization if $p \not \in [2,\floor{m/2}]$.
\end{theorem}

\subsection{Explicit  parametrization of semi-simple representations of algebras} 
\label{ssemisimple}
Next, we show (cf. Theorem~\ref{tpararepr}) that the existence of a separating e.s.o.p. for $K[V]^G$,
with $V=M_{m}(K)^r$ and $G=SL_m(K)$   as above, 
 implies explicit parametrization of semi-simple representations of any finitely generated algebra.

Let $R$ be a finitely generated  associative algebra over $K$, specified by its generators   $f_1,\ldots,f_r$, and relations 
among them. We assume that the coefficients in the relations are in a finite extension of
$\Q$, if the characteristic is zero, or $F_p$, if the characteristic is $p$.
Let $\rho: R \rightarrow M_m(K)$ be an  $m$-dimensional representation of $R$. 
It can be identified 
with the  $r$-tuple $A=(A_1,\ldots, A_r)\in V = M_m(K)^r$ of $m\times m$ matrices, where $A_i=\rho(f_i)$. 
The set  $W_m=W_m(R)$ of the $r$-tuples corresponding to $m$-dimensional representations 
of $R$ is a closed $G$-subvariety of $V$. 
Two representations of $R$ are isomorphic iff they
lie in the same $G$-orbit, where $G=SL_m(K)$ acts on $V$ by the adjoint action as before. 
By Theorem~\ref{tartin}, a representation is semi-simple iff its $G$-orbit
is closed.
Thus the isomorphism classes of $m$-dimensional semi-simple representations of $R$
can be identified with the closed $G$-orbits in $W_m$. Let $n=r m^2$.

We say that semi-simple representations of $R$ of dimension $m$
have an {\em explicit parametrization} if 
there exists a set $S$ of $\poly(n)$ homogeneous invariants of $\poly(n)$ degree
in $K[V]^G$ such that 
(1) the image $\psi_S(W_m)$ of $W_m$ under the map $\psi_S$, defined in  Section~\ref{sclosed},  is closed,
(2)  for any $x \in \psi_S(W_m)$, $\psi_S^{-1}(x)$ contains a unique closed $G$-orbit in $W_m$, and
(3) given $m$ and the specification of $R$, the specification of $S$, consisting of a circuit of $\poly(n)$  bit-size for every 
element in it, can be computed in  time polynomial in $n$ and the bit-length of the specification
of $R$. The constants in these circuits are rational if the
characteristic $p$ of $K$ is zero. Otherwise, they are in a finite  extension field $F_{p^l}$ with $l=O(\log(m))$.

In this case, the points of the variety $\psi_S(W_m)$ are in one-to-one correspondence
with the isomorphism  classes of $m$-dimensional semi-simple representations of $R$,
and given any $r$-tuple $A \in V$ 
of matrices specifying an $m$-dimensional 
representation of $R$, $\psi_S(A)$ can be computed in $\poly(n)$ arithmetic operations over $K$.

\begin{theorem} \label{tpararepr} \label{trepr1}
For any $m$, the $m$-dimensional semi-simple representations of $R$ over $K$
have an explicit parametrization if $K[V]^G$ has a separating e.s.o.p.
\end{theorem}

This result follows from Theorem~\ref{texppara} and the following result.

\begin{prop} \label{palgebra}
Let $S$ be as in Lemma~\ref{lderksenpara}, with $V$ and $G$ as above.
Then $\psi_S(W_m)$ is a closed subvariety of $K^k$.
\end{prop} 

\proof 
By the generalization of Theorem~\ref{tmumfordnew} (c) to
 arbitrary characteristic \cite{mumford},
$Y=\pi_{V/G}(W_m)$ is a closed subvariety of $V/G$. 
As shown in the proof of Lemma~\ref{lderksenpara}, $\psi_S'$ in 
 (\ref{eqfactorpi}) is a finite morphism.
Since the image of a closed variety under a finite morphism is closed
(cf. Section 5.3. in \cite{shafa}), 
the image $\psi_S'(Y)=\psi_S(W_m)$ is closed.
\qed 

\noindent {\em Remark:} The set $S$ giving the explicit parametrization in Theorem~\ref{tpararepr}
depends only on $m$ and $r$, the number of generators of $R$, but  not on the  relations among the
generators of $R$.

Theorem~\ref{tpararepr}, in conjunction with 
Theorem~\ref{tmatrixarbi} (a), implies:

\begin{theorem} \label{trepr2}
For any $m$, the $m$-dimensional semi-simple representations of $R$ over $K$ have a quasi-explicit parametrization,
if  $p \not \in [2,\floor{m/2}]$.
\end{theorem}

\subsection{Other extensions in positive characteristic} \label{sposcharequi}
Next, we briefly explain how  the results in Sections~\ref{snnldeter} and \ref{sexplicit}
can be extended to   positive characteristics.

The strengthened black-box derandomization problem for low-degree polynomial
identity testing over an algebraically closed field $K$ of positive 
characteristic $p$ is defined just as in characteristic zero (cf. Section~\ref{sstrongblack}).
The hitting set against low-degree circuits over $K$ of size $\le s$ 
is assumed to be a subset of $F_{p^l}^n$, $n$ the number of variables, for 
a large enough $l=O(\log s)$. The phrase ``infinitesimally close'' is  
interpreted in the Zariski topology. The definition of NNL 
is  extended from characteristic zero to positive characteristics similarly in a straightforward way.

The following result extends Theorem~\ref{tequisditdet}
to positive characteristics. 

\begin{theorem}  \label{tequisditpos}
\noindent (a) The variety 
$\Delta[\det,m]$ has a strict e.s.o.p. in any characteristic
iff the strengthened black-box derandomization hypothesis for  symbolic determinant identity testing holds.

\noindent (b) The variety 
$\Delta[\det,m]$ has a strict e.s.o.p. over an
algebraically closed field of  $\Omega(2^{(\log m)^a})$ characteristic, 
for a large enough positive constant  $a$, iff, ignoring a quasi-prefix, there exists
a family $\{f_n(x_1,\ldots,x_n)\}$ of 
exponential-time-computable (cf. the remark after Theorem~\ref{trussell}), 
multi-linear, integral 
polynomials such that $f_n$ 
cannot be approximated infinitesimally closely over an algebraically
closed field of $\Omega(2^{n^\delta})$ characteristic 
by circuits   of  
$O(2^{n^\epsilon})$ size,  for some constants $\delta, \epsilon > 0$,
as $n \rightarrow \infty$.
\end{theorem}


Analogous result holds for the explicit variety  $\Delta[H(Y)_m,k,m]$ 
associated with the low-degree universal circuit in Section~\ref{shy}.
Similar extensions of  Theorems~\ref{tnnlexpfull} (a), \ref{tequisdit} (a),
 and \ref{tequisdit} (b) to arbitrary characteristics, 
and of Theorems~\ref{tnnlexpfull} (b) and \ref{tequisdit} (c)
to large enough characteristics also hold.

For the proof, we need the following results.

\begin{theorem}[Kaltofen and Lecerf]  \label{tkaltofenpositive}
(cf. \cite{kaltofenlec})
Suppose  $K$ is an algebraically closed field of positive characteristic $p$. Then:

\noindent (a) Given any polynomial $g \in K[x_1,\ldots,x_n]$ 
and a polynomial  $f \in K[x_1,\ldots,x_n]$ dividing $g$, there exists  a nonuniform circuit over $K$, with oracle gates for $g$, of 
$O((n \deg(g))^a)$ size, for some absolute positive constant $a$ not
depending on $n$ or $p$,  that computes the highest power  of $f$ of the form $f^{p^l}$, $l\ge 0$,  that divides $g$.

\noindent (b) In particular, 
given any polynomial $g \in K[x_1,\ldots,x_n]$, with  $\deg(g) < p$, and
a polynomial  $f \in K[x_1,\ldots,x_n]$ dividing $g$, there exists  a nonuniform circuit over $K$, with oracle gates for $g$,  of 
$O((n \deg(g))^a)$ size, for some absolute positive constant $a$ not
depending on $n$ or $p$,   that computes $f$.
\end{theorem}

The following is the analogue of Theorem~\ref{trussell2} in this setting.

\begin{theorem} \label{trussell2pos}
Suppose there exists a family $\{p_m(x_1,\ldots,x_m)\}$ of
exponential-time-computable,  multi-linear, integral 
polynomials  such  that $p_m$
cannot be approximated infinitesimally closely  
over  an algebraically closed field of $\Omega(2^{m^\delta})$ characteristic 
by  circuits  of 
$O(2^{m^\epsilon})$ size,  for some positive constants  
 $\delta$ and $\epsilon$, as $m \rightarrow \infty$.

Then polynomial identity testing  for low-degree 
circuits  of  size $\le s$  over an algebraically closed 
field of $\Omega(2^{(\log s)^a})$ characteristic,  for  a large enough 
positive constant $a$,  has $O(2^{\polylog(s)})$-time-computable 
strengthened black-box derandomization.
\end{theorem}

This is proved like  Theorem~\ref{trussell2}, using
Theorem~\ref{tkaltofenpositive} (b)  in place of Theorem~\ref{tkaltofen}. 
It can be checked that 
this replacement is possible, by choosing the constant $e$ in the proof of Theorem~\ref{trussell2} large enough, depending upon $\epsilon$ and $\delta$.
The analogue of this result also holds for exact computation
in place of infinitesimally close approximation, with a similar proof. 

\noindent {\em Proof of Theorem~\ref{tequisditpos}}:
All results, other than Theorem~\ref{trussell2},
used in the proof of Theorem~\ref{tequisditdet},
 namely, Theorem~\ref{theintz}, Noether's Normalization Lemma
(Lemma~\ref{lnnl0}),  Hilbert's Nullstellensatz,  and other standard facts from algebraic 
geometry hold in  arbitrary  characteristic. 

Hence, 
the proof of (a) is similar to that of Theorem~\ref{tequisditdet} (a).
The proof of (b) is similar to that of  Theorem~\ref{tequisditdet} (b),
using Theorem~\ref{trussell2pos} in place of Theorem~\ref{trussell2}. 
 \qed

\noindent {\em Remark 1:} The restrictions on the characteristics in
Theorem~\ref{tequisditpos} (b) (and its generalizations to arbitrary explicit varieties; cf. 
the remark after Theorem~\ref{tequisditpos})
can be dropped, and we can let the
base field be an algebraically closed field $K$ of any fixed 
characteristic $p$,
if we assume for the $f_n$ therein  that 
$f_n^{p^i}$, for any nonnegative $i=O(\poly(n))$, 
cannot be approximated infinitesimally closely 
by circuits over $K$  of  
$O(2^{n^\epsilon})$ size,  for some constant $\epsilon > 0$,
as $n \rightarrow \infty$.

\noindent {\em Remark 2:}
Theorem~\ref{tmonteexplicit} similarly holds in arbitrary characteristic.
Theorem~\ref{tclosedimage}   also holds in  arbitrary characteristic, since 
Theorem~\ref{tmumfordnew} holds in arbitrary characteristic \cite{mumford}.

\section{Discussion} \label{sdiscuss}
Finally, we discuss the difficulties that need to be overcome  to improve
the current best bound for NNL for $\Delta[\det,m]$
in  Theorem~\ref{tssopverifydet}.

Let $K$ now be an algebraically closed field of characteristic zero.
If  every polynomial  in $\Delta[\det,m]$ had  
a small circuit over  $K$ of $\poly(m)$ size, 
then the strengthened black-box derandomization problem for symbolic determinant identity 
testing would be  in PSPACE unconditionally, 
like the standard black-box derandomization problem (cf. Proposition~\ref{pstandard}), 
with essentially  the same proof. 
By (the proof of) 
Theorem~\ref{tnnldetblack}, NNL for $\Delta[\det,m]$  would then be in PSPACE unconditionally.

However, it may be  conjectured that 
the boundary of the orbit of the determinant 
in $\Delta[\det,m]$ contains points which 
do not have small circuits over $K$; 
cf.   Section 4.2  in \cite{GCT1}
for a preliminary investigation in this direction,
and \cite{GCTtutorial,GMQ} for  further investigation.
Formally, we conjecture 
that  $\overline {\mbox{VP}_{ws}} \not \subseteq \mbox{VP}$. Here 
 VP \cite{valiant2}
is the class of families of polynomials of small degree  having  circuits of polynomial
size, $\mbox{VP}_{ws}$ is the class of families of polynomials that can be computed by symbolic determinants
of polynomial size, 
 and  $\overline{\mbox{VP}_{ws}}$ \cite{burg2,manivel} is 
the class of families of polynomials that can be approximated infinitesimally closely
by symbolic determinants of polynomial size. 

This conjecture is  counter-intuitive, since 
one would have expected the complexity of infinitesimally 
close approximation of multi-linear polynomials by symbolic determinants 
to be polynomially related to that of exact computation.
As pointed out in B\"urgisser \cite{burg2} (cf. Lemma 5.6 (3) and Theorem 5.7 therein), 
this would be the case if every point in the boundary of the orbit of the determinant could be 
approached by a one-parameter deformation of the determinant of polynomial order. We conjecture that this is not the case.
However, for the VNP-complete  polynomials  such as the permanent, 
the complexity of infinitesimally 
close approximation can be conjectured to be polynomially related to that of exact computation.
At present, it is not even known  if  $\overline {\mbox{VP}_{ws}}  \subseteq \mbox{VNP}$, where
VNP \cite{valiant2} is the class of p-definable families of polynomials.

The conjectural   points with large circuit complexity  in $\Delta[\det,m]$ 
constitute  the main  obstacle 
to putting  NNL for $\Delta[\det,m]$ in PSPACE, or even EXP,  unconditionally with the existing techniques. 
(This obstacle is absent for explicit categorical quotients, 
as in Theorems~\ref{tintroexplicitmatrixnew} and \ref{tintroexplicitconstantnew},
 by Theorem~\ref{tmumfordnew} (a).)
In contrast,   the  Generalized Riemann Hypothesis assumption
in the current EXPH-bound in Theorem~\ref{tssopverifydet} may  be removed in 
the foreseeable  future (though, this by itself is a nontrivial problem).

Thus,  bringing  NNL for $\Delta[\det,m]$  from EXPH,
where it is currently assuming the Generalized Riemann Hypothesis, 
to  even EXP {\em unconditionally}  seems difficult with the existing techniques.
Theorem~\ref{tintronnldetnew} says that 
a sub-exponential algebraic  circuit-size lower bound for infinitesimally 
close approximation of the permanent would put NNL for $\Delta[\det,m]$  in  quasi-P.
Theorem~\ref{tintronnldetnew} (and Remark 2 after Theorem~\ref{tnnlexpfull}) 
 may thus explain why  the hardness hypothesis of  geometric complexity theory in \cite{GCT1} 
has turned out to be so difficult. 
(In the terminology above, this
hypothesis  is that $\mbox{VNP} \not \subseteq \overline {\mbox{VP}_{ws}}$; cf. Proposition 9.3.2 in \cite{manivel}.)
It is a reasonable thesis that any realistic approach to the 
$\mbox{VNP} \not \subseteq {\mbox{VP}_{ws}}$  conjecture in Valiant \cite{valiant2}
would also prove  this hypothesis. 
Indeed, 
all known lower bounds for the exact computation of the permanent
also hold for infinitesimally close approximation; eg. see
\cite{ressayre2,GrochowGCTUnifies}.
Hence, Theorem~\ref{tintronnldetnew} may  also explain why the
$\mbox{VNP} \not \subseteq {\mbox{VP}_{ws}}$  conjecture  in
\cite{valiant2} has turned out to be so difficult.

Theorem~\ref{tintronnldetnew} and the equivalence results  in this article (Theorems~\ref{tequivnnlnew} and \ref{tequisdit})
thus reveal that the fundamental  problems of geometry (NNL) and complexity theory 
(hardness) share a common root difficulty, 
namely,  the problem of overcoming the existing   EXPH vs. P gap (assuming the Generalized Riemann Hypothesis)  in the  complexity of NNL
for general explicit varieties,
or rather, the  EXPH vs. NC gap; cf. Remark 1 after Theorem~\ref{tnnlexpfull}.
We call this gap  the {\em geometric complexity theory (GCT) chasm}.
It may be viewed as the common {\em cause and measure}  of the difficulty of these  problems in geometry and complexity theory.

The superpolynomial lower bound in  \cite{GCTmaxflow}
for  additive approximation of 
the maxflow 
in the PRAM model without bit-operations,
which initiated geometric complexity theory (cf. the introduction of \cite{GCT1}), 
assumes special significance in view of this  chasm.
First, this lower bound is the main reason why the hardness hypothesis  
in  \cite{GCT1} is expected to hold, 
despite the conjectural non-containment of  $\overline{\mbox{VP}_{ws}}$ in VP. 
This is because  a lower bound akin to that in \cite{GCTmaxflow} 
for  additive approximation of the permanent of 
integral matrices (instead of the maxflow) implies  the  hardness hypothesis in \cite{GCT1}
for infinitesimally close approximation. Such a lower bound 
can be  expected since, in view $\#P$-completeness  \cite{valiant} of the permanent, the 
approximation of the permanent is expected to be harder than the approximation of
the maxflow.
Second, the lower bound in \cite{GCTmaxflow} is 
the only known arithmetic version of a foundational  conjecture
in complexity theory (in this case, the P $\not =$ NC conjecture) that holds unconditionally
in a natural and realistic model of computation. 
It is now likely to remain the only such lower bound in complexity theory, 
until the GCT chasm is crossed.

We conjecture that the strong form of 
NNL for every  explicit variety is in P, and hence, the GCT  chasm  can  be crossed, 
as suggested by Theorem~\ref{tnnlexpfull}.
By geometric complexity theory, we mean henceforth any approach to cross the GCT chasm using a synthesis of geometry and
complexity theory. One such   approach  will be described in the sequel \cite{GCT6}.

\noindent {\bf Acknowledgement:}
The author is grateful to Michael Forbes and Amir Shpilka   for bringing 
\cite{manindra,amir}
to his attention and for pointing out \cite{fs3} an error in the preliminary version \cite{GCT5focs} of this paper (cf. 
Section~\ref{sprooftech}), and to Jonah Blasiak,  Peter B\"urgisser, Josh Grochow, 
Joseph Landsberg, Nitin Saxena, Jimmy Qiao, and the referees   for helpful comments.

\bibliographystyle{abbrv}
\bibliography{main}

\end{document}